\documentclass{pasa}%
\usepackage{graphicx}
\usepackage{float}
\usepackage{subfigure}
\usepackage[dvipsnames]{xcolor}
\usepackage{pdflscape}
\usepackage{afterpage}
\usepackage{upgreek}
\newcommand\micron{\rm{\upmu m}}
\newcommand\ion[2]{#1$\;${\scshape{#2}}}
\usepackage{multirow}

\title[Mid-IR Spectroscopic Cosmological Surveys]{Mid-IR cosmological spectrophotometric surveys from space: Measuring AGN and star formation at the Cosmic Noon with a SPICA-like mission}

\author[L. Spinoglio et al.]{Luigi Spinoglio$^1$\thanks{E-mail: \textsf{\href{mailto:luigi.spinoglio@inaf.it}{luigi.spinoglio@inaf.it}}}, 
Sabrina Mordini$^{1,2}$, 
Juan Antonio Fern\'andez-Ontiveros$^1$, 
Almudena Alonso-Herrero$^3$, 
Lee Armus$^4$,  
Laura Bisigello$^5$, 
Francesco Calura$^5$,
Francisco J. Carrera$^6$,  
Asantha Cooray$^7$,
Helmut Dannerbauer$^{8,9}$,
Roberto Decarli$^5$,
Eiichi Egami$^{10}$,
David Elbaz$^{11}$,
Alberto Franceschini$^{12}$, 
Eduardo Gonz\'alez Alfonso$^{13}$,
Luca Graziani$^2$, 
Carlotta Gruppioni$^5$, 
Evanthia Hatziminaoglou$^{14}$, 
Hidehiro Kaneda$^{15}$, 
Kotaro Kohno$^{16}$,
\'Alvaro Labiano$^3$,
Georgios Magdis$^{17}$,
Matthew A. Malkan$^{18}$,
Hideo Matsuhara$^{19}$,
Tohru Nagao$^{20}$,
David Naylor$^{21}$,
Miguel Pereira-Santaella$^{22}$,
Francesca Pozzi$^{23,5}$, 
Giulia Rodighiero$^{12}$, 
Peter Roelfsema$^{24,25}$,
Stephen Serjeant$^{26}$, 
Cristian Vignali$^{23,5}$,
Lingyu Wang$^{24}$,
Toru Yamada$^{19,27}$
}

\jid{PASA}
\doi{10.1017/pas.\the\year.xxx}
\jyear{\the\year}

\bibpunct{(}{)}{;}{a}{}{,}
\usepackage{aas_macros}
\usepackage{hyperref}

\hypersetup{draft=True,colorlinks,citecolor=blue,linkcolor=blue,urlcolor=blue}

\usepackage[T1]{fontenc}
\usepackage{ae,aecompl}
\usepackage{newtxtext}
\usepackage{newtxmath}

\begin{document}

\begin{frontmatter}
\maketitle
\textit{(Affiliations can be found after the references)}

{\footnotesize \today \vspace{0.3cm}}

\begin{abstract}
{ We use the SPace Infrared telescope for Cosmology and Astrophysics (\textit{SPICA}) { project} as a template to demonstrate how deep spectrophotometric surveys covering large cosmological volumes over extended fields ($1$--$15\, \rm{deg^2}$) with { a mid-IR} imaging spectrometer ($17$--$36\, \rm{\micron}$) in conjunction with deep $70\, \rm{\micron}$ photometry with { a far-IR camera}, at wavelengths which are not affected by dust extinction can answer the most crucial questions in current galaxy evolution studies. A SPICA-like mission will be able for the first time to provide an unobscured three dimensional (3-D, i.e. \textit{x}, \textit{y} and redshift \textit{z}) view of galaxy evolution back to an age of the Universe of less than $\sim$ 2 Gyrs, in the mid-IR rest-frame.}
This survey strategy will produce a full census of the Star formation Rate (SFR) in the Universe, using Polycyclic Aromatic Hydrocarbons (PAH) bands and fine-structure ionic lines, reaching the characteristic knee of the galaxy luminosity function, where the bulk of the population is distributed, at any redshift up to $z \sim 3.5$. Deep follow-up pointed spectroscopic observations with grating spectrometers { onboard the satellite}, across the full IR spectral range ($17$--$210\, \rm{\micron}$), { would} simultaneously measure Black Hole Accretion Rate (BHAR), from high-ionization fine-structure lines, and SFR, from PAH and low- to mid-ionization lines in thousands of galaxies from solar to low metallicities, down to the knee of their luminosity functions. 
The analysis of the resulting atlas of IR spectra will reveal the physical processes at play in evolving galaxies across cosmic time, especially its heavily dust-embedded phase during the {\it activity peak} at the cosmic noon ($z \sim 1$--$3$), through IR emission lines and features that are insensitive to the dust obscuration.

\end{abstract}

\end{frontmatter}

{\noindent
\bf Preface}
\vspace{0.1cm}

\noindent
{The articles of this special issue focus on some of the major scientific questions that a future IR observatory will be able to address. We adopt the {\it SPace Infrared telescope for Cosmology and Astrophysics (SPICA)} design as a baseline to demonstrate how to achieve the major scientific goals in the fields of galaxy evolution, Galactic star formation and protoplanetary disks formation and evolution. The studies developed for the \textit{SPICA} mission serve as a reference for future work in the field, even though the mission proposal has been cancelled by ESA from its M5 competition.}

The mission concept of {\it SPICA} employs a 2.5\,m telescope, actively cooled to below $\sim$ 8\,K, and a suite of mid- to far-IR spectrometers and photometric cameras, equipped with state of the art detectors \citep{roelfsema2018}. In particular the {\it SPICA} Far-infrared Instrument (SAFARI) is a grating spectrograph with low (R\,$\sim200$--$300$) and medium (R\,$\sim3000$--$11\,000$) resolution observing modes instantaneously covering the $35$--$210\, \rm{\micron}$ wavelength range. The {\it SPICA} Mid-infrared Instrument (SMI) has three operating modes: a large field of view ($10' \times 12'$) low-resolution $17$--$36\, \rm{\micron}$ imaging spectroscopic (R\,$\sim50$--$120$) mode  and photometric camera at $34\, \rm{\micron}$ (SMI-LR), a medium resolution (R\,$\sim1300$--$2300$) grating spectrometer covering wavelengths of $18$--$36\, \rm{\micron}$ (SMI-MR) and a high-resolution echelle module (R\,$\sim29\,000$) for the $10$--$18\, \rm{\micron}$ domain (SMI-HR). Finally, B-BOP, a large field of view ($2'.6 \times 2'.6$), three channel, ($70\, \rm{\micron}$, $200\, \rm{\micron}$ and $350\, \rm{\micron}$) polarimetric camera complements the science payload.

\section{Introduction}
\label{sec:intro}
In the context of current standard cosmological models (e.g. $\Lambda$CDM) the evolution of the large scale structure in the Universe is governed by dark matter, which forms a cosmic network of filaments that connect gravitationally-bound regions, known as dark matter halos \citep[e.g.][]{white1978,bond1996}. The gas density follows, under the effect of gravity, the filamentary large-scale structure (filaments contain $\gtrsim 60\%$ of the gas mass at $z \sim 3$; \citealt{vogelsberger2014,umehata2019}), while baryonic physics regulate the formation and growth of both supermassive black holes (SMBHs) and galaxies inside the halos of this cosmic web \citep{springel2005,somerville2015}. The overall efficiency to produce stars from the available baryonic matter is very low and depends strongly on the mass of the halo, reaching a maximum value of $20$--$30\%$ for halos of $\sim 10^{12}\, \rm{M_\odot}$, typical of Milky Way like galaxies, with a significant drop towards lower- and higher-mass galaxies \citep{behroozi2019,wechsler2018}. Thus, the morphology and characteristics of galaxies are tightly linked to the environment where they formed \citep{blanton2007}. In the present day Universe, massive ellipticals reside preferentially in the innermost parts of galaxy clusters, whereas star-forming galaxies are preferentially found in the clusters outskirts or in the field. This diversity arises from the different formation histories of low and high mass galaxies. 

The extensive photometric surveys in the X-rays, UV, optical, and IR ranges performed over the last two decades have shown a common evolution of the two main energetic processes inside galaxies, i.e. star formation (SF) and black hole accretion (BHA), with a peak between $z\sim1$ and $z\sim3$, in an epoch known as the ``Cosmic Noon'' \citep{madau2014,heckman2014}. By the end of this epoch massive elliptical galaxies and the bulges of Milky Way like progenitors were already in place \citep{perezgonzalez2008}, while the global SF and BHA activity started a steep decline towards the present time. The bulk of galaxies at any epoch show a correlation between their stellar mass and star formation rate. This relation, evolving with redshift, is called ``main sequence" \citep{noeske2007, daddi2007, elbaz2007, rodighiero2011}, which is abandoned by massive galaxies in less than $\sim 1\, \rm{Gyr}$, reducing their SFR by a factor 10 to become members of the ``red and dead'' sequence.
%, while their morphology changed from late- (spirals) to early-types \citep[ellipticals and S0;][]{schawinski2014}. 
In this scenario, energetic feedback from massive stars though stellar winds and supernovae (SNe), and active galactic nuclei (AGN) are required to quench the SF and explain the low efficiency of transforming baryons into stars \citep{behroozi2019}. In particular, current simulations require a strong contribution of feedback from accreting BHs to reconcile the galaxy mass function with that of dark matter halos \citep{silk2012}, and to explain the local scaling relations between the BH masses and host galaxy properties \citep{kormendy2013}. This feedback is mainly ascribed to quasar outflows and jets in AGN, however robust observational evidence probing such scenario at high-z is still missing.

A complete study of galaxy evolution, that includes the history of massive galaxies, has to account for the environment where they originated and thus has to incorporate the large-scale structure, which requires the mapping of large cosmological volumes. This requires deep spectroscopic surveys to locate in space and in time the presence of star forming galaxies and AGN and to assess their activity. During the peak epoch of SF in galaxies, the UV and optical light is absorbed by dust grains and re-emitted in the IR and therefore nearly 90\% of the energy emitted by young stars emerges in the infrared \citep[e.g.,][and references therein]{madau2014}. Observation of the UV rest-frame continuum associated with young massive stars ($1400$--$1700$\AA{}), in spite of being one of the most popular methods to measure the SFR density \citep{lilly1995,madau1996,cucciati2012,bouwens2015}, is highly unreliable in obscured environments such as those at the cosmic noon. { Therefore, rest-frame ultraviolet (UV) and optical observations cannot access these crucial dust-obscured regions where these massive stars form.}  Alternative methods based on massive star tracers, such as gamma-ray bursts (GRBs; e.g. \mbox{\citealt{greiner2015}}) show a strong bias towards subsolar metallicity low-mass galaxies at redshift $z<1$ ($Z \lesssim 0.5\, Z_{\odot}$ and $M_{\star} < 10^{10}\, \rm{M_\odot}$; \mbox{\citealt{vergani2015}}) and miss the dust obscured fraction of the starburst population at redshift $z > 2$ \mbox{\citep{gatkine}}. Similarly, the derivation of the black hole accretion rate (BHAR) in galaxies from X-ray observations is affected by large uncertainties, due to the large bolometric corrections (factors of tens to hundreds) and the large dispersion in their determinations, required to derive the accretion luminosity from the observed X-ray luminosity \citep[e.g.][]{vasudevan2009, lusso2012, ueda2014, duras2020}. Moreover, even hard-X-rays surveys, which directly probe the hot gas around the accreting BH's, have limitations in sensitivity that often miss faint, Compton thick AGN, whose population constitutes $\sim 40\%$ of the AGNs in the local Universe, or underestimating their luminosities \citep{georgantopoulos2019,lambrides2020}.

%IR data, on the other hand, are largely unaffected by dust extinction and can detect the warm dust signature for both obscured and unobscured AGNs \citep{delvecchio2014}, tracing SF in galaxies. 
An alternative determination of the SFR and BHAR densities that does not suffer from the above cited limitations, %are %theThe most reliable current determinations of the SFR and the BHAR densities up to redshift of $\sim 2$ 
was derived from the mid- to far-IR photometric observations of {\it Spitzer} \citep{werner2004} and {\it Herschel} \citep{pilbratt2010} of deep cosmological fields \citep{schreiber2015,delvecchio2014}, based on the bolometric luminosities of galaxies (\mbox{\citealt{lefloch2005}}, \mbox{\citealt{gruppioni2013}}). However, infrared photometric surveys are not immune to limitations or assumptions about the intrinsic spectra, since the measured light is usually reprocessed emission from dust heated by a central power source.
%these estimates have intrinsic limits and have to be taken with caution, 
%because, due to the lack of reliable spectral emission line indicators to measure SFR and BHAR, the physical conditions in galaxies responsible for the observed emission  remain unclear.  %are based on the observed integrated luminosities, while the contributions of SF and BHA were not separated through observed physical quantities, but through modelling of the spectral energy distributions (SEDs), relying on model-dependent assumptions and local templates, with large uncertainties and degeneracies.
Indeed {\it Spitzer}- and {\it Herschel}-based studies have been successful in estimating number counts and galaxy luminosity functions (LF) \citep{lefloch2005,perezgonzalez2005,rodighiero2010a,oliver2012,gruppioni2013,magnelli2013,lapi2011}, { but only through statistical techniques applied to the bulk of the galaxy population as a function of redshift. Here we refer to the assumptions that are necessarily made to derive from the observed number counts at some specific wavelength (e.g., 24 $\mu$m for {\it Spitzer}, 100 and 160$\mu$m for {\it Herschel}, to give two examples) a well defined luminosity. This in turn is determined by the assumed spectral energy distribution of galaxies and the measure of the galaxy redshift. The luminosities associated to a galaxy population are then used to build up the luminosity functions, well defined only for complete statistical samples, and therefore the LFs still suffer quite significant uncertainties.} 

Only wide-area spectroscopic surveys, immune to dust, are the key to advancing the field. The mid- to far-IR spectral range contains a great variety of ionic, atomic, molecular lines, as well as  continuum features that can be used as spectral diagnostic tools in dust-obscured environments \citep{spinoglio1992, armus2007, nardini2010, imanishi2011}. In particular, the mid- to far-IR ionic fine-structure lines, besides being significantly less affected by extinction than UV, optical or near-IR lines, are characterised by having their emissivities only weakly dependent on the electron temperature, making them very good tracers of the gas phase physical conditions. {This is a consequence of the atomic level energies involved in the IR fine-structure transitions, which are much closer to the ground state with respect to optical and UV transitions. For instance, the emissivity of IR lines such as [\ion{Ne}{ii}]${\rm 12.8 \mu m}$, [\ion{Ne}{iii}]${\rm 15.6 \mu m}$, [\ion{O}{iii}]${\rm 52,88 \mu m}$, [\ion{O}{iv}]${\rm 25.9 \mu m}$, or [\ion{N}{ii}]${\rm 122,205 \mu m}$ changes at most by a factor $\lesssim 2.7$ in the $3\,000 < T_{\rm e} < 50\,000\, \rm{K}$ and $10 < n_{\rm e} < 1000\, \rm{cm^{-3}}$ ranges, where $T_{\rm e}$ and $n_{\rm e}$ are the electron temperature and density  (based on \textsc{pyneb} estimates; \citealt{luridiana2015}), while the intensities of optical transitions such as [\ion{N}{ii}]${\rm \lambda \lambda 6548,6584}$ or [\ion{O}{iii}]${\rm \lambda \lambda 4959,5007}$ show variations of a factor 270 and 2800, respectively, that is $100$--$1000$ times larger for the same temperature and density ranges. The strong dependence of the optical emissivities with the temperature becomes a major source of uncertainties when the nebula show temperature stratification or inhomogeneties that are not spatially resolved \citep[e.g.][]{vermeij2002,dors2013}, which is typically the case in galaxies observed at high-$z$}. 

Using IR spectroscopy of these lines, we can therefore physically separate the processes due to stellar formation and evolution from those due to AGN accretion, because of the different energies involved in line ionization and excitation processes (\citealt{ho2007,tommasin2010,zhuang2019,xie2019} for SFR and \citealt{rigby2009, tommasin2010, diamondstanic2012} for BHAR). An IR spectroscopic investigation will lead to a complete census of SF and obscured BHA in galaxies from low-mass/low-metallicity galaxies to heavily obscured high-mass starburst galaxies. This will establish a new reference point for cosmological models providing access to obscured star formation, i.e. the most fundamental key for galaxy formation theories. Indeed, the rest-frame mid- to far-IR spectrum of each observed galaxy will allow us to measure directly redshifts, SFRs, BHARs, metallicities and physical properties of gas and dust in galaxies, thus changing our current knowledge of galaxy evolution and, in particular, metallicity evolution, which could be heavily biased from the available observations of only the UV and visible lines, as well as optical and UV photometry.

A description of the power of rest-frame mid-IR spectroscopy for galaxy evolution studies with { the SPICA project} can be found in \citet{spinoglio2017}, while a study of the SPICA mid-IR photometric surveys can be found in \citet{gruppioni2017}.
{ In this paper, we present the assessment of the mid-IR cosmological spectrophotometric surveys that has been prepared for the SPICA mission. Because the SPICA proposal has been cancelled by ESA from the Cosmic Vision 2015-2025 M5 competition, we consider in this article the SPICA mission as a template case that in the future could be accomplished by a new IR observatory project, such as the {\it Origins Space Telescope} (OST), currently under study by NASA \citep{meixner2019}. 
}
In this paper, we follow up on the work presented in \citet{kaneda2017} and \citet{gruppioni2017}, which describe deep and ultra-deep SPICA spectrophotometric and photometric surveys, respectively, with the SPICA Mid-IR Instrument SMI \citep{kaneda2016}. These papers are focused on the detection of the mid-IR features from the polycyclic aromatic hydrocarbons (PAHs) in SF galaxies and on the rest-frame mid-IR continuum in AGN, respectively. Besides the bright PAH features, here we also consider the detection of the mid-IR rest-frame ionic fine structure lines in both solar and low-metallicity SF-dominated galaxies and AGN, adopting the same two surveys approach presented in \citet{kaneda2017}. Moreover, we assess the feasibility of pointed observations with SMI ($17 < \lambda < 37\, \rm{\micron}$) using the medium resolution mode (R\,$\sim\,1300$--$2300$) and SAFARI at longer wavelengths ($35 < \lambda < 210\, \rm{\micron}$) in the low resolution mode. We also include in this paper a new calibration for both the PAH features and the ionic fine structure lines to the total IR luminosity, which is derived from observations of samples of active, starburst and dwarf low-metallicity galaxies in the local Universe.

Throughout this paper, we adopt the following cosmological parameters: $H_{0} = 67.8\, \rm{km\,s^{-1}\,Mpc^{-1}}$, $\Omega_{M} = 0.3$ and $\Omega_{\Lambda} = 0.7$ \mbox{\citep{planck_col_2016}}.

\section{SPICA cosmological surveys}\label{sec:strategy}

Through its SMI spectrophotometric surveys, {SPICA would have uniquely supplied} for the first time a full census of the hidden SF and BHA over 90\% of cosmic time in all environments, from voids to cluster cores. In order to reach this goal, the SPICA %will be used to obtain:
observing strategy of cosmological fields { would have adopted} the following sequence of observations and sample definition: %aimed at the determination of the basic astrophysical quantities in galaxies through rest-frame IR spectroscopic observations, along their evolution in the last $12\, \rm{Gyrs}$, at redshifts up to $z \sim 4$. This sequence can be summarized as follows:

\begin{enumerate}
\item Deep and ultra-deep spectrophotometric surveys with SMI at wavelengths of 17-36\,$\mu$m and photometric surveys with the {\it SPICA} B-BOP  photopolarimetric camera \citep{andre2019} at $\lambda$=70\,$\mu$m, over large cosmological volumes of 1\,deg$^2$ and 15\,deg$^2$, respectively.

\item Definition of a sample of galaxies based on the outcome of the above surveys. From the detected spectral features and lines in the low-resolution SMI spectra, %and the slope of the $34-70\, \rm{\micron}$ continuum from the photometric data, 
we { would have been able} to measure the redshifts and SFRs for \textit{tens to hundreds thousand galaxies} (see \mbox{\citealt{kaneda2017}}) at the cosmic noon. %For those galaxies already known by 2030, 
Thanks to the coming extragalactic surveys, the new ancillary multi-frequency data at shorter (near-IR) and longer (submillimeter and millimeter) wavelengths { would} be used %together with the SPICA data 
to fully characterize them in terms of stellar and dust masses and bolometric luminosities.

\item Pointed spectroscopic follow-up observations with SAFARI and SMI at medium- to high-spectral resolution ($R\sim 300$--$11000$) of thousands of galaxies, drawn from the sample defined above, { would} allow a complete characterization of the physical processes driving the evolution of these galaxies. %\textit{SPICA} will establish the connection between BHAR, the impact of massive and energetic outflows, and the chemical evolution of galaxies.
\end{enumerate} 

In this paper we describe the \textit{SPICA} observing strategy by: (1) assessing the potential of the low-resolution spectroscopic SMI surveys with the predicted detections of solar and low-metallicity SF and AGN in galaxies (see Sections \ref{sec:help_results} and \ref{sec:lumfun_results}); (2) outlining how a sample, suitable to study galaxy evolution at its cosmic noon, can be defined (see Sect.\,\ref{sec:sample}) and (3) assessing how the SAFARI and SMI follow-up pointed observations (see Sect.\,\ref{sec:safari}) would be able to measure SFR and BHAR of galaxies of lower luminosities, reaching the knee of the luminosity and mass functions, at various redshifts. { Reaching the knee of the luminosity function includes 50\% or more of the total IR light of galaxies up to a redshift of $z$ = 2.5 and slightly less above (at $z$=3: 48\% and at $z$=4: 46\%), while going to $+1\, \rm{dex}$ above the knee means losing 55-60\% of the light and, conversely, going $-1\, \rm{dex}$ fainter would add 60-70\% of the remaining total luminosity. Because the SFR is proportional to the total IR luminosity, these numbers can also be applied to the SFR fractions. For what concerns the mass function, if we limit the study to the Main-Sequence (M.-S.) galaxies, the same proportions will be maintained.}

In order to proceed with the above assessment,  first of all, we have revised, by recomputing them with the available observations in the Local Universe, the calibrations, with respect to their total IR luminosities, of a number of mid-IR fine-structure emission lines as well as a few bright PAH spectral features (see Section\,\ref{sec:line_calibrations}).

Then we have used the photometric IR catalogue which includes the deepest {\it Herschel} observations of galaxies, the {\it Herschel} Extragalactic Legacy Project (HELP; \citealt{vaccari2016,malek2020,shirley2019}) as input catalog to count how many galaxies, which have been detected photometrically with {\it Herschel}, would be detected spectroscopically with the SMI deep and ultra-deep spectrophotometric surveys (see Section\,\ref{sec:help}), using the calibrations of Section\,\ref{sec:line_calibrations}.  

Finally, we have made predictions of the expected number of galaxies that \textit{SPICA} would be able to detect with the SMI spectrophotometric surveys at different redshifts from the IR luminosity functions derived from the {\it Herschel} far-IR surveys (see Section\,\ref{sec:counts}), using again the line calibrations presented in Section\,\ref{sec:line_calibrations}.

\subsection{Calibrations of the IR lines to the total IR luminosities}\label{sec:line_calibrations}

Following the calibration of the IR lines presented in \citet{spinoglio2012, spinoglio2014}, we have chosen the fine-structure lines discussed in this paper, but limited to the rest-frame spectral range from $10$--$35\, \rm{\micron}$, and added to these the [\ion{Ne}{vi}] line at $7.7\, \rm{\micron}$. We limited to this particular spectral range because it is the most suitable to study the behaviour of galaxies at the cosmic noon, i.e. at redshift of $z = 1$--$3$, and it is the best range where we can easily discriminate the different ionizing continua from AGN and SF galaxies, from solar to subsolar metallicities, and thus quantify the proportions of each emission line produced by each mechanism \citep[see, e.g., the case of NGC\,1068 in ][]{spinoglio2005}.

We include a new calibration for low metallicity galaxies based on the {\it Herschel} Dwarf Galaxies Sample (DGS) catalogue \citep{madden2013,cormier2015}, which includes galaxies with the largest range in metallicities achievable in the local Universe ($\sim$7 $\lesssim$ 12 + log O/H $\lesssim$ 8.5) {to extend the parameter space including sub-solar metallicities. This is motivated by the detection of massive galaxies ($\sim 10^{10}\, \rm{M_\odot}$) in optical surveys with sub-solar metallicities during the cosmic noon, which apparently experience a fast chemical evolution above $z \gtrsim 2$ (\citealt{maiolino2008,mannucci2009,troncoso2014,onodera2016,sanders2020a}). These changes are expected to have an impact on the ISM structure of high-$z$ galaxies, favouring stronger radiation fields \citep[e.g.][]{steidel2016,kashino2019,sanders2020b} and a more porous medium similar to that observed in dwarf galaxies \citep{cormier2019}.}

%possible apparently fast decrease in significantly below the at high redshift, {\color{red} There are in fact some observations of galaxies with sub-solar metallicities at $z \gtrsim 2.2$ to $3$ \citep[e.g.][]{onodera2016}, which however need further study to verify their representativeness.[referee.4]} Such observations would therefore suggest a drop in the metallicity with redshift, at least as measured by optical lines.

%the $\sim 0.8\, \rm{dex}$ drop in metallicity observed in galaxies from $z \gtrsim 2.2$ to $3$ \citep[e.g.][]{onodera2016}. Thus, a low-metallicity ISM could be more representative to predict the line fluxes of galaxies at high-z.

A more general analysis of a revised calibration of all brightest fine-structure lines will be presented in a specific paper (Mordini et al. 2021, in prep.). In this work, which specifically addresses the observing feasibility with the \textit{SPICA} spectrometers, we considered the fine structure lines, in order of wavelength: [\ion{Ne}{vi}] at $7.7\, \rm{\micron}$,  [\ion{S}{iv}] at $10.51\, \rm{\micron}$, [\ion{Ne}{ii}] at $12.81\, \rm{\micron}$, [\ion{Ne}{v}] at $14.32\, \rm{\micron}$, [\ion{Ne}{iii}] at $15.56\, \rm{\micron}$, 
[\ion{S}{iii}] at $18.71\, \rm{\micron}$, [\ion{Ne}{v}] $24.32\, \rm{\micron}$, [\ion{O}{iv}] at $25.89\, \rm{\micron}$, 
[\ion{S}{iii}] at $33.48\, \rm{\micron}$, [\ion{Si}{ii}] $34.81\, \rm{\micron}$. For an early discussion of the power of the IR fine-structure lines to trace AGN and star forming galaxies, we refer to the work of \citet{spinoglio1992}. 	

\begin{figure}
    \includegraphics[width=\columnwidth]{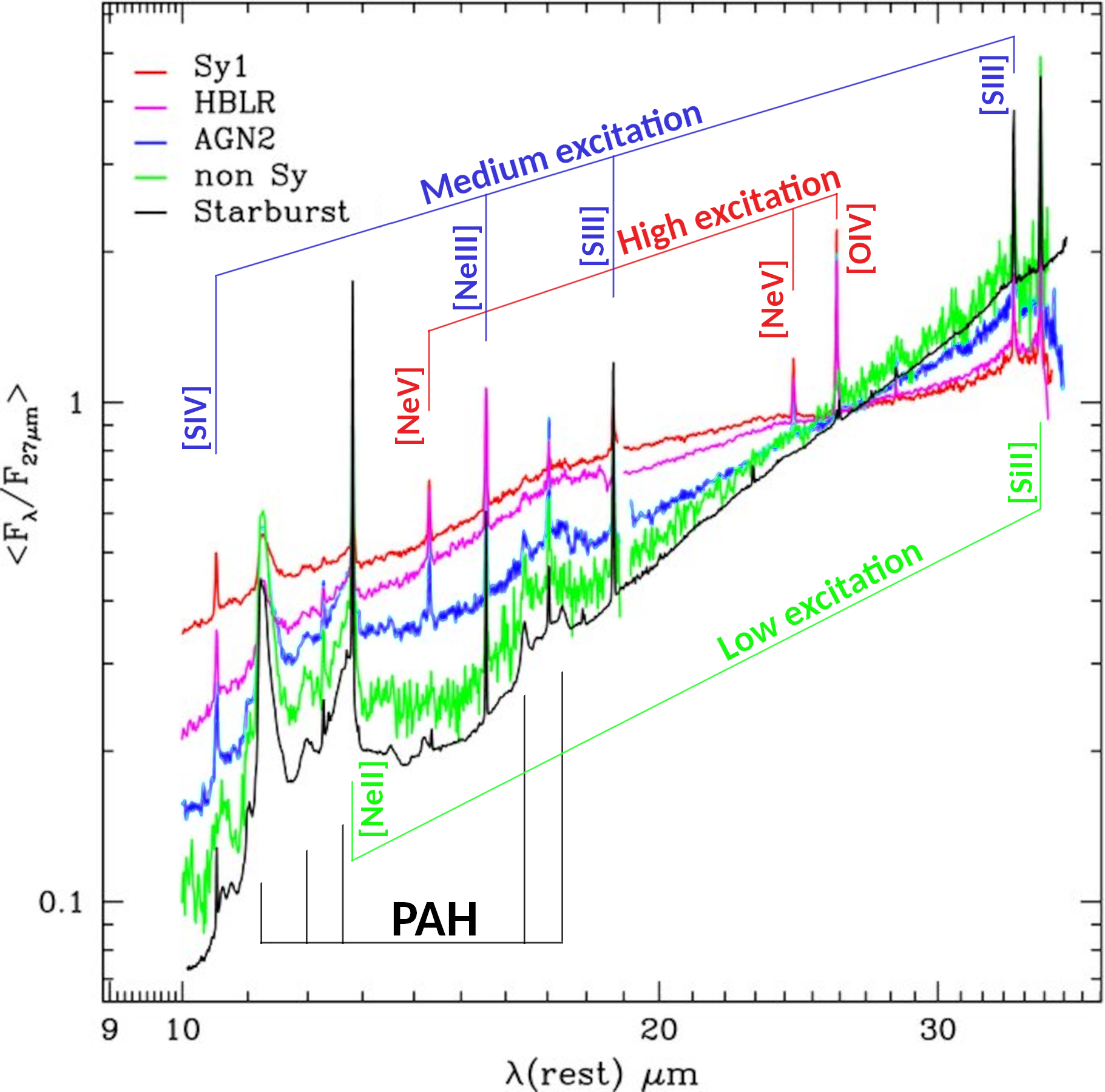}
    \caption{Mid-IR rest-frame spectra of different galaxy types, in a sequence from high to zero nuclear activity from the AGN, from type 1 Seyfert galaxies, through Hidden Broad Line Region Galaxies (HBLR), AGN type 2, weak active galaxies (non Sy) to Starburst galaxies as observed in the local Universe from the Spitzer IRS high resolution (R$\sim$600) spectrometer. The black line shows the average spectrum of the starburst galaxies \citep{bernardsalas2009}, showing the strong PAH features. The figure has been adapted from \citet{tommasin2010} and the positions of the brightest fine-structure lines (divided in low-, medium- and high-excitation) and features are indicated.}\label{fig:spitzer_spectra}
\end{figure}

Moreover, we included five PAH \citep{puget1989} spectral features  ($6.2\, \rm{\micron}$, $7.7\, \rm{\micron}$, $8.6\, \rm{\micron}$, $11.25\, \rm{\micron}$, and $17\, \rm{\micron}$, that are sensitive tracers of the star formation rate \citep{wu2005, brandl2006, smith2007}.  These latter features are attributed to C-C stretching and C-H bending modes and are very luminous and very broad, with extended Lorentzian wings, well matching the spectral resolution of the SMI-LR ($R \sim 50$--$120$) spectrometer,  thus allowing the detection of very faint distant galaxies, their classification and measure of the redshift. Moreover, PAHs are characteristic of photo-dissociation regions (PDRs; \citealt{lutz2008,shipley2016}). %, and provide a useful tool to trace SF activity
{ Recent detailed models of PAH emission by \citet{draine2020} explored the sensitivity of PAH emission spectra to the illuminating starlight, ranging from the bright far-UV starburst spectra to the red spectra of evolved stellar populations, to the PAH size distribution, and to the PAH charge distribution. These studies will be essential for the correct interpretation of the future observations at high redshift.}
We notice here that PDRs by themselves are an important probe of the gas in galaxies, because all neutral atomic hydrogen gas and a large fraction of the molecular gas in the Milky Way and in external galaxies lie in these regions, which are the origin of most of the nonstellar infrared and the millimeter CO emission from a galaxy \citep{hollenbach1997}. The PAH features in star forming galaxies also correlate with CO emission, so they can be used also as molecular gas tracers \citep{cortzen2019}. 
{ Furthermore, \citet{mckinney2020} showed that the study of star formation at high redshift could benefit from the use of multiple tracers, such as the combination of the mid-IR PAH features, tracing photoelectric heating in PDR regions, and [CII]158$\mu$m emission, a major ISM coolant, which can be detected with ALMA at redshift above z$\sim$1.2. Moreover, \citet{kirkpatrick2017} studied how to identify star forming galaxies, AGN and composite AGN+SF galaxies through {\it JWST}-MIRI color-selection techniques at z$\sim$1-2, covering the 6.2 and 7.7 $\mu$m PAH features and the 3–5 $\mu$m stellar minimum, which are robust tracers of star formation. They also show that the 7.7$\mu$m feature needs to be corrected for AGN contamination, before converting to an SFR.}
We show in Fig. \ref{fig:spitzer_spectra}  the average spectra of different galaxies types, from AGNs to Starburst galaxies, showing a sequence of decreasing nuclear non-thermal activity \citep{tommasin2010}, where the most relevant IR lines and features are indicated. 

The calibration of the different IR fine structure lines and features has been computed using mid-IR spectroscopic observations, mainly from the Infrared Spectrograph (IRS) \mbox{\citep{houck2004}} onboard {\it Spitzer}, but also from the Infrared Space Observatory (ISO; \mbox{\citealt{kessler1996}}) Short Wavelength Spectrometer (SWS; \mbox{\citealt{degraauw1996}}).

In particular, for SF galaxies we have selected both the sample of ``classical starburst galaxies'' observed by \citet{brandl2006} and \citet{bernardsalas2009} from which we have used the PAH spectral features and the fine structure lines, respectively, and the sample of star forming galaxies contained in \citet{goulding2009}, which is based on the 60$\mu$m Galaxy Sample \citep{sanders2003}, limited to distances below 15 Mpc for the fine-structure lines. For AGN we considered the complete 12\,$\mu$m selected active galaxies sample \citep{rms1993}, which is the brightest sample of Seyfert galaxies in the local Universe and was systematically observed by the {\it Spitzer} IRS spectrometer at low spectral resolution from \citet{gallimore2010} from which we derived the PAH emission fluxes and at high spectral resolution from \citet{tommasin2008,tommasin2010} from which we have taken the fluxes of the fine structure lines.
Data from \citet{sturm2002} were considered only for the [\ion{Ne}{vi}] line at $7.65\, \rm{\micron}$. Lastly, to compute the calibrations of the fine-structure lines in low metallicity galaxies, we have used the data from the DGS \citep{madden2013}, which have been published in \citet{cormier2015}.
The total IR luminosities (L$_{8-1000\mu m}$) of the selected galaxies have been taken from the cited papers. For the objects without a given luminosity, we calculated L$_{IR}$ using the IRAS point source catalog fluxes \mbox{\citep{iras1984}} and the formula by \mbox{\cite{sanders1996}}. 

We have derived a linear correlation between the logarithms of the total IR luminosity of the galaxies and the luminosity of the lines using the least square method, adopting the following equation:

\begin{equation}
\log(L_{\rm line}) = (a \pm \delta a)\log(L_{\rm IR}) + (b \pm \delta b)
\end{equation}

In the following relations, the luminosities are expressed in units of $10^{41}\, \rm{erg\,s^{-1}}$.
We report the best-fit parameters for the revised line calibrations in Table\,\ref{tab:linecal}, where for each linear relation we give the number of objects \textit{N} used to calculate the linear regression, and the regression coefficient \textit{r}.

An analysis of the IR lines and features calibrations of Table\,\ref{tab:linecal} shows that: 
\begin{itemize}
\item by far the brightest features are PAH features especially in star forming galaxies \citep[see, e.g.,][]{wu2005}. As shown by the normalization constants in Table  \ref{tab:linecal}, their equivalent widths are typically 10 times higher than the strong fine-structure lines.

%, but also to some extent in AGN. %To be noticed that the angle of the correlation is relatively flat for the shortest wavelength features (for the PAH at $6.2$, $7.7$ and $8.6\, \rm{\micron}$, a $\sim 0.5$--$0.7$);
\item the mid-ionization fine-structure lines in AGN are all of about the same brightness (and their luminosities increase linearly with the total IR luminosity) compared to solar-like star forming galaxies, while the higher ionization lines of [\ion{O}{iv}]$\rm 25.9\micron$ and the [\ion{S}{iv}]$\rm 10.5\micron$ lines are five times and more than one order of magnitude brighter, respectively. The mid-ionization lines are and significantly brighter, and more prominent with respect to the continuum, in low metallicity galaxies due to: \textit{i)} the stronger ionisation spectrum from massive stars in the $30$--$50\, \rm{eV}$ range at low metallicities \citep[see discussion in][]{fernandez2016}; and \textit{ii)} the lower dust content which results in a fainter continuum emission. The [\ion{Ne}{vi}]$\rm 7.65\micron$ and the [\ion{Ne}{v}]$\rm 14.3\micron$ and $\rm 24.3\micron$ are only present in AGN, because of the harder ionizing spectra of these objects.
\item a comparison of the AGN and low metallicity galaxies (LMG) shows that the [\ion{S}{iv}]$\rm 10.5\micron$ and the [\ion{Ne}{iii}]$\rm 15.6\micron$ lines are more than one order of magnitude brighter in LMG compared to AGN.
\end{itemize}
The above comparisons show that the differences in the ionizing spectra among the three classes of galaxies are very significant and therefore they can be easily discriminated by adequate line ratio diagrams, like those suggested in \citet{fernandez2016} --\,the so-called IR BPT diagrams.

\subsection{Predictions of SMI spectroscopy of the deepest \textit{Herschel} photometric fields}\label{sec:help}
As anticipated in Section \ref{sec:strategy}, we have exploited the IR photometric catalog of the {\it Herschel} Legacy Extragalactic Program \citep[HELP, ][]{vaccari2016}, as input data to determine how many galaxies, that have been detected with the {\it Herschel} photometric surveys, { would} be detected with the SPICA SMI spectroscopic surveys at each redshift and luminosity. 

In particular, we have used the data already analyzed with the Code Investigating GALaxy Emission (CIGALE; \mbox{\citealt{cigale}}), which are available on the HELP (\footnote{available at: \url{https://herschel-vos.phys.sussex.ac.uk}}) service with estimated total IR luminosity, stellar mass and SFR, for a total of about 15,000 objects selected in the whole COSMOS field \citep{scoville2007}, covering an area in the sky of $2\, \rm{deg^2}$. All the selected sources have 24$\mu$m fluxes from MIPS \citep{rieke2004} onboard  \textit{Spitzer}, 100$\mu$m and 160$\mu$m fluxes from \textit{Herschel}-PACS \citep{poglitsch2010} and 250$\mu$m fluxes from \textit{Herschel}-SPIRE \citep{griffin2010}. Because the parent sample of our selection of HELP galaxies is the COSMOS field, the homogeneity completeness is ensured.

Starting from the total infrared luminosity, we determined the expected line luminosity for each of these objects through our calibrations and whether they would be detected in the ultra-deep and deep SMI surveys. We note here that the use of the HELP catalog has two limitations: first of all, the estimates will suffer the completeness limits of the HELP catalog, therefore they will be biased against the low luminosity galaxies that {\it Herschel} did not detect: secondly, there is a bias against high luminosity galaxies, which could have been missed because of the limited area of the considered {\it Herschel} surveys. Nevertheless, the use of the HELP catalog is valuable because it is based on the knowledge of the galaxies as a function of cosmic time that we do have now from {\it Herschel} and at the same time we can demonstrate how many of the galaxies detected photometrically by {\it Herschel} { could} be detected spectroscopically by a { SPICA-like observatory}.
The results of this analysis are presented in Section\,\ref{sec:help_results}.

\subsection{Predictions from the far-IR galaxy luminosity functions}
\label{sec:counts}

In this section, we outline the method used to make the predictions of spectral lines and features from the luminosity functions. 
First, we compute the number of galaxies in a given sky area for each redshift interval we want to probe, namely from $z = 0.5$ to $z = 4$, using the IR luminosity functions. Then, through the calibrations of the IR fine structure lines and PAH features in the mid-IR (see Section \ref{sec:line_calibrations}), we compute the line emission luminosity functions from the IR continuum luminosity functions. The results about the line/feature detectability have been obtained from these derived line luminosity functions and assuming the sensitivity of the SMI spectrometer at the observed wavelength \citep{kaneda2017}. This approach follows the one already presented in \citet{spinoglio2012} and  \citet{gruppioni2016}, but with the new calibration of the IR lines, updated galaxies luminosity functions, as well as updated characteristics of the SPICA focal plane instruments.

Following the work presented in \mbox{\citet{kaneda2017}} and in \mbox{\citet{gruppioni2017}} our reference surveys, besides the two SMI/B-BOP spectrophotometric surveys presented there, include also a hyper-deep survey down to the confusion limit expected at 34$\mu$m for a 2.5\,m size telescope (in the range of 2-3$\mu$Jy at 3$\sigma$) (see Table \ref{tab:surveys} for the details): 
\begin{enumerate}
\item a hyper-deep SMI survey over an SMI-LR field of  10'$\times$ 12' (= 120 arcmin$^2$ = 0.033 deg$^2$)  down to the spectroscopic flux limit of $\sim\,4.6 \times 10^{-21}\, \rm{W\,m^{-2}}$  at 34\,$\mu$m,  corresponding to 0.5$\mu$Jy in the continuum (at 5$\sigma$); 
\item an ultra-deep SMI survey over a field of  $1\, \rm{deg^2}$ down to the spectroscopic flux limit of $2.8 \times 10^{-20}\, \rm{W\,m^{-2}}$ at 34\,$\mu$m,    corresponding to 3$\mu$Jy in the continuum (at 5$\sigma$), 
complemented with a B-BOP photometric survey at $70\, \rm{\micron}$ at a flux limit of $\sim$ $60\, \rm{\mu Jy}$ (at 5$\sigma$); 
\item a wide field deep survey over $15\, \rm{deg^2}$ to the spectroscopic flux limit of $1.2 \times 10^{-19}\, \rm{W\,m^{-2}}$  at 34\,$\mu$m (at 5$\sigma$), corresponding to 13$\mu$Jy in the continuum (at 5$\sigma$),complemented with the B-BOP  observations at $70\, \rm{\micron}$ at a flux limit of $\sim$ $100\, \rm{\mu Jy}$ (at 5$\sigma$). 
\end{enumerate}

\begin{table*}
\caption{SMI/B-BOP Spectrophotometric surveys parameters}
\begin{center}
\footnotesize
\begin{tabular}{lrrr}
\hline\hline
Parameters & Deep & Ultra-deep & Hyper-deep \\
\hline
Survey area (deg$^2$)  & 15  & 1 & 0.033 (10'$\times$ 12') \\
\hline
Number of SMI fields (10'$\times$12') & 450  & 30 & 1 \\
%Limiting flux at 34$\mu$m (5 $\sigma$)                  &  13\,$\mu$Jy  &  3\,$\mu$Jy    \\
%Limiting line flux at 24$\mu$m (5 $\sigma$) (10$^{-19}$ W m$^{-2}$)		& 7.1 &  1.2 & 0.39 \\
Limiting cont. flux at 34$\mu$m (5 $\sigma$, R = 2.5) ($\mu$Jy)		& 13 &  3.0 & 0.5 \\
%SMI exposure per source                                       & 74\,s & 28.2\,min & 4.9\,hrs. \\
SMI Time per field  (hr.)                                      & 1.07            & 20.2 & 260.6\\
Total SMI integration time             (hr.)               & 475.5               & 605.0 & 260.6 \\
Total SMI time (including overheads)   (hr.)       &  483.0              & 605.0 & 260.7\\

\hline
Number of B-BOP fields ($2.6'\times2.6'$)               & 7935             & 529 &  --- \\
Limiting flux at 70$\mu$m (5 $\sigma$) ($\mu Jy$)  &  100                & 60 & ---\\
B-BOP Time per field  (min.)                                     & 3.46                & 9.41 & ---\\
Total B-BOP integration time                    (hr.)          & 458.0                 & 83. &--- \\
Total B-BOP time (including overheads)   (hr.)         &  482.0                  & 102. & ---\\

%On-source time (sec) & 82  & 1071 \\
\hline\hline
\end{tabular}
\end{center}
\label{tab:surveys}
\end{table*}

Predictions based on the {\it Planck} proto-clusters Luminosity Function evolutionary model \citep{negrello2017} indicate that proto-clusters \citep[see, for a definition, e.g.][]{sunyaev1972}, the rarest source population, would be detected in abundance in the \textit{SPICA} ultra-deep survey over $1\, \rm{deg^2}$ with SMI, from hundreds to thousands of protocluster members. However, to be able to fully assess the effects of clustering and environment on galaxy evolution, we also planned the $15\, \rm{deg^2}$ survey, at a spectroscopic depth that will allow the determination of the redshifts through the strong PAH features.

The number of sources expected to be detected in the wide-field SMI spectral surveys, and their nature, has been estimated first in each $10^{\prime}\times 12^{\prime}$ field of view of the SMI instrument in low spectral resolution mode, as a function of redshift. We adopted the luminosity functions derived by \citet{wang2019} %\citet{gruppioni2013} 
from the {\it Herschel} far-IR surveys, in which the whole population of galaxies is considered and no attempt to classify the galaxies in terms of different populations is performed. 
From the luminosity function, we derived the expected number of objects per $\rm{Mpc^3}$ in every luminosity interval, for each redshift. Then, from each of the three different line calibrations (see Section\,\ref{sec:line_calibrations}) for AGN, Star Forming galaxies and Low Metallicity galaxies, we predicted the expected flux of each spectral line or feature for a given IR luminosity and redshift and therefore the fraction of the total population of galaxies that we would be able to classify in the three classes of galaxies considered.

\begin{figure*}
  \centering
  \includegraphics[width=0.34\textwidth]{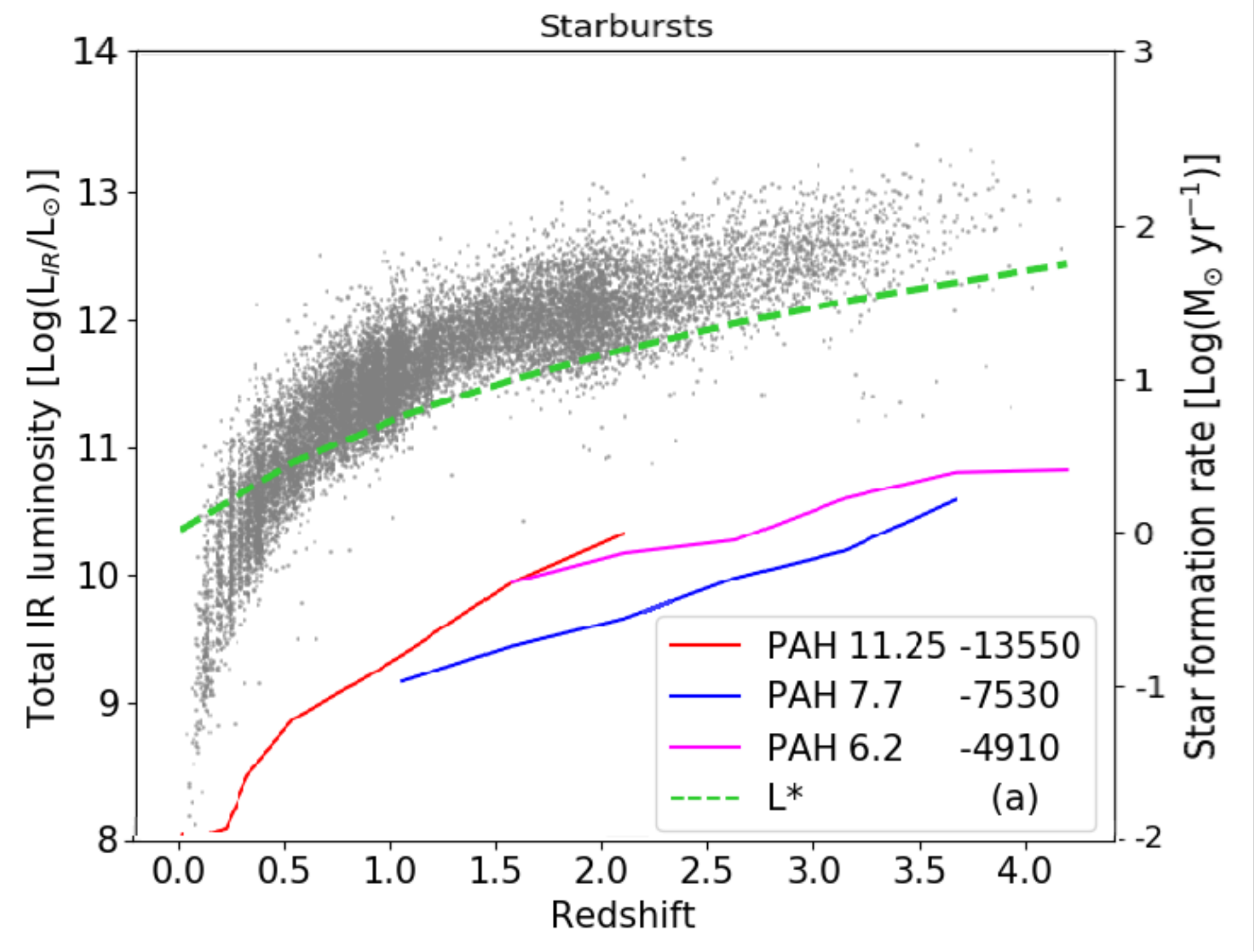}~
  \includegraphics[width=0.32\textwidth]{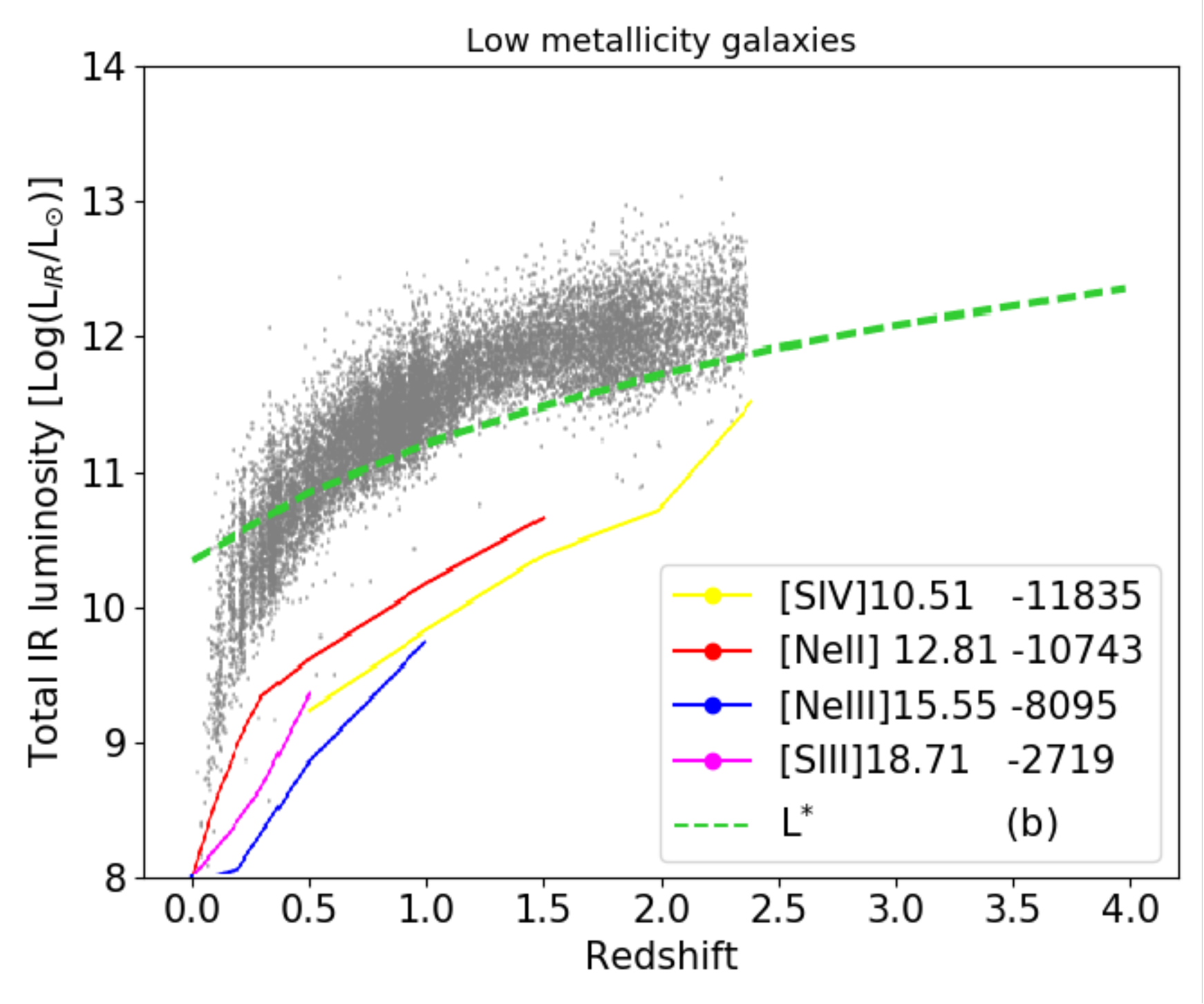}~
  \includegraphics[width=0.32\textwidth]{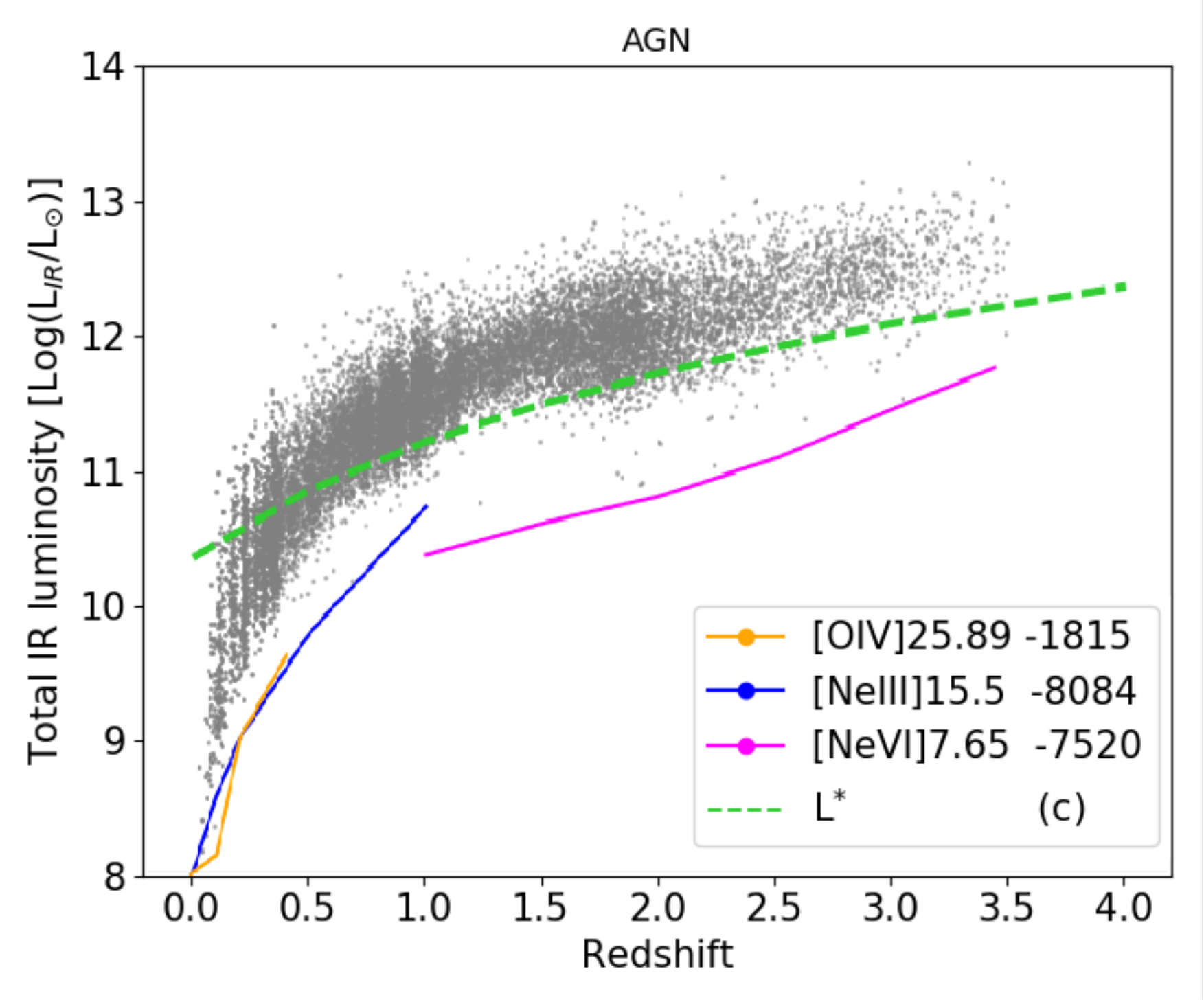}
  \caption{Redshift-luminosity diagrams simulating the SMI 120~arcmin$^2$ (0.033 deg$^2$) hyper-deep survey using the HELP database. In all three panels, the grey dots represent the whole galaxy population as detected photometrically by {\it Herschel}, while each continuum coloured line represent the detection limit of SMI at the given line or feature, as indicated by the legend, which also gives the total number of detections in that particular line or feature. The green broken line shows the knee of the luminosity function, as a function of redshift for a $10^{10.7}\, \rm{M_\odot}$ galaxy in the Main-Sequence (M.-S.; \citealt{scoville2017}). {\bf (a: left):} SF galaxies detectable with SMI in the PAH features. On the right-hand y-axis the SFR correspondent to the total L$_{IR}$ is also given, adopting the conversion factor from \citet{kennicutt2012}. {\bf (b: center):} SF galaxies, adopting the line calibration of low-metallicity galaxies ($\sim 1/5\, \rm{Z_\odot}$), detectable with SMI in intermediate ionization  fine-structure lines. {\bf (c: right):} AGN detectable with SMI in high-ionization fine-structure lines.}\label{fig:z_lum_help_hyperdeep}
\end{figure*}

\begin{figure*}
  \centering
  \includegraphics[width=0.34\textwidth]{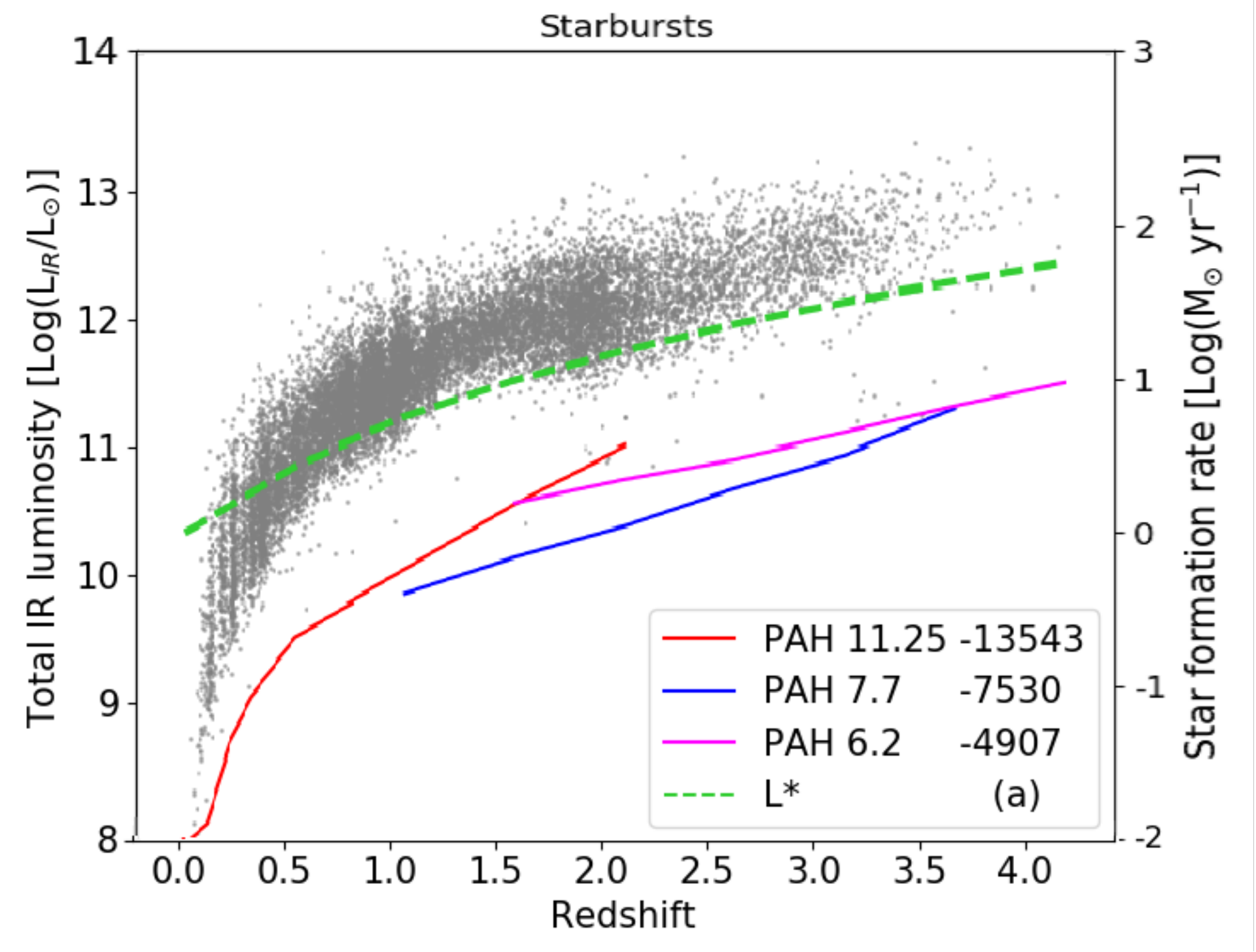}~
  \includegraphics[width=0.32\textwidth]{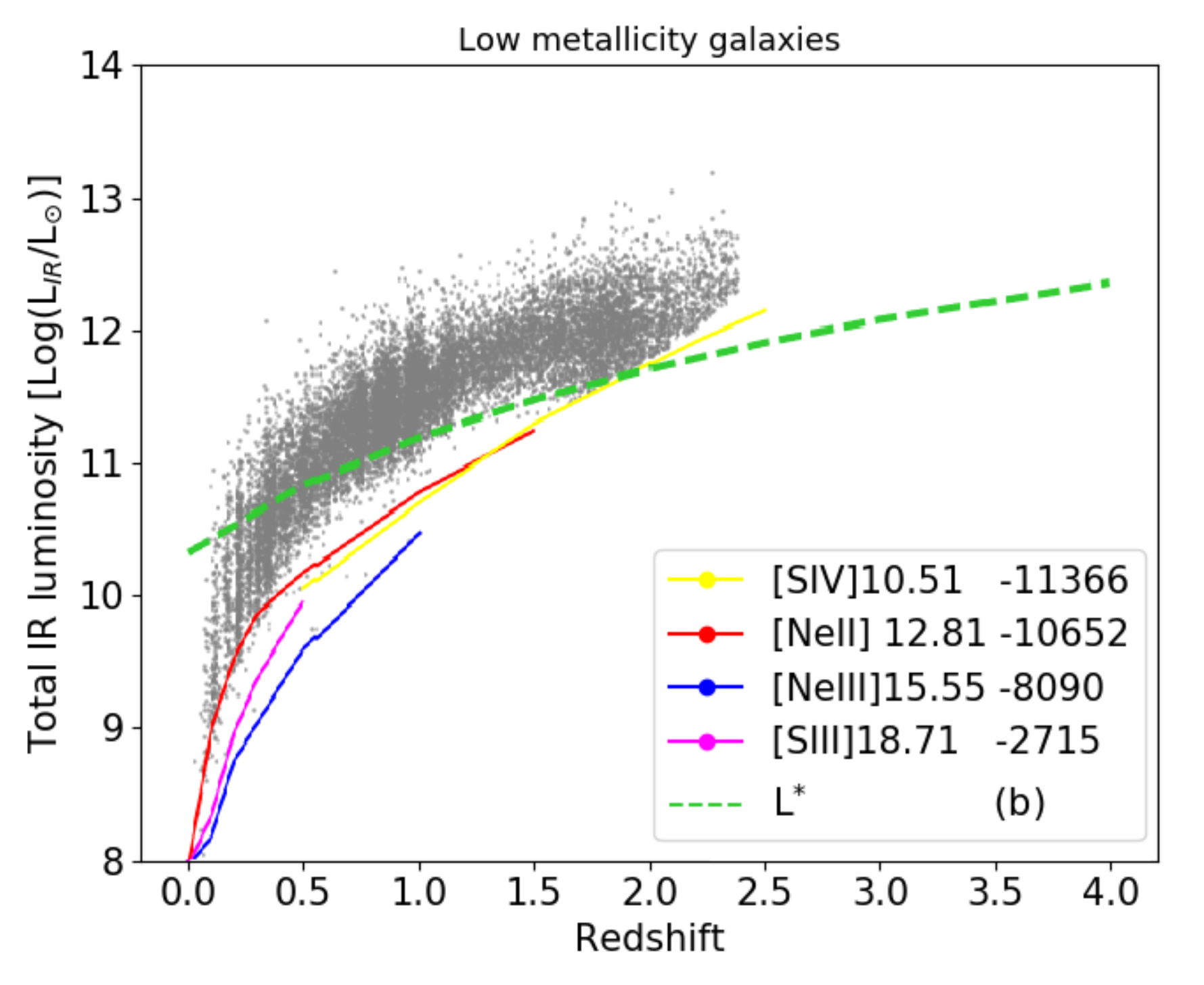}~
  \includegraphics[width=0.32\textwidth]{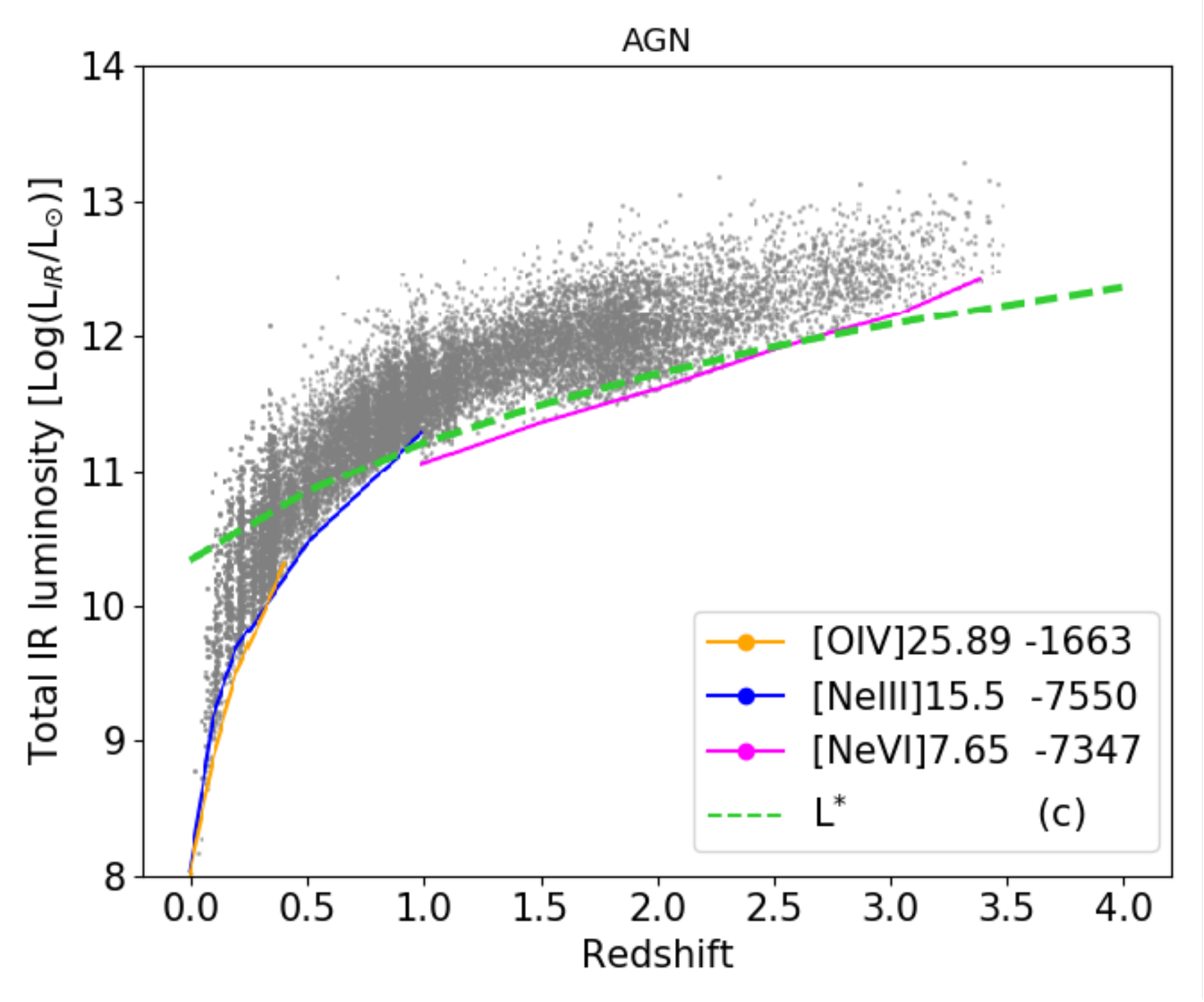}
  \caption{Redshift-luminosity diagrams simulating the SMI 1~deg$^2$ ultra-deep survey using the HELP database. We refer to Fig. \ref{fig:z_lum_help_hyperdeep} for the lines coding and legends  in each frame. {\bf (a: left):} SF galaxies detectable with SMI in the PAH features. On the right-hand y-axis the SFR correspondent to the total L$_{IR}$ is also given, adopting the conversion factor from \citet{kennicutt2012}. {\bf (b: center):} SF galaxies, adopting the line calibration of low-metallicity galaxies ($\sim 1/5\, \rm{Z_\odot}$), detectable with SMI in intermediate ionization  fine-structure lines. {\bf (c: right):} AGN detectable with SMI in high-ionization fine-structure lines.}\label{fig:z_lum_help}
\end{figure*}

\begin{figure*}
    \includegraphics[width=0.7\columnwidth]{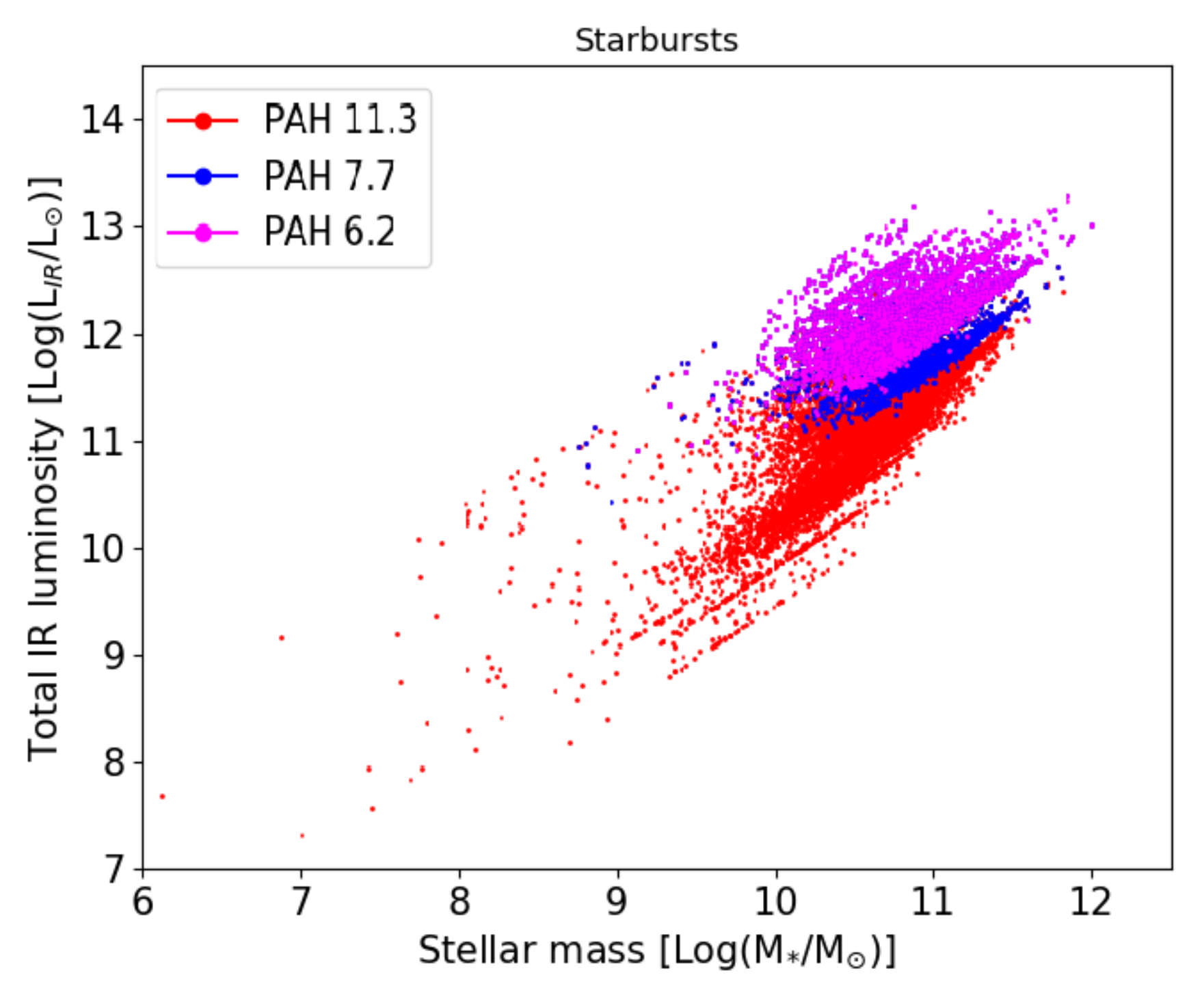}
    \includegraphics[width=0.7\columnwidth]{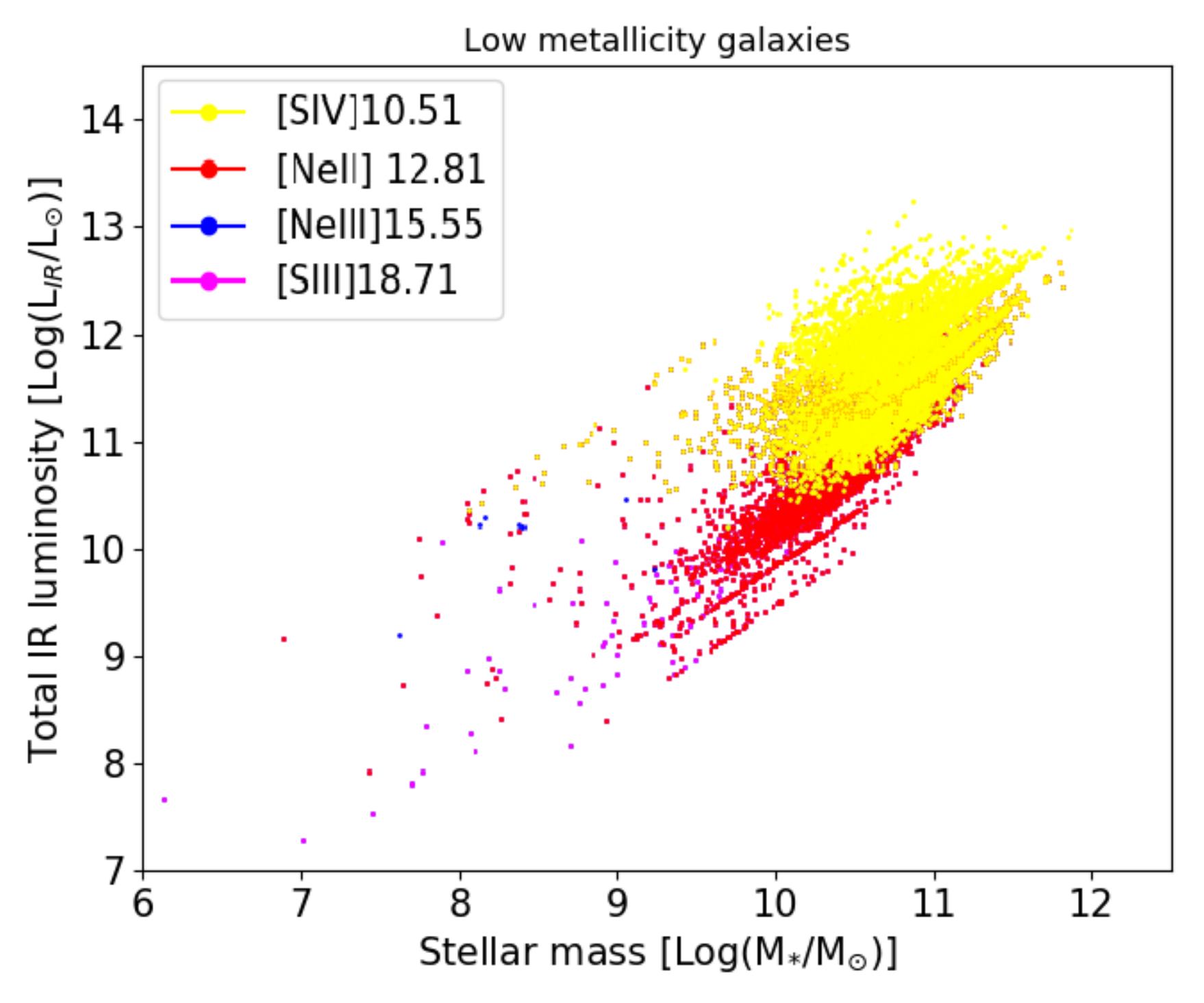}
    \includegraphics[width=0.7\columnwidth]{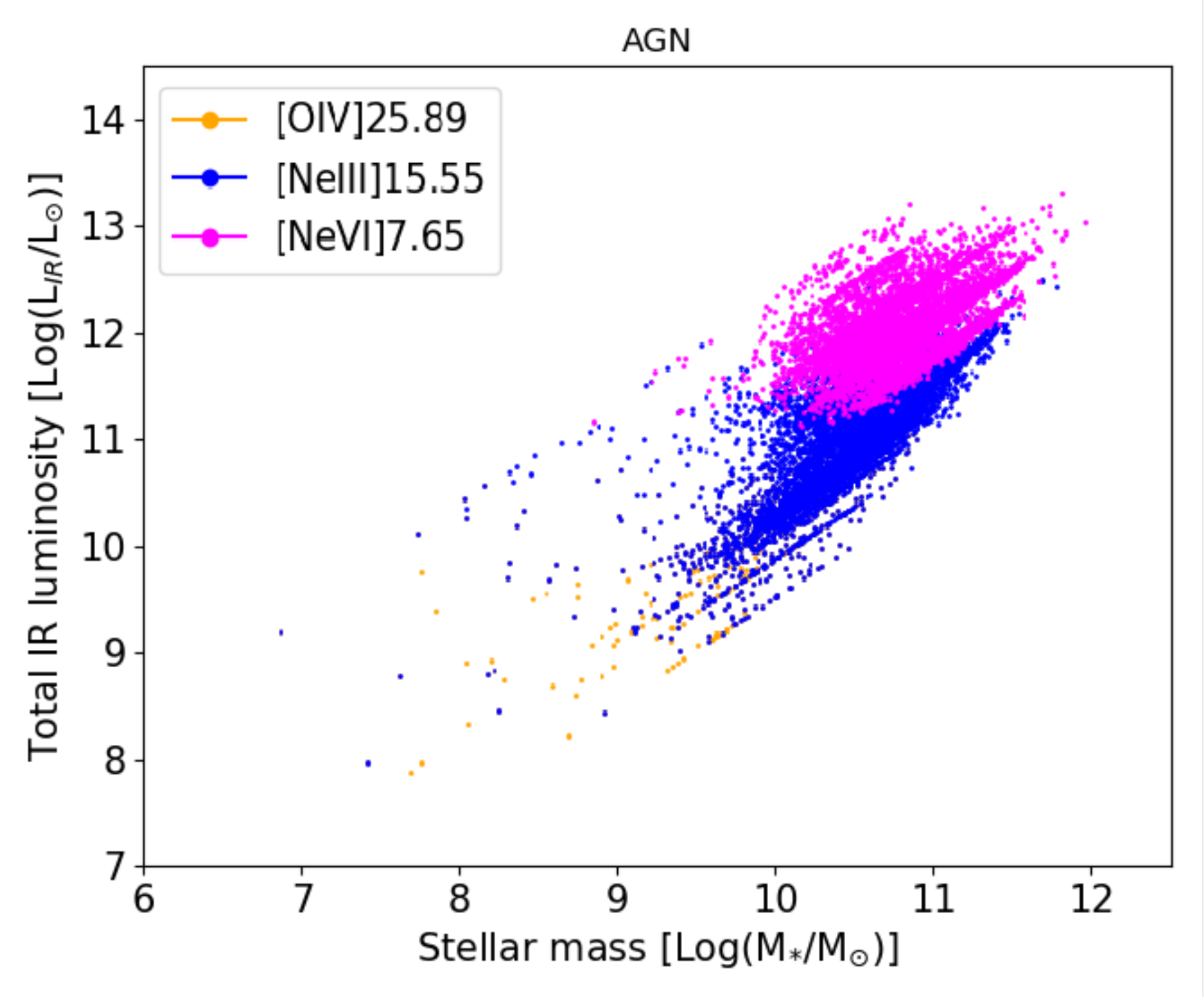}

    \caption{Stellar Mass -- Luminosity diagrams of galaxies detectable in the SMI 1~deg$^2$ ultra-deep survey  from the HELP database.
    {\bf (a: left)}  Star Forming galaxies.
    {\bf (b: center)} Low Metalicity Galaxies.
    {\bf (c: right)} AGN.
    }\label{fig:mass_lum_help}
\end{figure*}

\section{Results}

\subsection{SMI spectroscopic detections in the deepest {\it Herschel} fields}\label{sec:help_results}

The approach used in this paper is to adopt a single luminosity distribution function without assuming a particular separation among the different types of galaxies. Then, we derive the fraction of sources recovered if the whole population consists of either SF galaxies, low metallicity galaxies or AGN. This simple method allows us to derive the extreme situations, without relying on model assumption of separation of different galaxy populations.

We first show graphically the results of the simulation of the deepest \textit{Herschel} photometric fields (HELP) for the SMI-LR hyper-deep survey of 120 arcmin$^2$ (0.033 deg$^2$): in Fig.\,\ref{fig:z_lum_help_hyperdeep}(a) the detections of starburst galaxies through the PAH emission features show that the SMI-LR sensitivity of this survey is about two orders of magnitude below the knee of the luminosity function, as a function of redshift, that has been assumed for a $10^{10.7}\, \rm{M_\odot}$ galaxy residing in the Main-Sequence (M.-S.; \citealt{scoville2017}). For the low metallicity galaxies (LMG) presented in Fig.\,\ref{fig:z_lum_help_hyperdeep}(b) the detection threshold of SMI-LR is about one order of magnitude below the knee of the luminosity function, but only at redshift %z$<$1.5, while the [SiIV]10.5$\mu$m line can detect the bulk of the {\it Herschel} HELP population 
up to z$\sim$2.2. Finally, for the AGN population, as shown in Fig.\,\ref{fig:z_lum_help_hyperdeep}(c), SMI-LR can detect all objects at about 0.5-0.8 dex below the knee of the luminosity function up to redshift z$\sim$3.5. %2, while for 2.0$<$z$<$3.5, the detections through the [NeVI]7.65$\mu$m line will be $\sim$0.2-0.5 dex above the knee.

The results of our simulations of the ultra-deep $1\, \rm{deg^2}$ SMI spectroscopic survey, using the HELP database, are shown in Fig.\,\ref{fig:z_lum_help}(a) the detections of the expected PAH features in SF galaxies with solar metallicity, in Fig.\,\ref{fig:z_lum_help}(b) the detections of fine structure lines in low-metallicity galaxies, and in Fig.\,\ref{fig:z_lum_help}(c) the detections of fine structure lines typical of  AGN. For solar-like metallicity galaxies, the PAH features cover almost the whole redshift interval up to $z \sim 3.0$, at a luminosity level which about 0.5 dex or more below the knee of the luminosity function,up to redshift of $\sim$\,3.5, indicating that the bulk of the star forming galaxies, which were detected photometrically by {\it Herschel} { would} be easily detected spectroscopically by { a mid-IR camera such as SMI}. 

The situation is different for AGN, because of the lack of a tracer as bright as the PAH features, and Fig.\,\ref{fig:z_lum_help}(c) shows that the various fine-structure lines can cover the redshift range up to z$\sim$2.5, however at luminosities which are about 0.5 dex or more above the knee of the luminosity function at each redshift above $z \sim 0.5$. For redshift above $z \sim 1$, the best tracer for AGN is the [\ion{Ne}{vi}]7.65$\mu$m line, while at lower ($0.5 < z < 1$) redshift the best tracer is the [\ion{Ne}{iii}]15.6$\mu$m, %\ion{Ne}{v}]14.3$\mu$m and [\ion{S}{iv}]10.5$\mu$m lines can be used, 
and at $z < 0.5$ the [\ion{O}{iv}]25.9$\mu$m line would be visible in the SMI spectral range. About the detection of the high-ionization line of [\ion{Ne}{vi}]7.65$\mu$m, we are aware that on one hand, in the co-existence in the host galaxy of strong starburst emission, at the low resolution of the SMI imaging spectrometer, this line can be overwhelmed by the strong PAH feature at 7.7$\mu$m, however, on the other hand, it can be very bright in strong AGN, such as the prototypical Seyfert type 2 galaxy NGC1068 in the local Universe, which does not show PAH emission features \citep{lutz2000}.

\begin{figure*}
    \includegraphics[width=0.74\columnwidth]{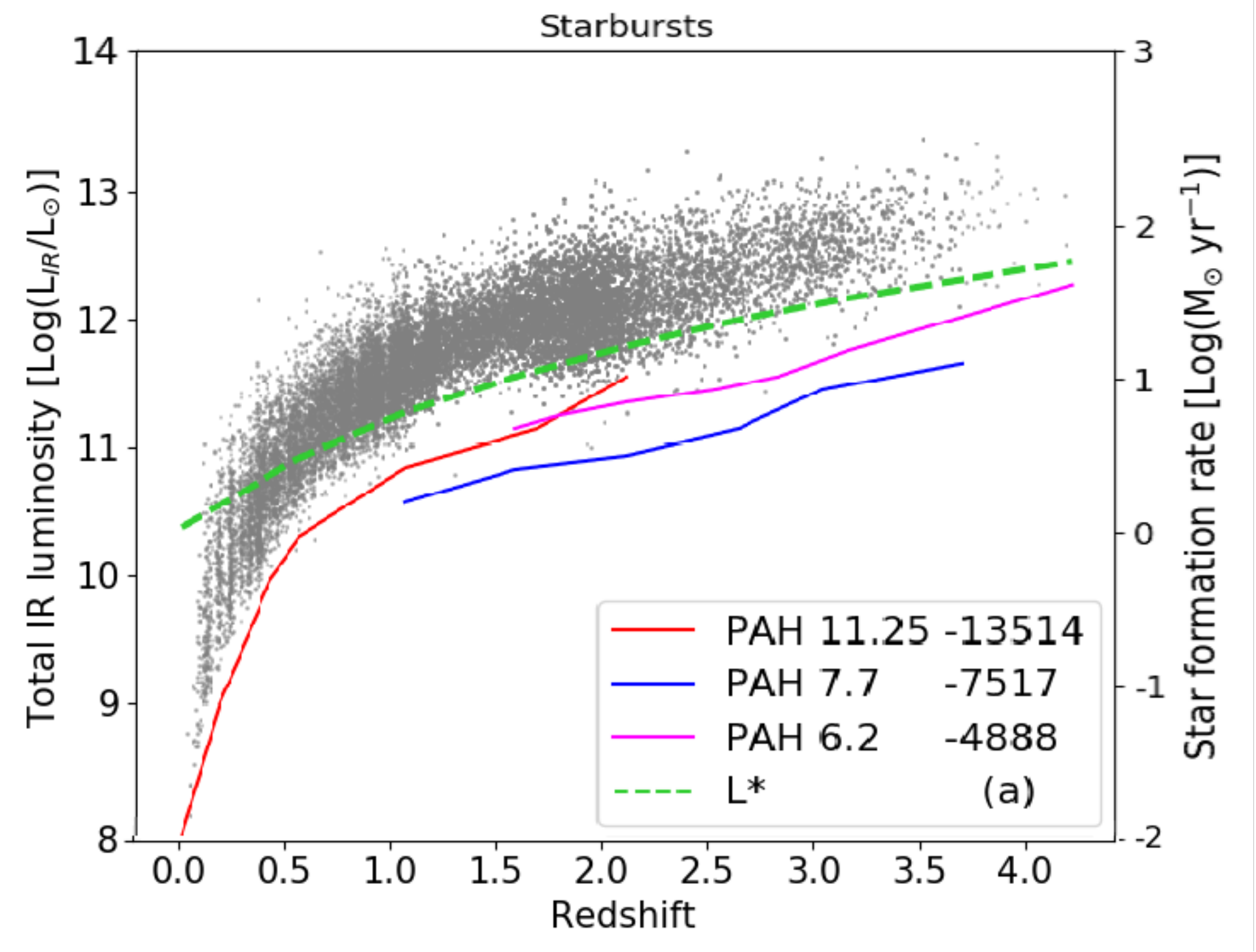}
    \includegraphics[width=0.67\columnwidth]{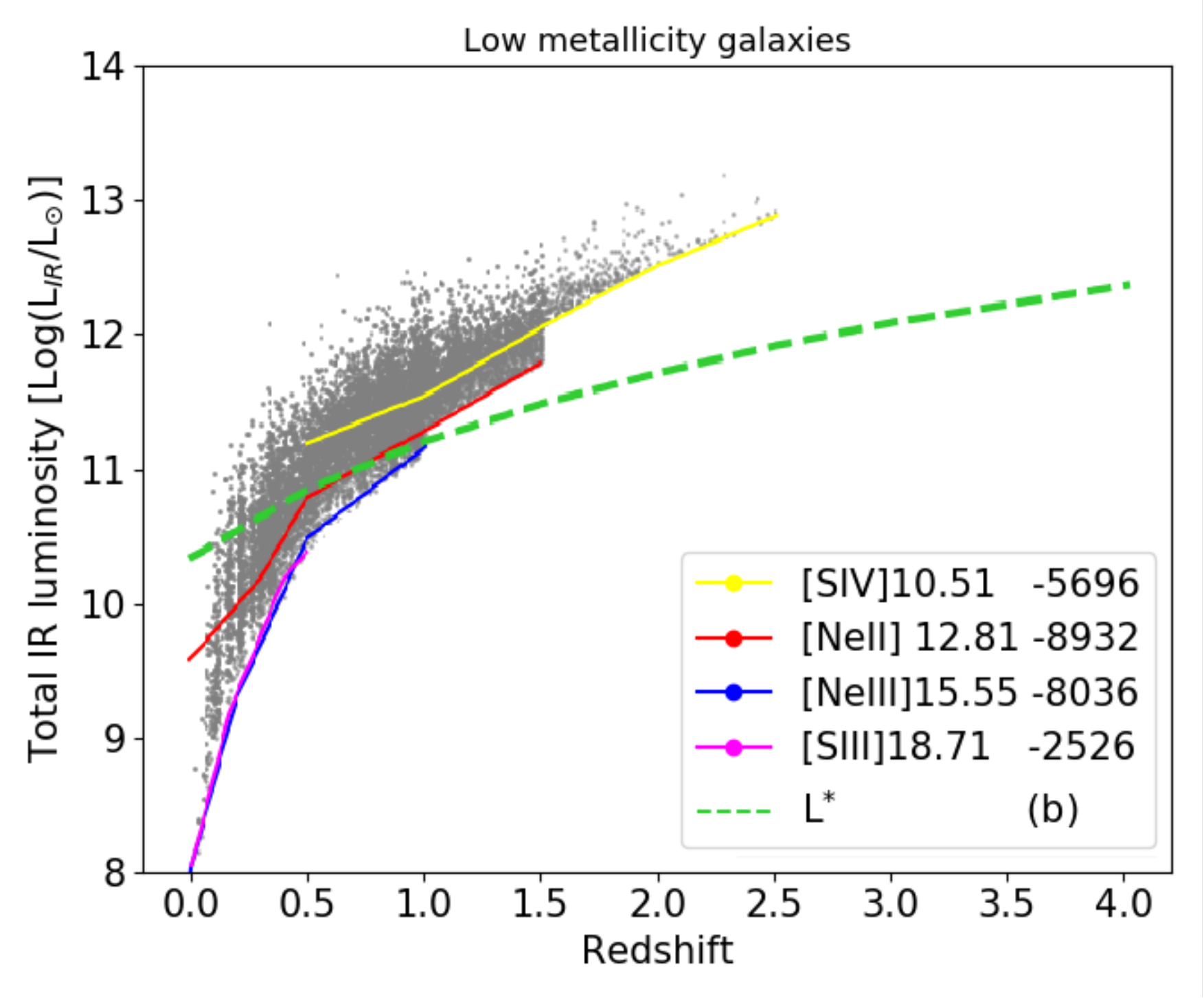}
    \includegraphics[width=0.67\columnwidth]{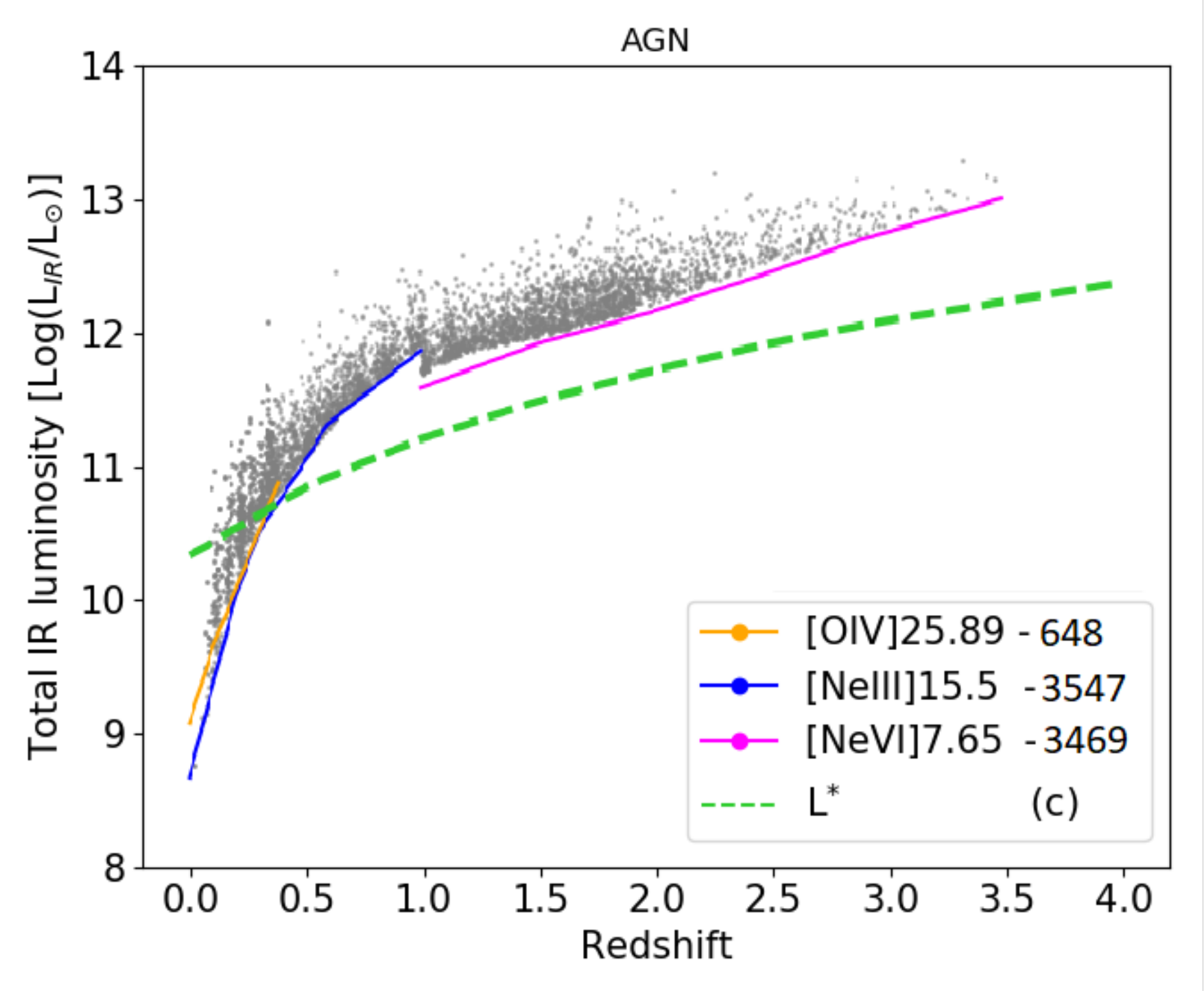}
    \caption{Redshift-luminosity diagrams simulating the SMI 15~deg$^2$ deep survey using the HELP database. We refer to Fig. \ref{fig:z_lum_help_hyperdeep} for the lines coding and legends in each frame.
 {\bf (a: left):} SF galaxies detectable with SMI in the PAH features.  {\bf (b: center):} SF galaxies assuming the line calibration at low metallicities, detectable with SMI in the intermediate ionization fine-structure lines. %, with [\ion{S}{iv}]$\rm 10.5 \micron$ (yellow), [\ion{Ne}{ii}]$\rm 12.8 \micron$ (red), [\ion{Ne}{iii}]$\rm 15.5\micron$ (blue) and [\ion{S}{iii}]$\rm 18.7 \micron$ (magenta).
 {\bf (c: right):} AGN detectable with SMI in the high-ionization fine-structure lines.%, with [\ion{O}{iv}]$\rm 25.9 \micron$ (yellow),  [\ion{Ne}{iii}]$\rm 15.5\micron$ (blue) and [\ion{Ne}{vi}]$\rm 7.7\micron$ (magenta).
 }    \label{fig:z_lum_help_deep}
\end{figure*}

\begin{figure} 
{\includegraphics[width=\columnwidth]{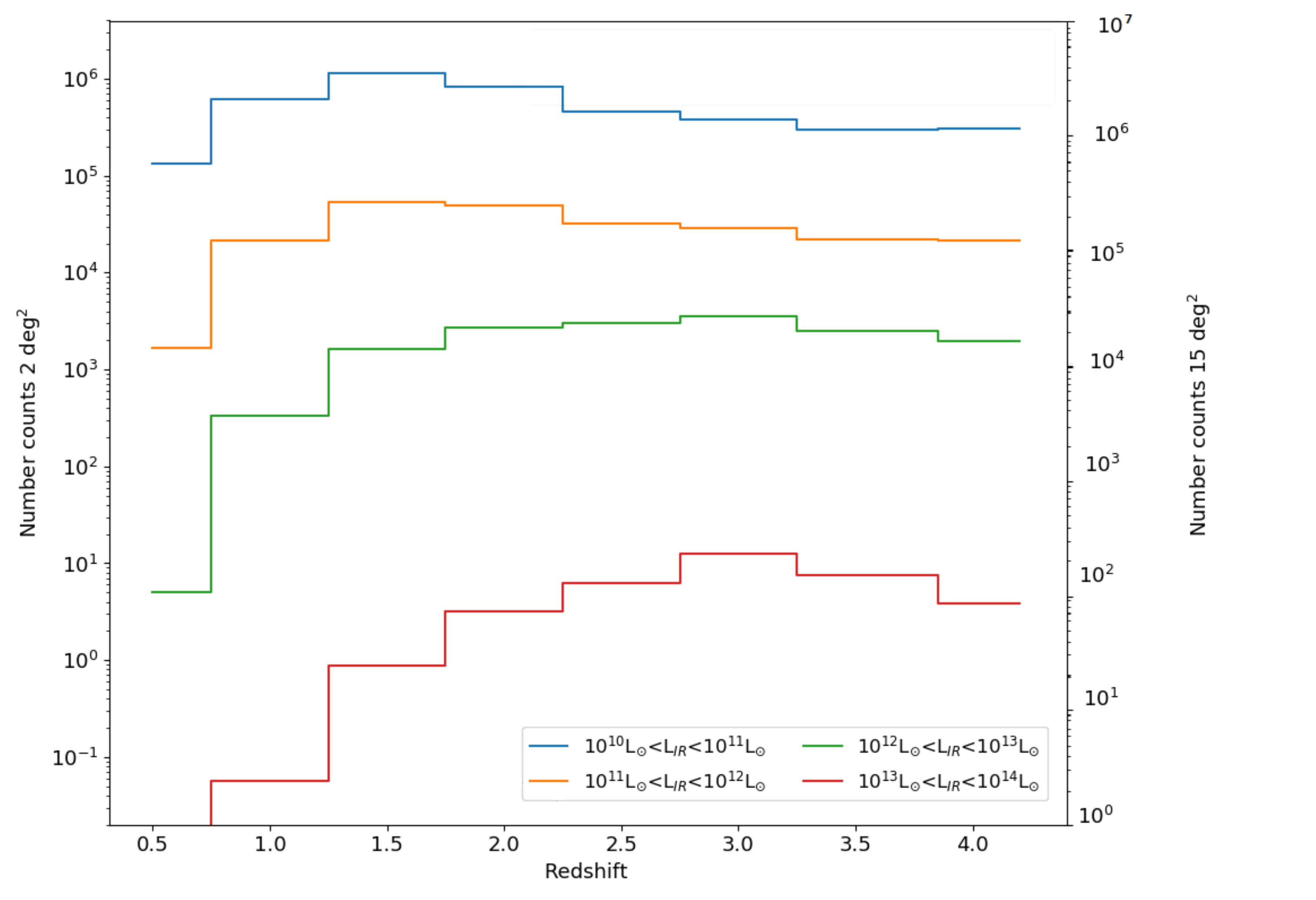}}
%{\includegraphics[width=\columnwidth]{LF_wang_15deg.pdf}}
\caption{
Expected galaxy number counts in the two fields  of the SMI spectroscopic surveys, as derived from the \citet{wang2019} luminosity functions. On the {left y-axis} are shown the counts as a function of redshift for different luminosities for the ultra-deep survey of 1 deg$^2$. On the {right y-axis} the same, for the deep survey of 15 deg$^2$.}
\label{fig:number_counts}
\end{figure}

\begin{figure*}[ht]
	\centering
	\includegraphics[width=0.34\textwidth]{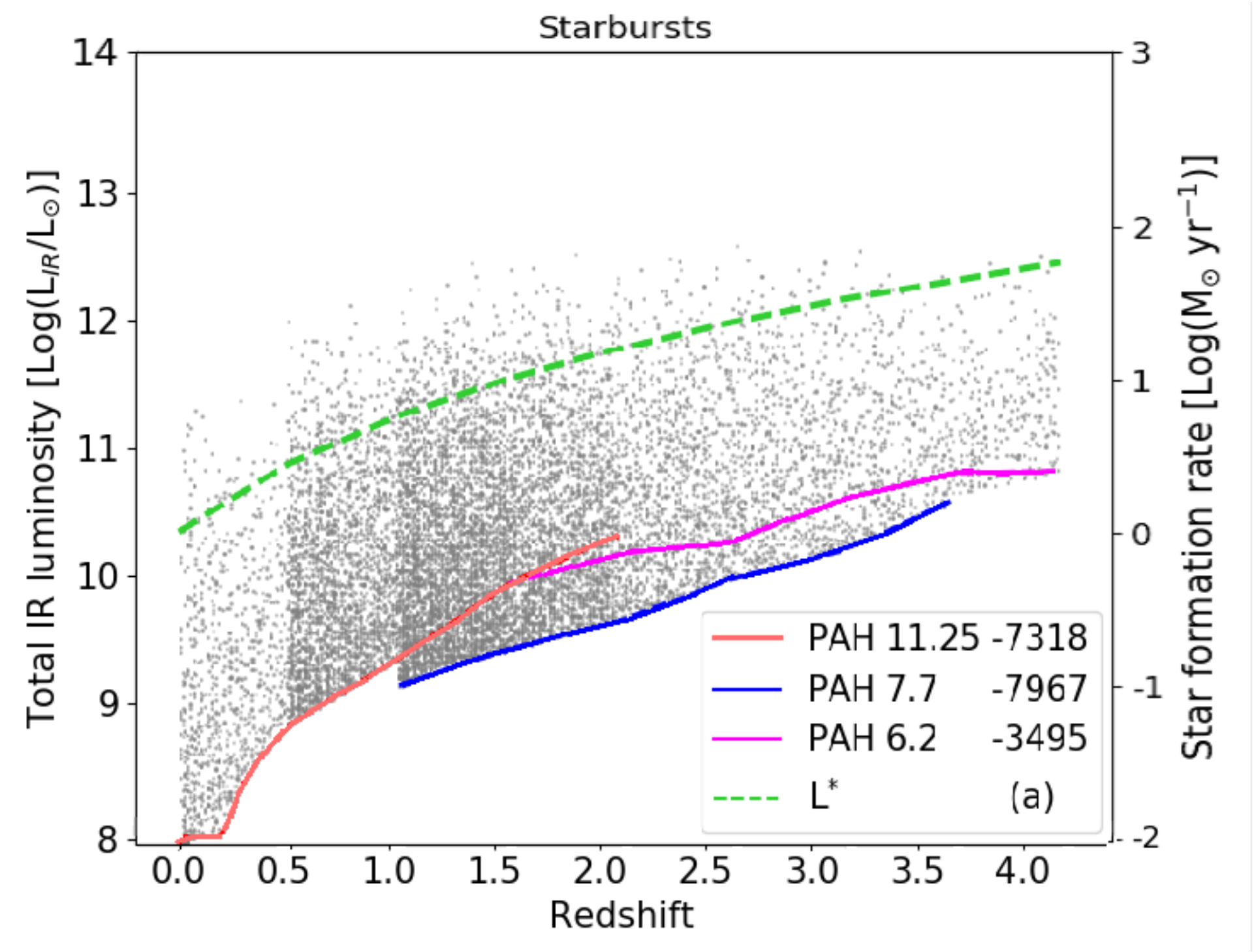}~
	\includegraphics[width=0.32\textwidth]{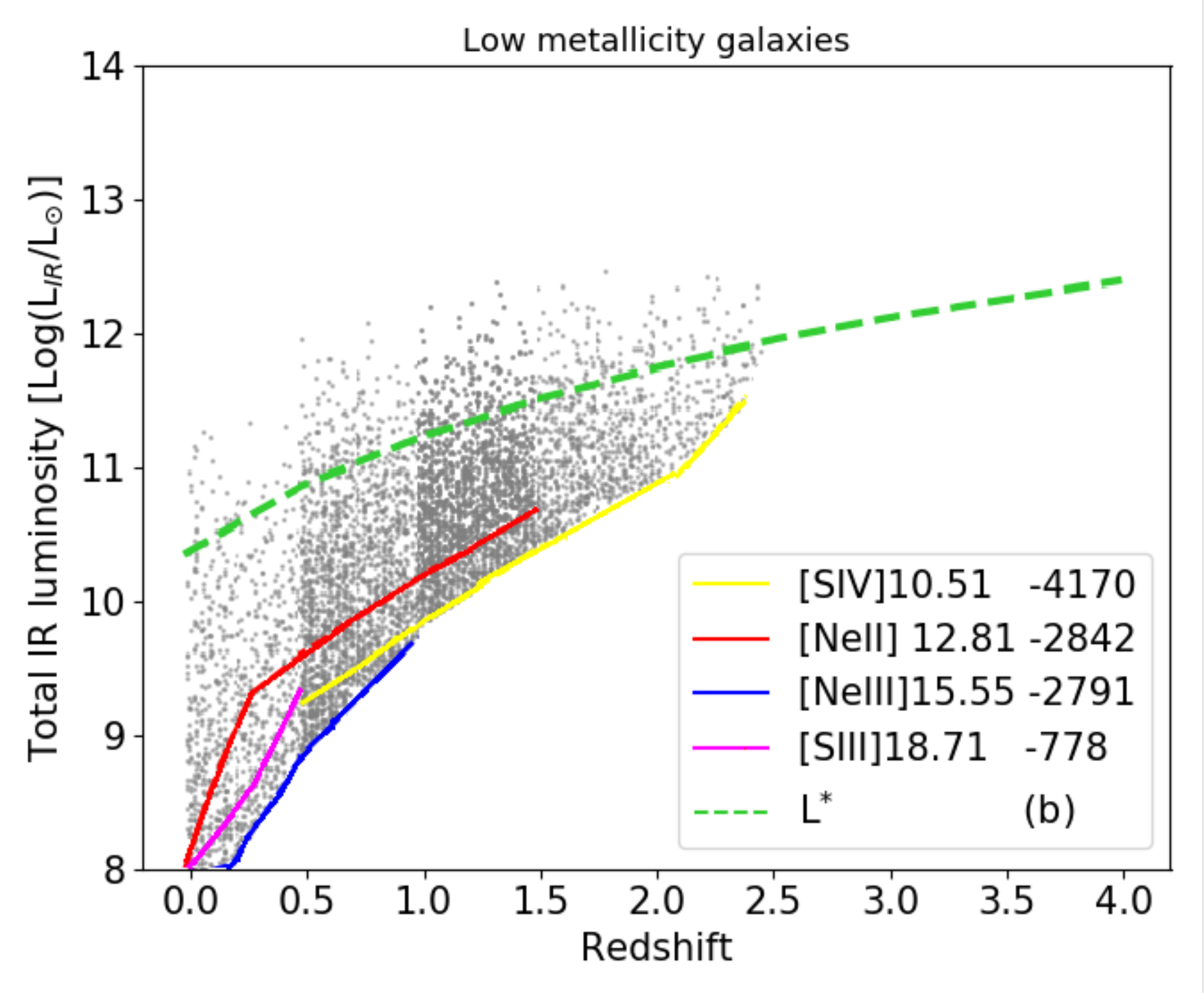}~
	\includegraphics[width=0.32\textwidth]{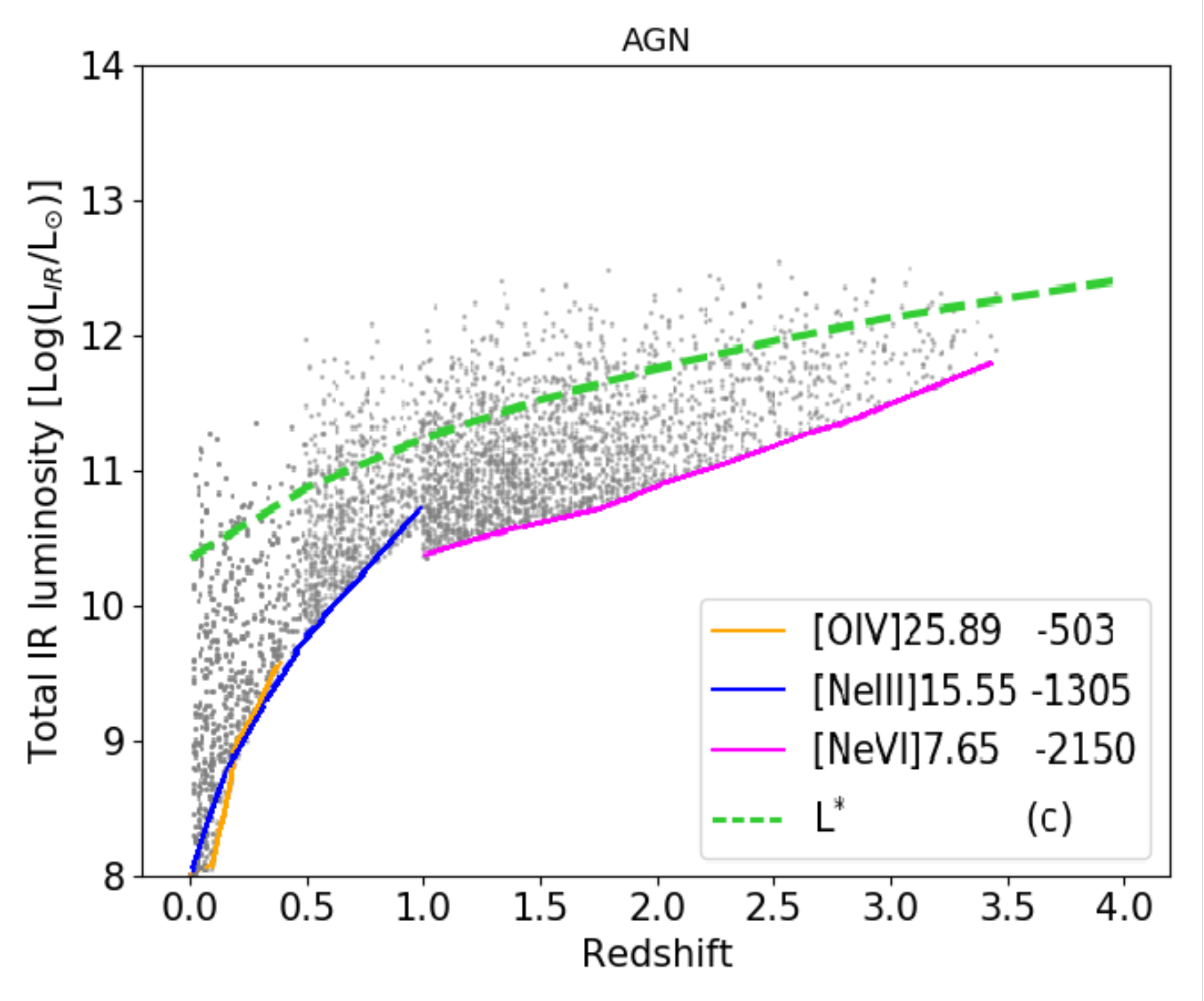}
\caption{Redshift-luminosity diagrams simulating the SMI hyper-deep survey using the predictions from the far-IR Luminosity Functions \citep{wang2019}. We refer to Fig. \ref{fig:z_lum_help_hyperdeep} for the lines coding and legends in each frame. {\bf (a: left)} SMI simulations of SF galaxies. {\bf (b: center)} SMI simulations for SF galaxies assuming the low metallicity calibration in Table\,\ref{tab:linecal}. {\bf (c: right)} SMI simulations of AGN.}\label{fig:lumfun_pred_hyperdeep}
\end{figure*}

\begin{figure*}[ht]
	\centering
	\includegraphics[width=0.34\textwidth]{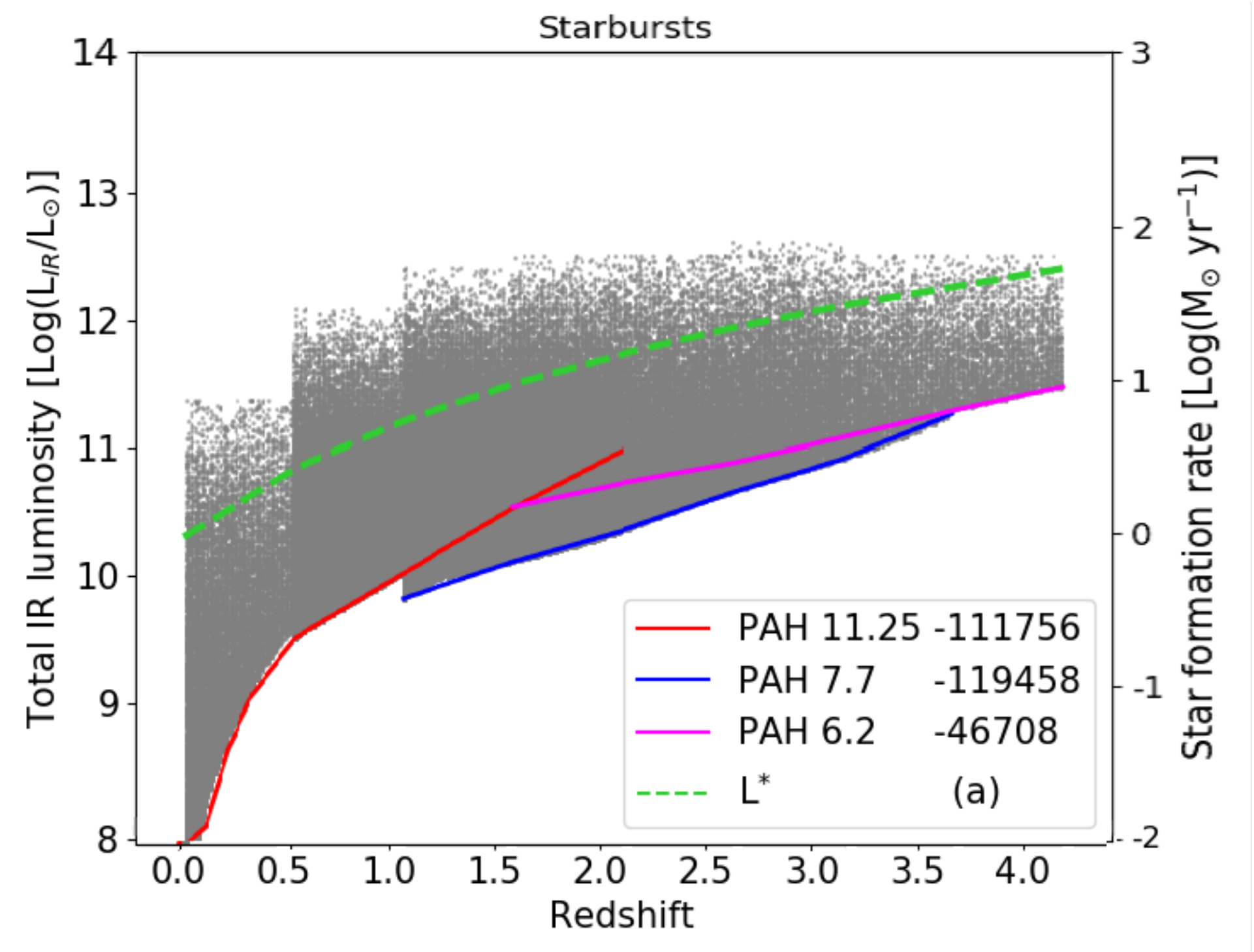}~
	\includegraphics[width=0.32\textwidth]{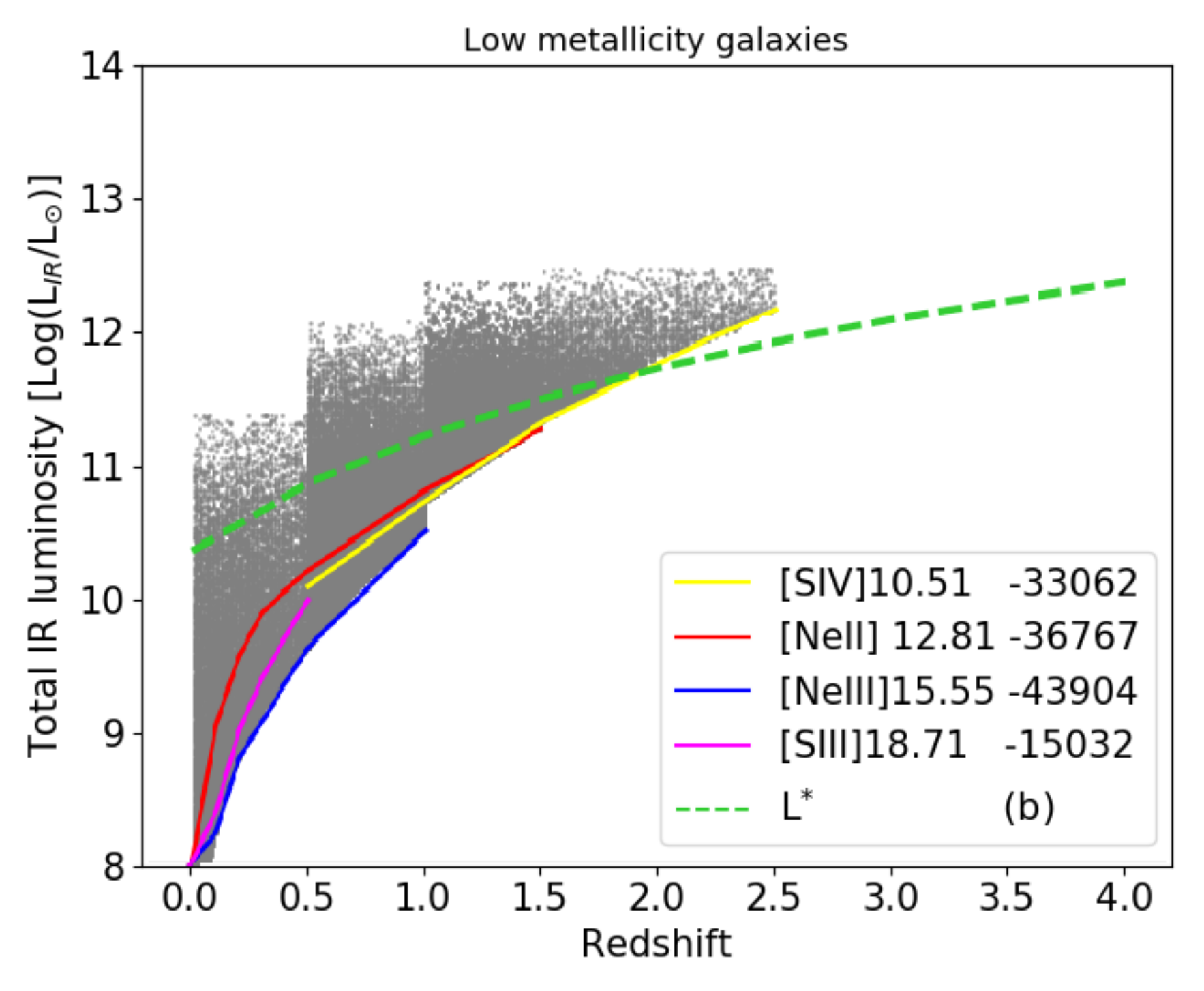}~
	\includegraphics[width=0.32\textwidth]{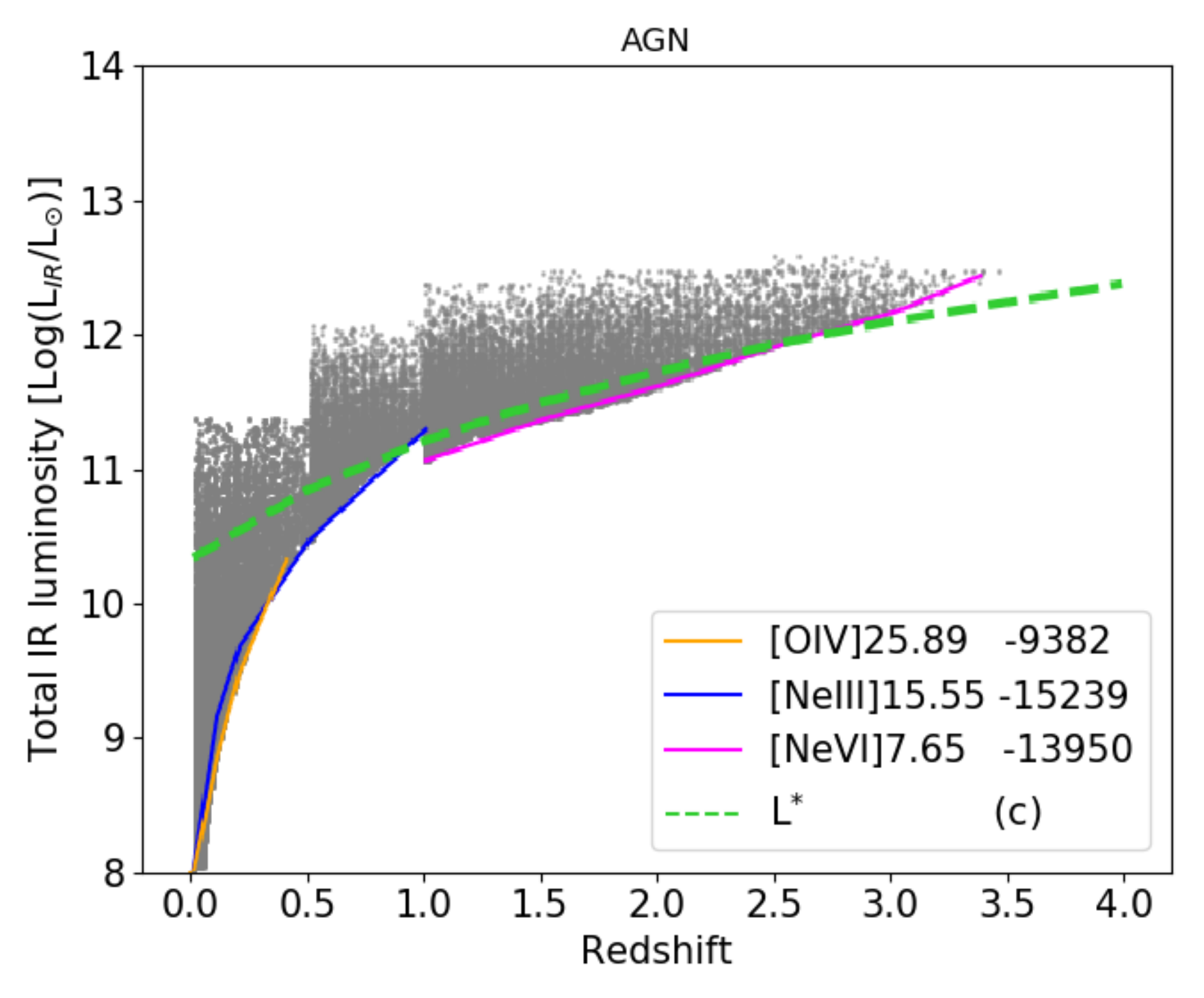} 
\caption{Redshift-luminosity diagrams simulating the SMI ultra-deep survey using the predictions from the far-IR Luminosity Functions \citep{wang2019}. We refer to Fig.\,\ref{fig:z_lum_help_hyperdeep} for the lines coding and legends  in each frame. {\bf (a: left)} SMI simulations of SF galaxies. {\bf (b: center)} SMI simulations for SF galaxies assuming the low metallicity calibration in Table\,\ref{tab:linecal}. {\bf (c: right)} SMI simulations of AGN.}\label{fig:lumfun_pred}
\end{figure*}

For the Low Metallicity galaxies, shown in Fig.\,\ref{fig:z_lum_help}(b), the situation is somewhat intermediate to what we have found for the SF galaxies and the AGN: the SMI ultra-deep survey can go as deep as the knee of the galaxy luminosity function up to $z\sim1$, about $0.3\, \rm{dex}$ above the knee  at $1.0 < z < 1.5$ and about one order of magnitude above the knee for $1.5 < z < 2.2$.

The simulation with the {\it Herschel} data allowed us to also analyze the stellar mass interval of the galaxies that can be detected, shown in figures \ref{fig:mass_lum_help}(a) for the Star Forming galaxies, in Fig.\,\ref{fig:mass_lum_help}(b) for Low Metallicity galaxies and in Fig.\,\ref{fig:mass_lum_help}(c) for AGN. The SF galaxies detected in the PAHs cover well the whole mass-luminosity interval analyzed in the HELP project, the LMG, detected through fine-structure lines, show a narrower region of the plane stellar mass - luminosity and the AGN, because of the lower detection rate, show only a sparse filling of highest masses and luminosities. 

%while the AGN detected in the high-ionization fine-structure preferentially cover the region of the most massive objects with $M{\geq}10^{10} M_{\odot}$. The bottom part of the mass-luminosity diagram includes AGN with detections in the [\ion{O}{iv}]25.9$\mu$m and [\ion{Ne}{v}]24.3$\mu$m lines, while the top part will be covered mostly using the [\ion{Ne}{vi}]7.7$\mu$m line.

The results for the simulations of the deep ${\rm 15\, deg^{2}}$ SMI survey of the same \textit{Herschel} photometric fields are shown in Fig.\,\ref{fig:z_lum_help_deep}(a) for the SF galaxies, %adopting the line calibration for solar-like metallicity galaxies, 
in Fig.\,\ref{fig:z_lum_help_deep}(b) for LMG, %galaxies adopting the line calibration of low-metallicity galaxies, 
and in Fig.\,\ref{fig:z_lum_help_deep}(c) for the AGN. As can be seen from this figure, the SMI deep survey would be able to reach Star Forming galaxies at 0.5-1.0 dex below the knee of the luminosity functions at the various redshifts. In particular, for the SF galaxies, the PAH features cover the whole redshift interval up to redshift $z \sim 3.5$. For the AGN, the SMI deep survey would allow detections below the knee of the luminosity functions up to redshift z\,$\sim$\,0.5, while for larger redshifts the detections of AGN would be at luminosities 0.5-0.8 dex above the knee up to a redshift of z$\sim$2.5 and up to 1 dex above for 2.5$<$z$<$3.5.

As for the case of the ultra-deep survey, the situation for the low metallicity galaxies is intermediate between the other two galaxy populations, with detections of galaxies below the knee of the LF up to a redshift of z$\sim$1, up to 0.3 dex above for $1.0 < z < 1.5$ and 1 dex above for $1.5 < z < 2.2$.
%only ULIRGs ($L>10^{12} L_{\odot}$) above z$\sim$1 and LIRGs ($10^{11}L_{\odot}<L<10^{12}L_{\odot}$) at 0.5$<$z$<$1.0.

To summarize, the use of the HELP catalog demonstrated that %the \textit{SPICA} SMI 
{ a hyper-deep 120 arcmin$^2$ survey with a mid-IR camera, such as the SMI instrument proposed for SPICA, could} detect all HELP galaxies and in particular galaxies two orders of magnitude fainter that the knee of the luminosity function (LF) at any redshift up to z$\sim$4. For low metallicity galaxies, this survey would detect objects about one order of magnitude fainter than the knee of the LF up to z$\sim$2, while for AGN the hyper-deep survey would detect galaxies 0.5-0.8 dex below the knee of the LF up to z$\sim$3.5.

{ A ultra-deep 1 deg$^2$ spectrophotometric survey, with the same camera as above, would} detect PAH features in the spectra of the complete population of galaxies observed by {\it Herschel} in its deepest cosmological fields, allowing a complete census of the SFR density from spectral features. %Furthermore, the different PAH features detected will provide additional information on the dust properties and the excitation mechanism in the ISM of high-z galaxies. 
On the other hand, the same {\it Herschel} galaxies, if they harbour an AGN, { could} be detected by { mid-IR} spectroscopic surveys at a threshold of about 0.5\,dex above the knee of the luminosity function at each redshift above z $\sim$ 0.5. Therefore, follow-up pointed { mid- to far-IR} %SAFARI and SMI 
spectroscopic observations will be needed to provide a complete census of the BHAR density as a function of redshift. 
For the low metallicity galaxies the ultra-deep survey will be complete, i.e. down to the knee of the luminosity function, at z$<$1.0, 0.2-0.3\,dex above the knee at 1.0$<$z$<$1.5 and 1\,dex above the knee at z $>$ 1.5. %The \textit{SPICA} SMI deep surveys of 
{ A 15 deg$^2$ mid-IR deep survey} will be 0.5-1.0\,dex below the knee of the luminosity functions at any redshift for the Star Forming galaxies, at the knee of the LF up to z$\sim$1 for Low Metallicity Galaxies and about 1 dex above the knee of the LF for AGN at z $>$ 0.5.

% XXXX THINK ABOUT THIS {\color{red} {\bf Asantha's comment:}  What is meant here by a complete census of BHAR density? down to what level is BHAR density can be measured with SMI? follow-up is only a smaller sample, there will be larger uncertainties from any extrapolation. If density is not a measurable quantity it should be not be described in a document like this as a science goal.}

\subsection{SMI spectroscopic detections from simulations derived from the luminosity functions}\label{sec:lumfun_results}

After having assessed, in the previous section, what \textit{SPICA} spectroscopy { would} be able to cover, in terms of type of sources (SF- or AGN-dominated), luminosities and redshift in the fields that have been observed photometrically by {\it Herschel}, we want now to go deeper and compute the limiting line fluxes that a { SMI-type instrument would} reach using the predictions from the luminosity functions  \citep{wang2019}.  Following the method outlined in Section \ref{sec:counts}, we have computed the number of galaxies that can be detected by the SPICA SMI  deep and ultra-deep surveys, through the SF features in both the solar and low-metallicity cases and via the high-ionization fine structure lines typical of AGN.

In Fig.\,\ref{fig:number_counts} we show the evolution with redshift of the number of galaxies for different luminosity bins expected to be present as a function of redshift in the ultra-deep survey of the area of $1\, \rm{deg^2}$, and in the deep survey of $15\, \rm{deg^2}$. %We note a significant drop in the SF class above $z \sim 2$. This is due to both the spiral and starburst populations fading at higher redshifts.
We report in Table\,\ref{tab:deep_counts} the number of detections for AGN, star forming galaxies, and low-metallicity galaxies as a function of redshift and total IR luminosity for the deep survey. Table\,\ref{tab:ultradeep_counts} shows the same results for the ultra-deep surveys and Table\,\ref{tab:hyperdeep_counts} those of the hyper-deep survey.

The results of the SMI hyper-deep and ultra-deep survey predictions from the luminosity functions are also shown graphically in Figure \ref{fig:lumfun_pred_hyperdeep} and Figure \ref{fig:lumfun_pred}, respectively, where Fig. \ref{fig:lumfun_pred_hyperdeep}(a) and Fig. \ref{fig:lumfun_pred}(a) (left panels) show the detections of SF galaxies through the PAH features; Fig. \ref{fig:lumfun_pred_hyperdeep}(b) and Fig. \ref{fig:lumfun_pred}(b) (central panels), the detections of LMG through fine-structure lines and Fig. \ref{fig:lumfun_pred_hyperdeep}(c) and Fig. \ref{fig:lumfun_pred}(c) (right panels)  show the high-ionization lines detections in AGN. %, together with the PAH detections of the host galaxies.

From these results (Fig. \ref{fig:lumfun_pred_hyperdeep} and Fig. \ref{fig:lumfun_pred}) it emerges that the PAH features, used as SF tracers, { would} allow us to reach SF galaxies about 2\,dex in the hyper-deep survey and more than 1\,dex  in the ultra-deep survey below the knee of the luminosity function up to a redshift of $z \sim 4$. For the case of AGN detections through fine-structure lines, the SMI hyper-deep survey { would} be able to detect objects $\sim$1\,dex fainter than the knee of the LF up to redshift z$\sim$3, while for the ultra-deep survey the detection of AGN { would} reach the knee of the luminosity function at any redshift between 1.0\,$<$\, z\,$<$\,3.0 and below the knee of the LF below z\,$<$\,1.0. 

Considered in their entirety, the three cases will allow us to detect several times $\sim 10^{3}$ and $\sim 10^{4}$ sources in the hyper-deep and ultra-deep survey, respectively, and up to $\sim 10^{5}$ in the deep survey, thus providing a thorough data set for follow up surveys with the %SAFARI and with the SMI in 
pointed spectroscopic mode. %The numerical results in Table\,\ref{tab:deep_counts} are shown as luminosity-redshift distributions in Fig.\,\ref{fig:lumfun_pred}(a) -(b) and -(c), which represent the ultra-deep surveys for SF, AGN, and SF adopting the low-metallicity calibration in Table\,\ref{tab:linecal}, respectively. 
It emerges that the detection of AGN at the knee of the LF can be guaranteed not only by the deepest survey (the hyper-deep), but also in the ultra-deep survey. %in the other shallower surveys it is limited by the chosen tracers, which do not cover the full spectral range with relatively bright spectral lines. 
In Star Forming galaxies, the PAH features, for both their intrinsic brightness and because they cover the whole spectral range, can be detected well below the knee of the luminosity function, thus permitting and in-depth analysis of the SF process in a very faint population of galaxies, reaching luminosities as low as ${\rm L{\sim}10^{10} L_{\odot}}$ at redshift $z \sim 2$ and lower than ${\rm L{\sim}10^{11} L_{\odot}}$ at redshift $z \sim 3.5$.

\subsection{Definition of a representative sample of galaxies}\label{sec:sample}
The {\it blind} spectrophotometric surveys carried out with a { SMI-type mid-IR camera} --\,as shown in Sections \ref{sec:help_results} and \ref{sec:lumfun_results}\,-- { would} allow us to preliminarily classify the galaxies in terms of AGN-dominated or SF-dominated based on the spectral features and lines, and the continuum slope measured in their low-resolution mid-IR spectra. This will provide also accurate redshift estimates on a great number of objects, as well as IR luminosities using complementary {\it Herschel} or {\it ALMA} continuum data. { With respect to what we already know from {\it Spitzer} and {\it Herschel}, a SPICA-like mission would perform deep unbiased spectrophotomectric surveys at mid-IR wavelengths that will go deeper in photometric limit and at the same time characterize, for the first time, the observed galaxies spectroscopically in the rest-frame mid-IR. With respect to the {\it HST}, a {\it SPICA}-like mission would  be able to detect those galaxies which most actively contribute to the buildup of stellar mass at cosmic noon (z$\sim$1-2) containing large amounts of dust \citep[e.g.,][and references therein]{murphy2011,madau2014}, which have so far been missed by the optical surveys. With respect to what {\it JWST} will be able to discover, a 2.5\,m class mid- to far-IR telescope has two advantages: it will be able to cover larger sky areas (of order of square degrees), while the {\it JWST} is limited to square arc-minutes fields, and it will have the mid-IR spectral range, which is very rich in ``dust-unaffected'' physical tracers, positioned at the {\it cosmic noon} (z$\sim$1-3) while {\it JWST} will explore lower redshift in the mid-IR (see also Section\,\ref{syn}). 
%(e.g., Murphy et al. 2011; Madau & Dickinson 2014, and references therein)
}
%{\color{red} no mention is made on the various ways of how a deeper sensitivity and a broader wavelength range, such as those offered by SICA, will improve upon existing similar selections by e.g. HST, Spitzer and Herschel. Also, no mention is made by what will be soon contributed by JWST, I feel that these issues should be discussed in this section.}

With our scientific goal, i.e. to physically characterize the evolution of galaxies through IR spectroscopy and measure directly redshifts, SFRs, BHARs, metallicities and the other physical properties of gas and dust in galaxies (e.g. gas densities, ionizing continuum slope, dust features from silicates and PAH), we need to provide an atlas of galaxy spectra as a function of cosmic time, covering the main physical parameters of galaxies. The first task we need to accomplish is the definition of the suited sample of galaxies.

Following the outline in Section \ref{sec:strategy}, we will briefly define the criteria to select, from the above described { SMI-type} surveys, a representative sample of galaxies to perform further pointed observations { with mid- to far-IR grating spectrometers, such as those planned for SPICA,} SAFARI and SMI, which will be used to optimally study galaxy evolution, especially covering the cosmic noon.

\begin{figure*}
\centering
  \includegraphics[width=0.5\textwidth]{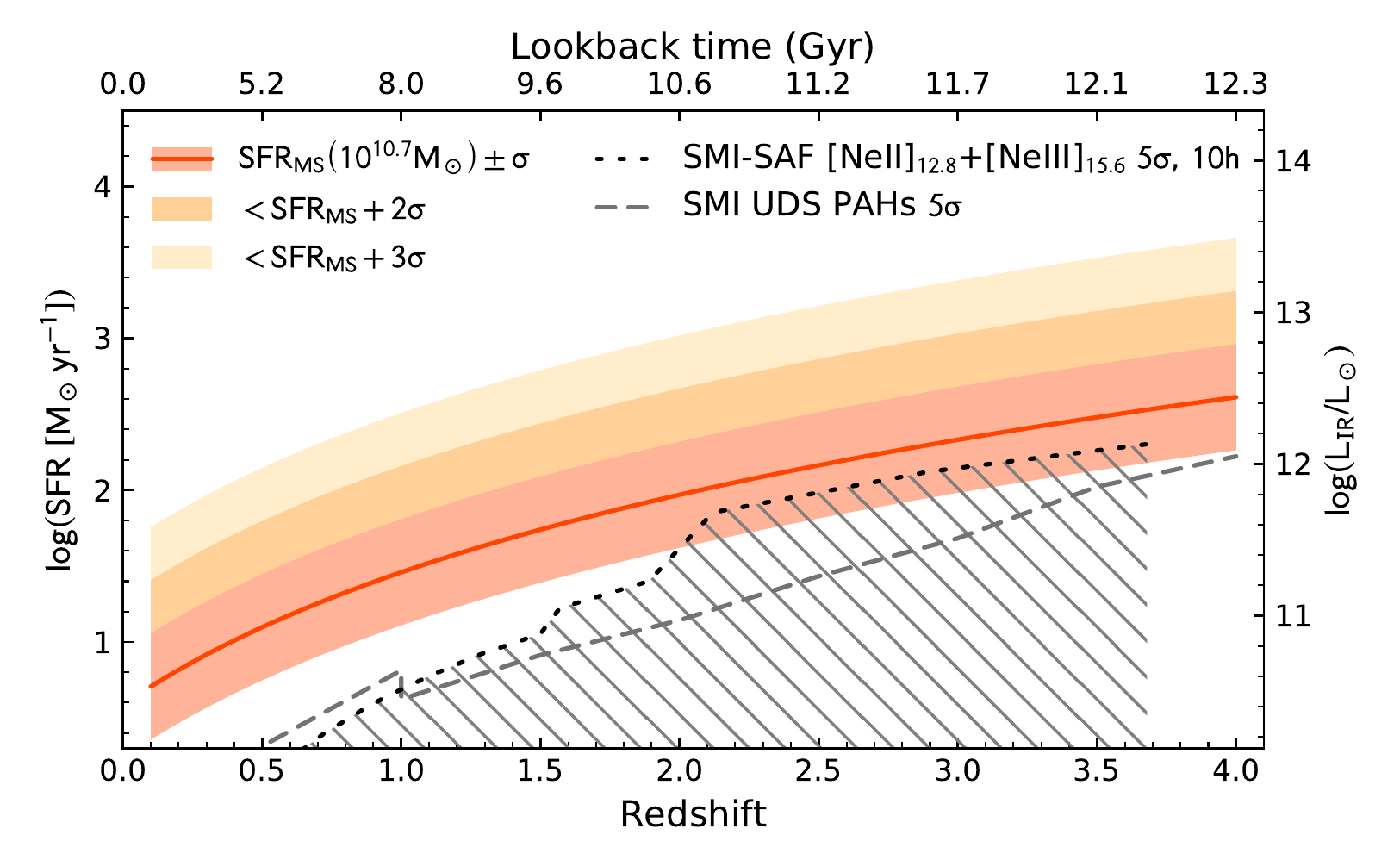}~
  \includegraphics[width=0.5\textwidth]{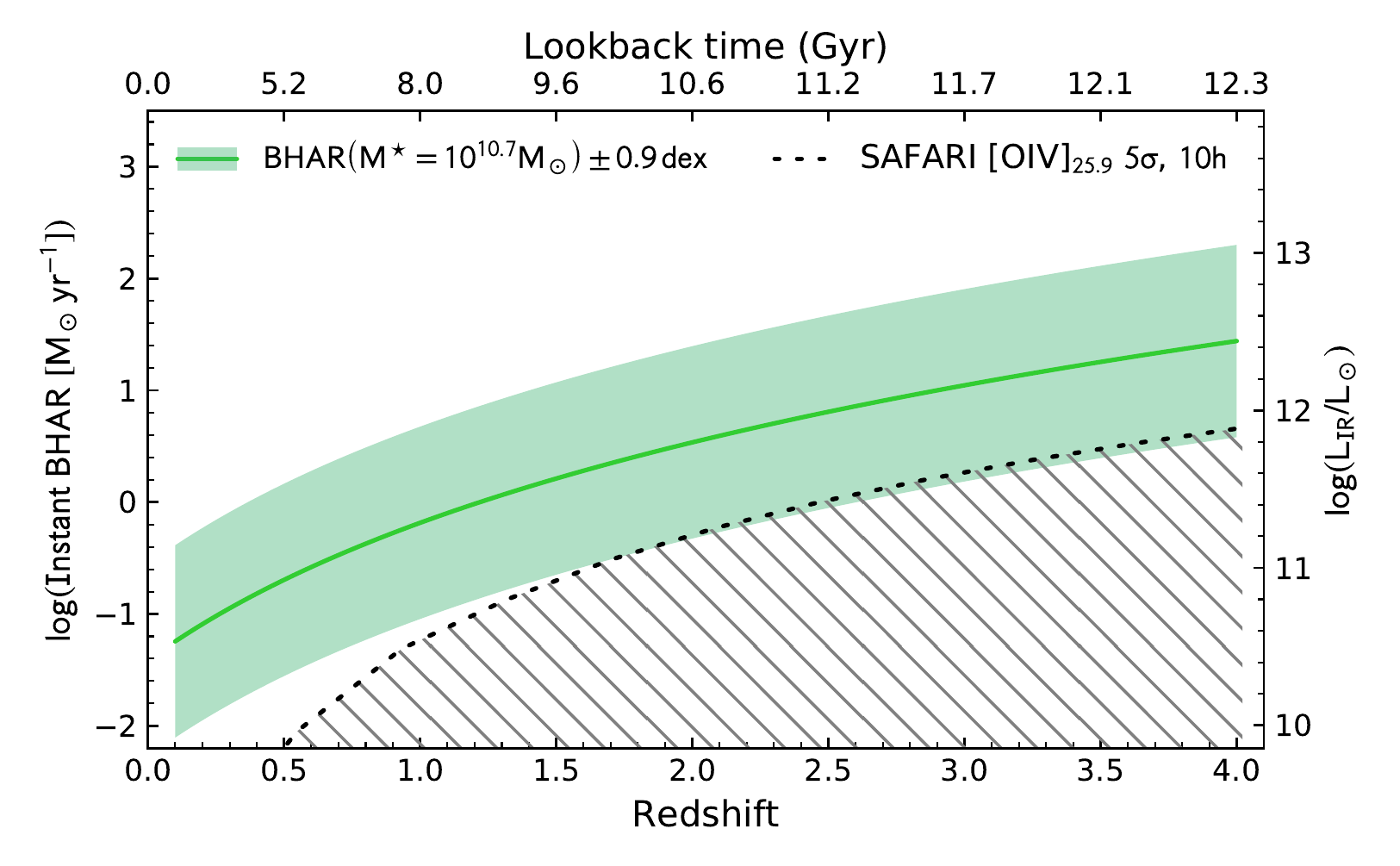}
  \caption{Sample definition for SAFARI spectroscopic follow-up. {\bf (a: left)} Star Formation Rate (SFR in $\rm{M_\odot \, yr^{-1}}$) as a function of redshift for a $10^{10.7}\, \rm{M_\odot}$ galaxy in the Main-Sequence (M.-S., red solid line; \citealt{scoville2017}). The red shaded area shows the $0.35\,\rm{dex}$ intrinsic scatter around the M.-S. \citep{schreiber2015}, while the dark and light orange shaded areas indicate the location of galaxies whose SFR is $+2\sigma$ and $+3\sigma$ above the M.-S., respectively. To derive the detectability of galaxies, two extreme scenarios are considered: solar-like metallicities where PAH emission is expected to be dominant in the mid-IR spectrum, and low metallicity galaxies where the brightest features should be the mid-ionization fine-structure lines. The dashed grey line corresponds to the $5\sigma$ sensitivity limit in the SFR derived from the PAH emission of star-forming galaxies detected in the SMI ultra-deep survey, assuming the calibration given in Table\,\ref{tab:linecal}. Similarly, the black dotted line indicates the SMI and SAFARI $5\sigma$ sensitivities for a $10$\,hr. follow-up observation to derive the SFR from the main tracers at low metallicities, i.e. [\ion{Ne}{ii}]$\rm 12.8\micron$ and [\ion{Ne}{iii}]$\rm 15.6\micron$. {\bf (b: right)} The green solid line indicates the instantaneous BH accretion rate (BHAR) as a function of redshift expected for a M-.S. galaxy with a mass of $10^{10.7}\, \rm{M_\odot}$ during its active BH accretion phase at each epoch. \textit{SPICA} would measure the BHAR in these galaxies through the [\ion{O}{iv}]$25.9\, \rm{\micron}$ emission. The SFR-BHAR relation adopted was derived by \citet{diamondstanic2012} for a sample of Seyfert galaxies in the nearby Universe. The associated dispersion of this relation is shown as a green shaded area. %The grey dashed lines correspond to the $5\sigma$ sensitivity limit in the SMI ultra-deep survey to detect different PAH bands in AGN galaxies, according to the calibrations in Table\,\ref{tab:linecal}.
The black dotted line shows the $5\sigma$ limit to measure the BHAR in AGN through the [\ion{O}{iv}]$\rm 25.9\micron$ luminosity with SAFARI, according to the calibrations in Table\,\ref{tab:linecal}. The hatched areas in both figures indicate the location of galaxies that require observations larger than $10$\,hr.}\label{fig:SAFARI_sample}
\end{figure*}

In order to define the best sample to study galaxy evolution, we first of all need to specify what are the main physical parameters driving galaxy evolution that we want to measure and then how we want to cover them. The main parameters are the SFR and the BHAR, and in Figs.\,\ref{fig:SAFARI_sample}(a) and (b) we show the SFR-redshift diagram for star forming galaxies and the BHAR-redshift diagram for AGN. Of course the other main parameter is the stellar mass of the galaxies, however this will not be measured by a SPICA-like mission, this task relies on future observations in the rest-frame near-IR where the old stellar populations peak, {e.g. using \textit{Euclid} \citep{laureijs2011}, the \textit{Nancy Grace Roman Space Telescope} \citep{green2012}, the  Large Synoptic Survey Telescope \citep[LSST;][]{ivezic2019}}. For this reason, the best cosmological fields must be carefully selected, where information on the galaxy stellar masses { would be already available before the launch of the future space IR observatory}. %\textit{SPICA}. 

{To achieve the scientific goals discussed in this work, a reasonable accuracy to be required for the mass determinations would be better than the typical intrinsic scatter of the scaling relations of interest, for instance the $0.3\, \rm{dex}$ dispersion shown by galaxies around the main sequence \citep{schreiber2015}. Currently, the main source of uncertainty in the mass determinations at high-$z$ comes mainly from the assumptions adopted on the mass-to-light ratios, the IMF, and the star formation history, rather than the precision of the near-IR photometry. These assumptions result in a dispersion in the stellar mass values of the order of $40\%$ \citep[see][]{reddy2012}, that may improve over the next few years in parallel with the advances on the characterization of the stellar populations in high-$z$ galaxies.}
%{\color{red}include mass determination accuracy needed....} 

In Fig. \ref{fig:SAFARI_sample}(a) we show the SFR of the galaxies in the Main Sequence \citep[M.-S.; ][]{rodighiero2010, elbaz2011}, whose values in luminosities as a function of redshift have been taken from \citet{scoville2007}, assuming a galaxy stellar mass of $M^{\star} = 10^{10.7}\, \rm{M_\odot}$ and an associated dispersion around the M.-S. of $0.35\, \rm{dex}$ (red shaded area; \citealt{schreiber2015}). The relation between luminosity and SFR has been taken from \citet{kennicutt2012}. The grey dashed line indicates the $5\sigma$ sensitivity limit of the SMI ultradeep survey to detect and measure the SFR of galaxies using the different PAH bands in the mid-IR spectrum. PAH luminosities have been estimated using the correlations in Table\,\ref{tab:linecal}. Similarly, the black dotted line corresponds to the $5\sigma$ sensitivity limit for a $10$\,h follow-up observation with SMI or SAFARI to detect the two main SF tracers at low luminosities, i.e. [\ion{Ne}{ii}]$\rm 12.8\micron$ and [\ion{Ne}{iii}]$\rm 15.6\micron$ (see low metallicity calibration in Table\,\ref{tab:linecal}). 

Analogously, Fig.\ref{fig:SAFARI_sample}(b) shows the instantaneous BHAR --\,proportional to the AGN luminosity\,-- as a function of redshift for an AGN in a M.-S. star-forming galaxy with $M^{\star} = 10^{10.7}\, \rm{M_\odot}$, assuming the SFR-BHAR calibration derived by \citet{diamondstanic2012} for a sample of nearby Seyfert galaxies. The green shaded area corresponds to the dispersion of the SFR-BHAR relation, i.e. $\sim 0.9\, \rm{dex}$. Note that this estimate applies only for the fraction of galaxies in the M.-S. undergoing a BH accreting phase at each epoch, thus the instantaneous BHAR has not been averaged over the duty cycle of the active nucleus. The host galaxies of these AGN can be detected through the different PAH bands in the SMI surveys (grey dashed line). A measurement of the BHAR { would} be obtained through the [\ion{O}{iv}]$25.9\micron$ line with far-IR spectroscopic observations. We notice here that some caveats have been suggested about the use of this line to unequivocally identify AGN emission. We are aware that this line, having an ionization potential of $\sim 55\, \rm{eV}$, can also be excited by energetic starbursts \citep[e.g.][]{lutz1998} and in Luminous IR Galaxies (LIRG) galaxies without AGN \citep{alonso-h2012}, however the equivalent width of this line in AGN is ten times larger that in a starburst galaxy, and therefore the two types of excitation mechanisms, black hole accretion versus starburst, can be easily recognized. Thus, the [\ion{O}{iv}]$25.9\micron$ line is potentially useful to find the most obscured AGN in the Universe. The hatched areas in both panels indicate the region where observations longer than $10$\,hr. would be required.

By examining Fig. \ref{fig:SAFARI_sample}(a) and (b), it follows that a representative sample of galaxies to be observed by \textit{SPICA} pointed observations will need to cover the following space of the parameters:
\begin{enumerate}
\item  Redshift in the range of $z = 0$--$4$;
\item $\log(SFR/\rm{M_{\odot}\,yr^{-1}})$ from $0.5$ to $3$, depending on $z$; 
\item Instantaneous $\log(BHAR/\rm{M_\odot\,yr^{-1}})$ from $-2$ to $2$, depending on $z$; 
\item The above values in SFR and BHAR translate into total IR luminosities in the range $\sim 10^{10} < L_{IR} < 10^{13}\, \rm{L_\odot}$;
\item Even more importantly, the sample selection needs to take into account environmental over-densities, as well as sub-densities, by looking at the 3-D distribution of galaxies and use the appropriate algorithms to represent the most characteristic deviations from the (theoretical) average distribution.
\end{enumerate}

%Asantha notes: 
%This sample selection is missing a big input from why we need to go from SMI-LR to begin with. Which is selecting galaxies based on the environmental overdensity. This can be written as rho_PAH(over some volume) - mean(rho_PAH) / mean  where rho_PAH is the number density of PAH emitters at a given line say in 8 h^-1 Mpc volume. There are more complicated density estimators but the point is we have to use density/environment to sample a range of conditions, not just select galaxies at some SFR or BHAR. and also we said we will not get BHAR (and perhaps even highly uncertain SFR) from SMI-LR. So how do we select galaxies for follow-up without BHAR etc. The selection has to SFR and BHAR independent. simply based on PAH strengths, environment, and PAH line ratios.\0

We now estimate the size of the sample to be followed-up with { spectroscopic mid- to far-IR pointed observations}. A rough estimate can be made by defining the number of redshift bins from $z = 0$ to $z = 4$, by adopting a step of $\delta z = 0.5$ would require 8 bins.
%\footnote{\color{red} To be discussed: time intervals are very short at high-$z$, i.e. $\sim 400\, \rm{Myr}$ between $z  = 3$ and $3.5$. Shall we adopt a different sampling strategy?}
The number of bins in SFR (equivalent to $L_{IR}$ in SF galaxies) can be set to three: one would trace the evolution of the M.-S. with redshift for the majority of the galaxy population, i.e. reaching SF galaxies at the knee of their mass function ($\sim 10^{10.7}\, \rm{M_\odot}$), assuming as bin width the intrinsic scatter of galaxies around the M.-S. ($\sim 0.35\,\rm{dex}$). Major impact from feedback processes and most of the chemical enrichment is expected for galaxies above the M.-S., thus the second and third bins in SFR would follow the same shape with an offset at $+2\sigma$ and $+3\sigma$ above (see Fig.\,\ref{fig:SAFARI_sample}a).
% REPEATED: Similarly, Fig.\,\ref{fig:SAFARI_sample}(b) shows the expected BHAR for M.-S. galaxies based on the SFR-BHAR correlation from \citet{diamondstanic2012} (green solid line) with its associated $0.9\, \rm{dex}$ dispersion (green shaded area).
Following the argument suggested by \citet{bundy2015} in the definition of a representative sample of galaxies, to require a precision of $0.1\, \rm{dex}$ per decade on the determination of the three main parameters translates into a requirement on the significance of a detected difference between bins to be equal to $\sqrt{n/2}$, where $n$ is the number of galaxies in each bin. Thus, to obtain  detections in the range $4$ to $5\sigma$, we estimate a number of galaxies per bin in a range of $n = 35 - 50$. Therefore, the proposed sample would have a total number of galaxies in the range of \textit{800--1200 galaxies}, i.e. \textit{n} galaxies $\times~3$ luminosity bins $\times~8$ redshift bins. This is in agreement with the statistical demands of current studies addressing the chemical evolution of galaxies, perhaps the most demanding ones due to the larger number of parameters that need to be covered (stellar mass, SFR, metallicity). Also, metallicity tracers are likely more affected by the scatter in gas density (n$_{\rm H}$) and ionization parameter (log U) than SFR and BHAR tracers. For instance, in a recent study by \citet{onodera2016} using a sample of 43 galaxies in the $3 < z < 3.7$ range, the authors conclude that the size of their sample is still too small to overcome the uncertainties in the measurement errors, which dominate over the intrinsic scatter of the mass-metallicity relation (MZR). This scatter is directly associated with the timescales of the physical processes that drive the galaxies in and out of the average behaviour traced
by the MZR \citep[e.g.][]{torrey2019}, and therefore to quantify this dispersion is a fundamental step to characterise the physics of galaxy evolution. The minimum sample size proposed in this work is a factor of $\gtrsim 2$ larger than the sample from \citet{onodera2016}, with $\sim 100$ galaxies in the $3 < z < 3.5$ range.

Regarding the proposed bin sizes in redshift, the optical measurements suggest that changes in the metallicity of galaxies, especially above the knee of the mass function at $10^{10.7}\, \rm{M_\odot}$, accelerate with increasing redshift. From $z \sim 0$ to $z \sim 2.2$ (in $10.9\, \rm{Gyr}$) the metallicity of massive galaxies decreases by only $\sim 0.15\, \rm{dex}$ \citep{maiolino2008}. However, between $z \sim 2.2$ and $z = 3$ ($0.85\, \rm{Gyr}$) the drop in metallicity is $0.5$--$0.6\, \rm{dex}$ \citep{troncoso2014,onodera2016}. Thus, a relatively fine sampling in redshift is required to be able to follow these changes in the chemical abundances of galaxies. Moreover, taking into account that the changes in other main parameters, like the SFR and the BHAR show the opposite behaviour, i.e. their variation is higher at $z < 1$ \mbox{\citep{madau2014}}, we propose an observational strategy based on a constant bin size of $z = 0.5$ across the whole redshift range ($1 < z < 4$).

\subsection{Spectroscopic follow-up with SAFARI and SMI}\label{sec:safari}

In this section, we estimate the limits in terms of total IR luminosities and redshifts that can be reached with { the two template instruments of the SPICA project,} SAFARI and SMI, follow-up observations of the above defined sample of galaxies (see Section\,\ref{sec:sample}).
The SAFARI sensitivity has been taken from \citet{roelfsema2018}. %For technical reasons, to ensure the optimal stability of the whole observing chain, we propose to limit the maximum observing time for each observation to 19 hours, however %there is no major stop over to observe with SAFARI over this integration time, being 
%the only "hard limit" to the length of an observation the 19 hours of continuum operation allowed on each observing day to any of the SPICA instruments.
For the sake of simplicity of the simulation, we have limited the integration time to $10$\,hr. %, \textbf{which corresponds to the expected time available for science observations during one satellite orbit}.
%but that this is not really a limit in terms of the observatory or detectors.

\begin{figure*}
    \centering
    \includegraphics[width=0.8\columnwidth]{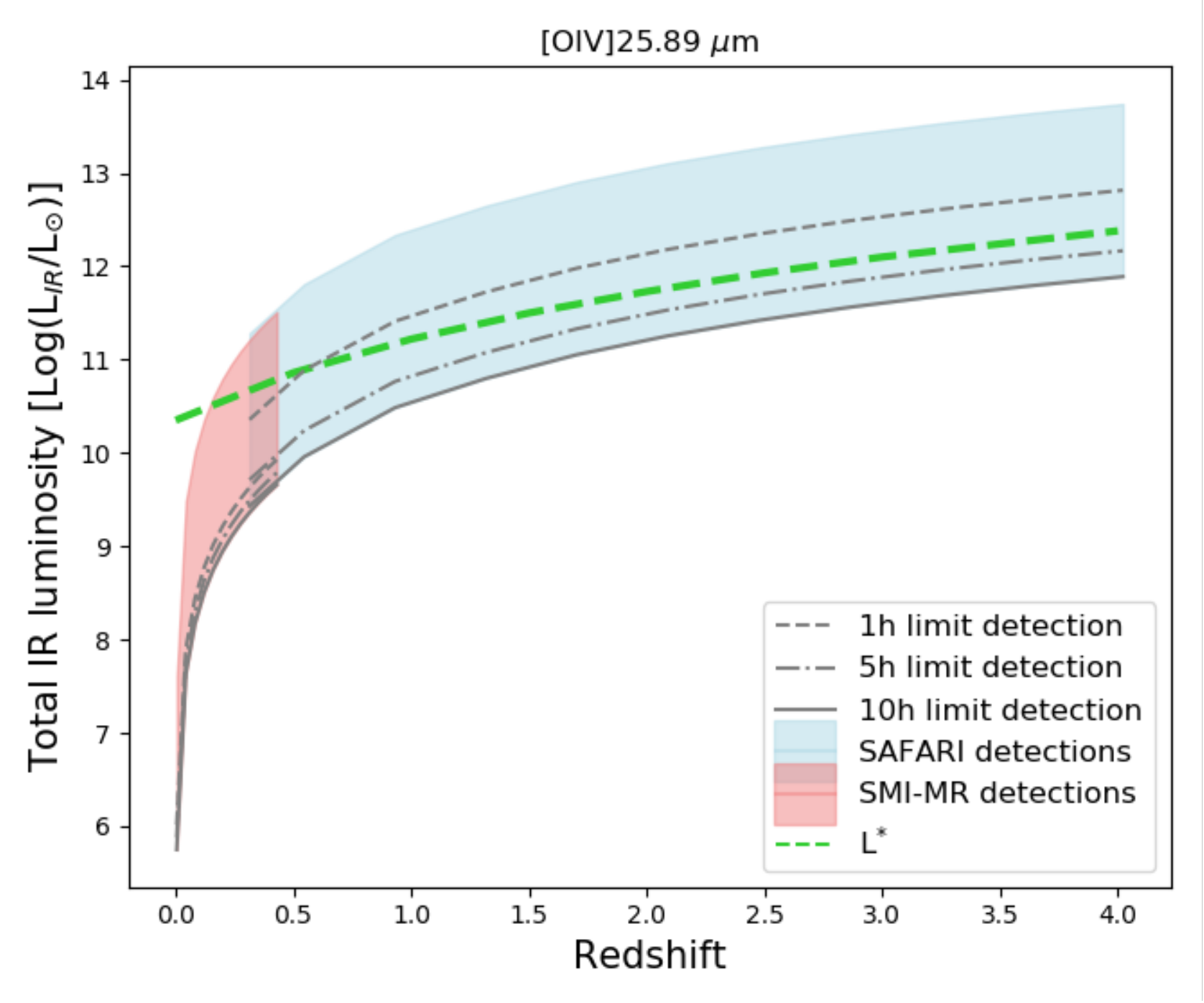}
    \includegraphics[width=0.8\columnwidth]{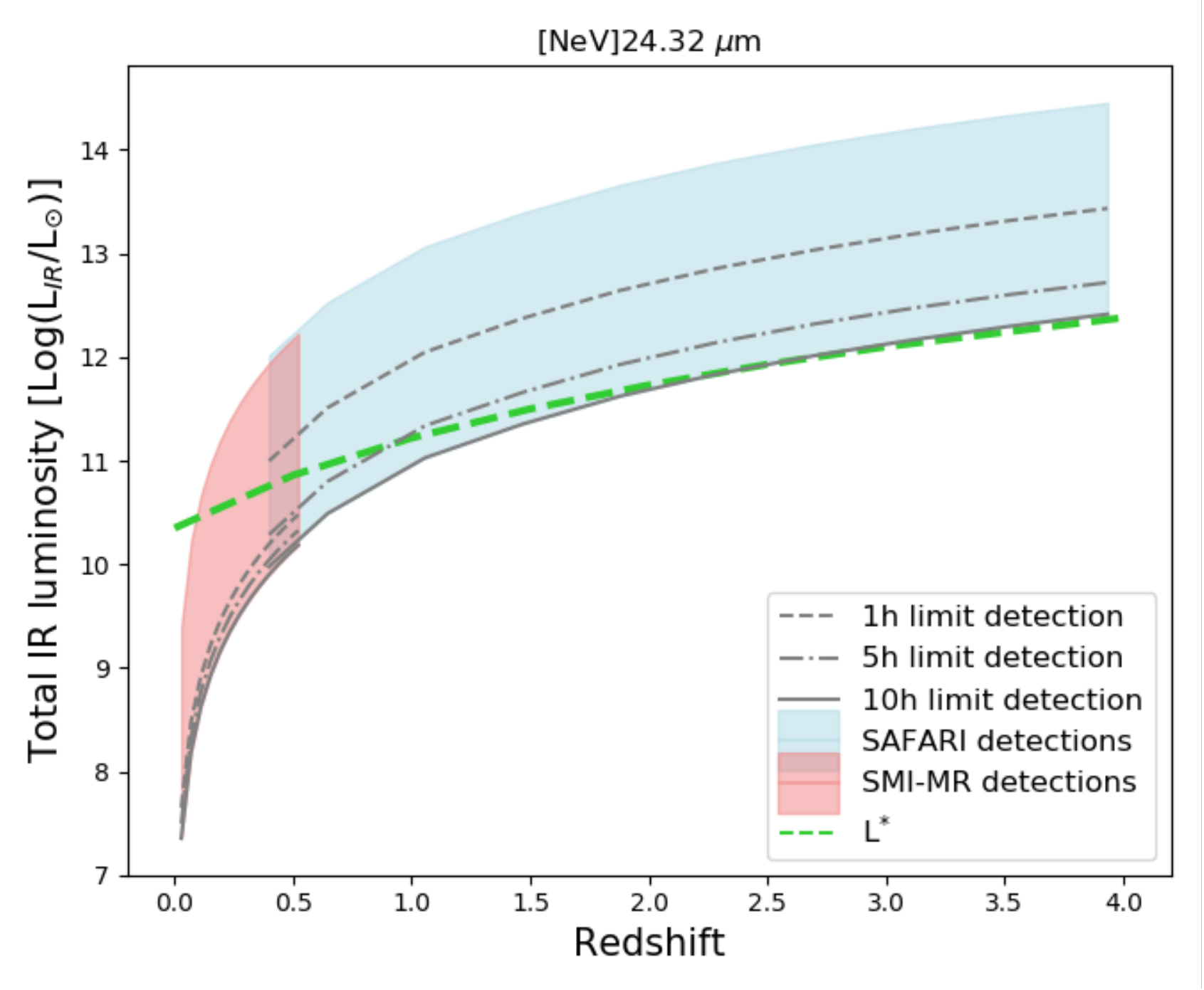}
    \caption{Redshift-luminosity diagrams simulating the AGN detections with SAFARI and SMI MR mode. {\bf (a: left:)} Detectability of the [\ion{O}{iv}]$\rm 25.89\micron$ line. The red shaded area represents the detectability with SMI-MR, while the blue area indicated the detectability with SAFARI; grey horizontal lines represent lower limits for different observational times, namely less than 1 hour for the dot-dashed lines, 5 hours for the dashed lines, and 10 hours for the solid lines. The horizontal green line indicates the knee of the luminosity function. {\bf (b: right:)} Detectability of the [\ion{Ne}{v}]$\rm 24.32\micron$ line.}
    \label{fig:SAFARI_AGN}
\end{figure*}

\begin{figure*}
    \centering
    \includegraphics[width=0.8\columnwidth]{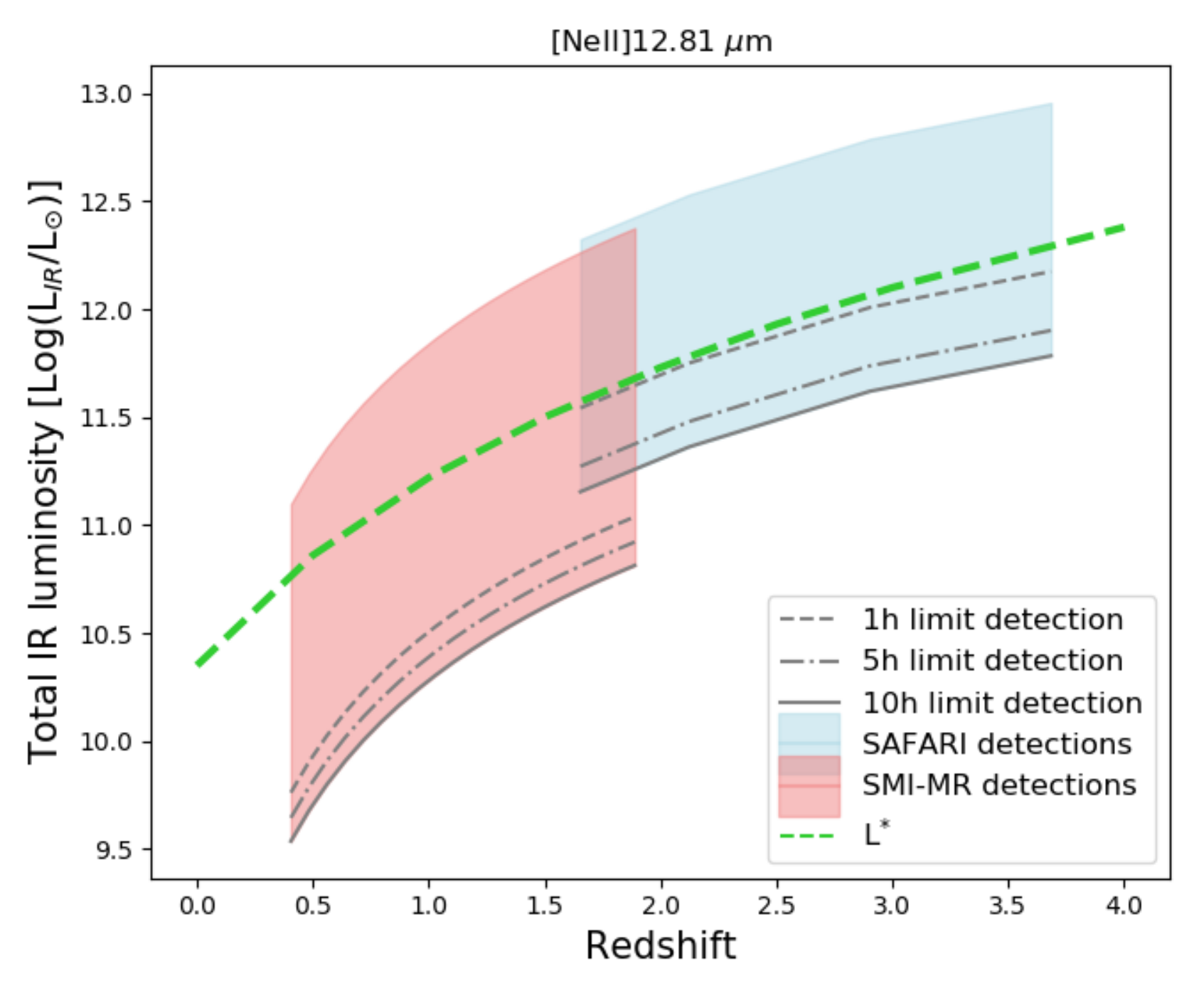}
    \includegraphics[width=0.8\columnwidth]{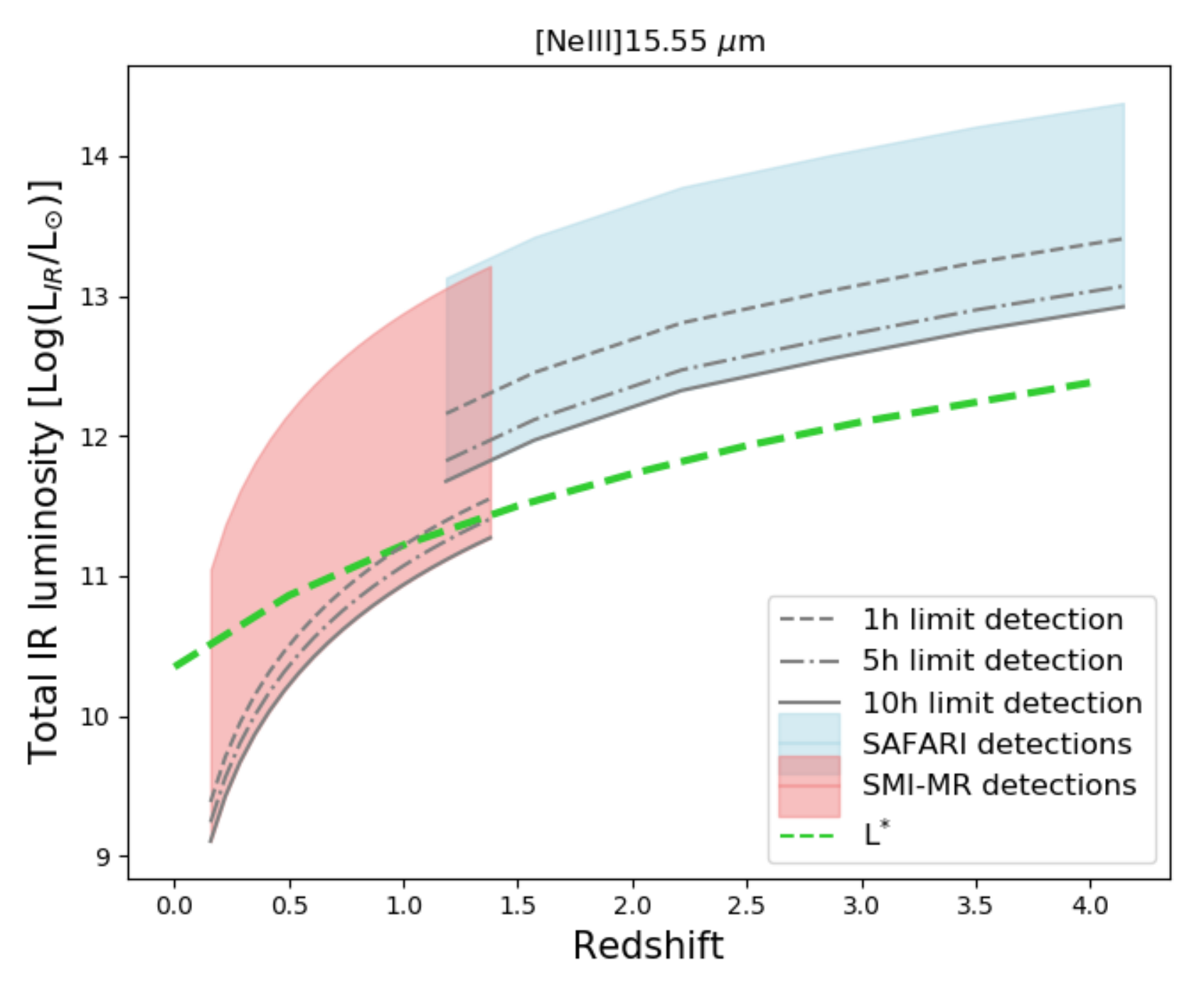}
    \caption{Redshift-luminosity diagrams simulating the SF galaxies detections with SAFARI and SMI MR mode. {\bf (a: left:)} Detectability of the [\ion{Ne}{ii}]12.81$\mu$m line. The same color code is applied as in Fig.\,\ref{fig:SAFARI_AGN}. {\bf (b: right:)} Detectability of the [\ion{Ne}{iii}]15.6$\mu$m line.}
    \label{fig:SAFARI_SFG}
\end{figure*}

\begin{figure*}
    \centering
    \includegraphics[width=0.86\columnwidth]{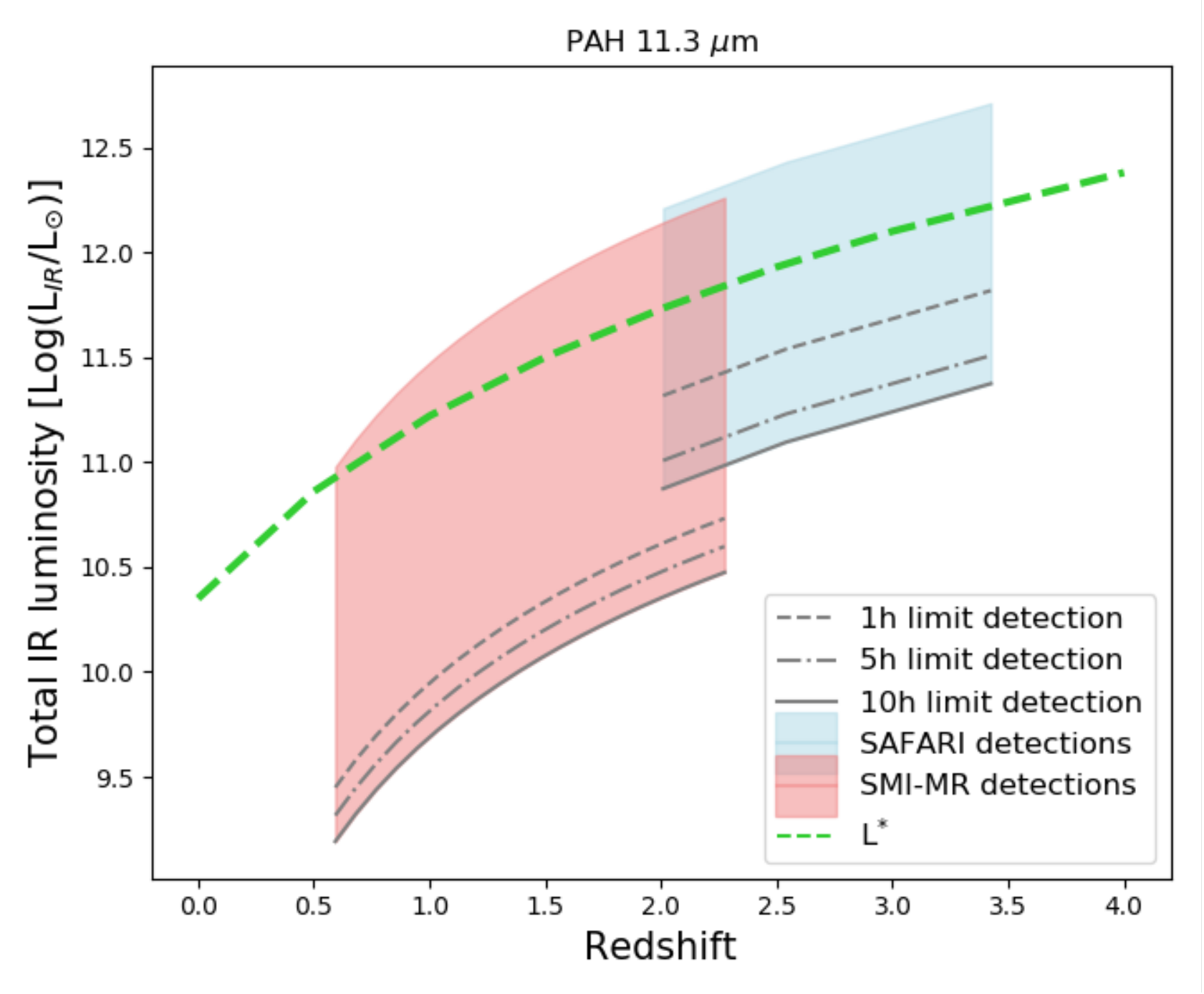}
    \includegraphics[width=0.8\columnwidth]{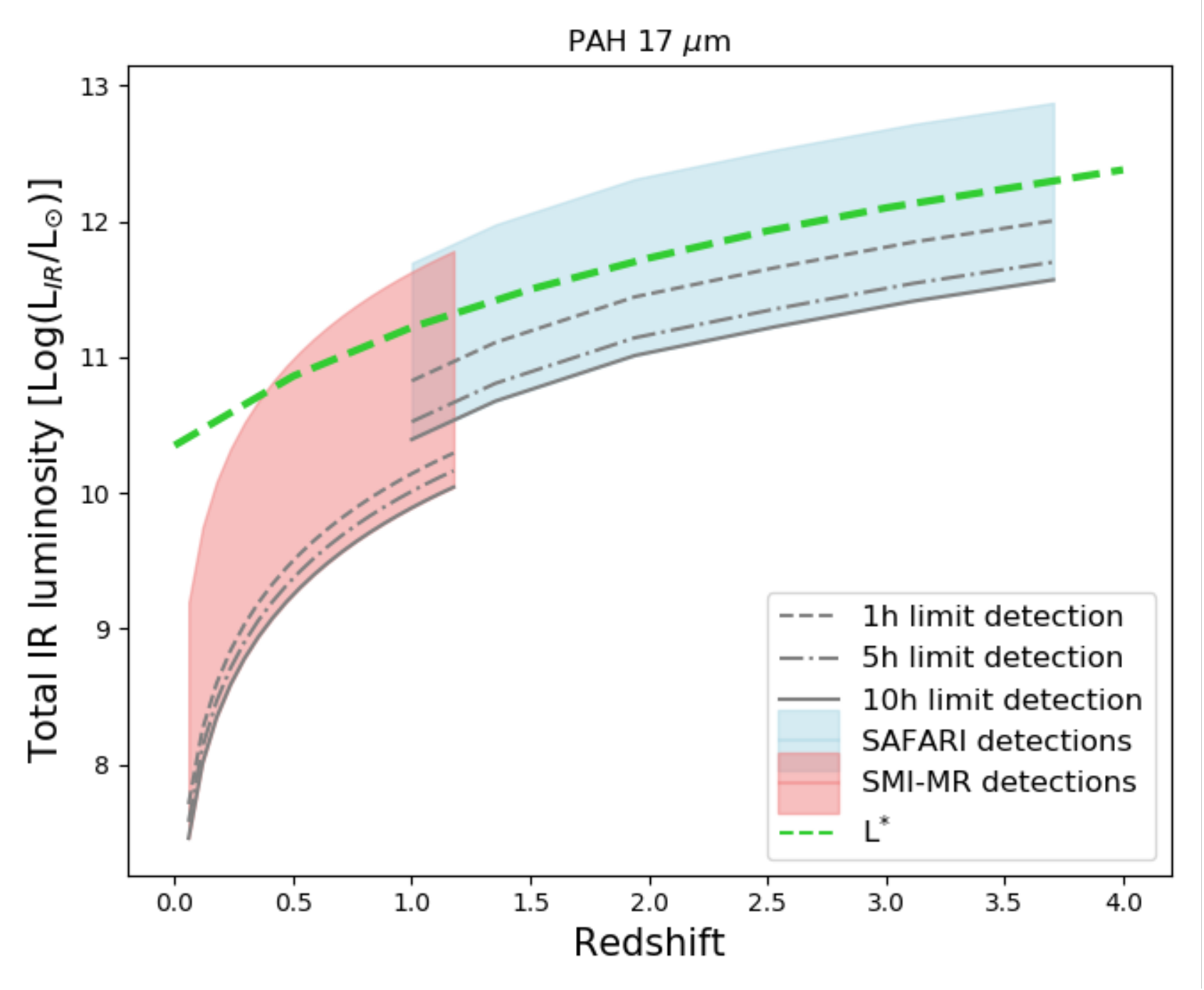}
    \caption{Redshift-luminosity diagrams simulating the SF galaxies detections with SAFARI and SMI MR mode. {\bf (a: left:)} Detectability of the PAH $11.3\, \rm{\micron}$ band. The same color code is applied as in Fig.\,\ref{fig:SAFARI_AGN}. {\bf (b: right:)} Detectability of the PAH $17.0\, \rm{\micron}$ band.}
    \label{fig:SAFARI_SFG_PAH}
\end{figure*}

\begin{figure*}
    \centering
    \includegraphics[width=0.85\columnwidth]{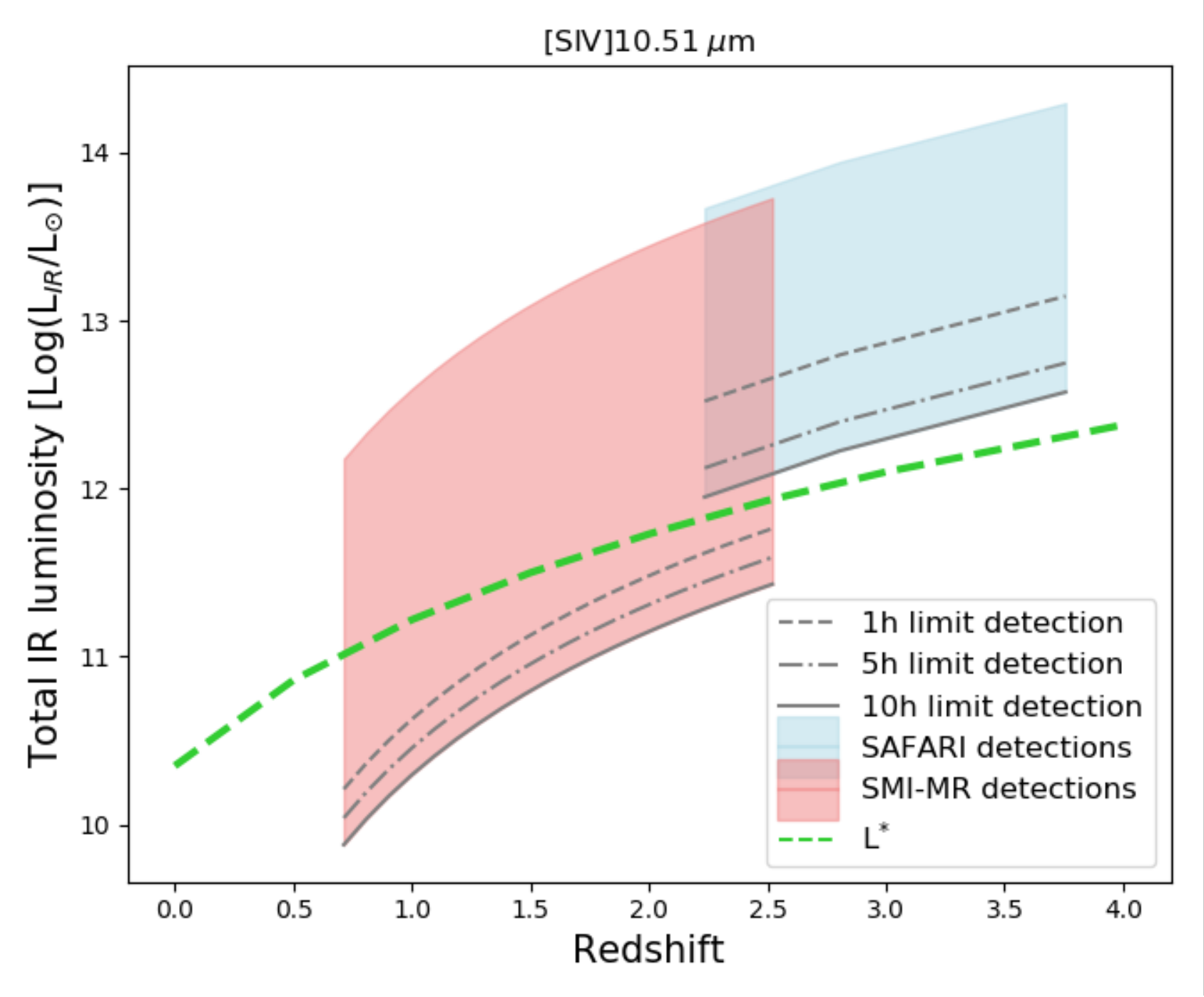}
    \includegraphics[width=0.84\columnwidth]{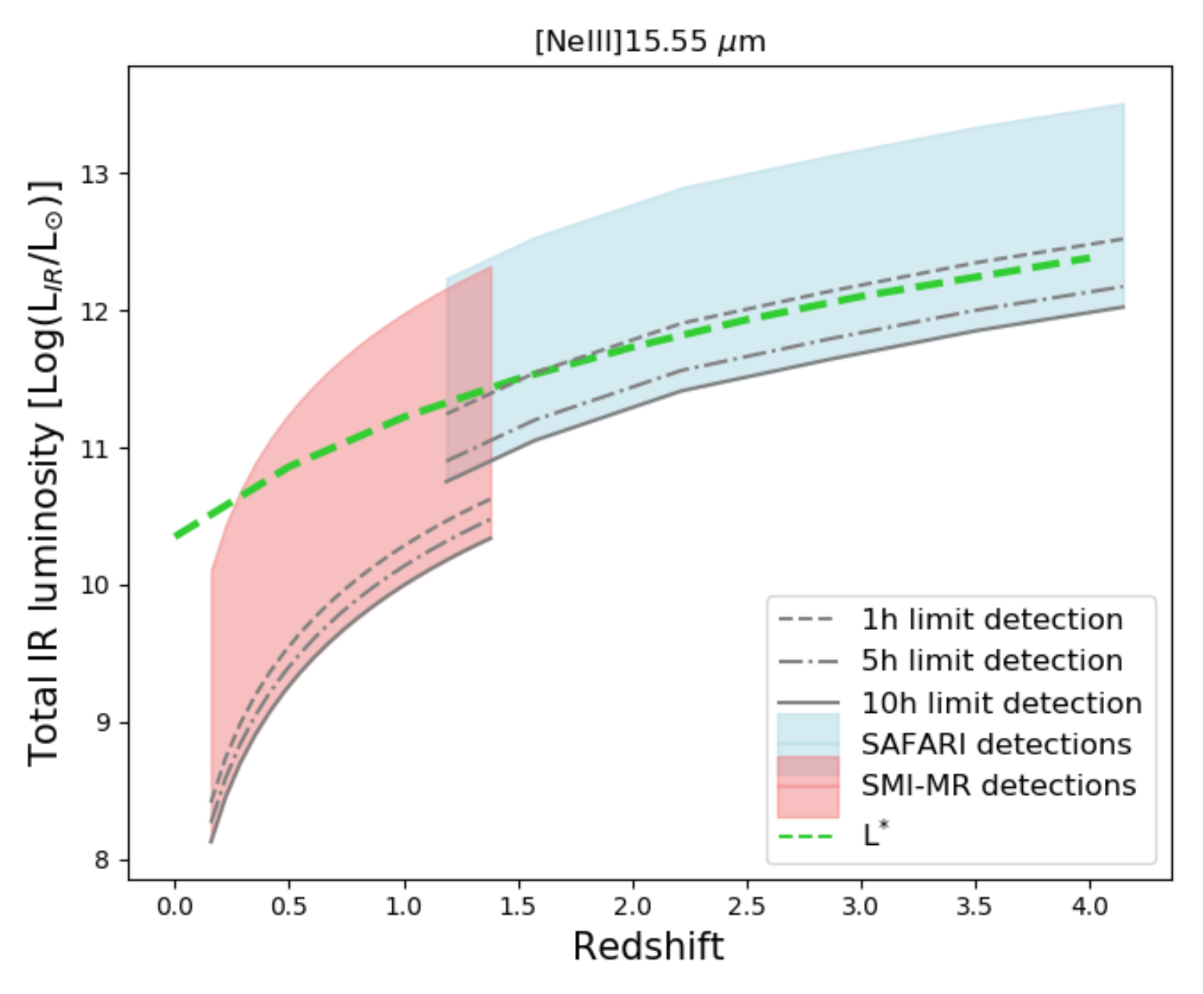}
        
    \caption{Redshift-luminosity diagrams simulating low metallicity galaxies detections with SAFARI and SMI MR mode. {\bf (a: left:)} Detectability of the [\ion{S}{iv}]10.51$\mu$m line. The same color code is applied as in Fig.\,\ref{fig:SAFARI_AGN}. {\bf (b: right:)} Detectability of the [\ion{Ne}{iii}]15.6$\mu$m line.}
    \label{fig:SAFARI_LMG}
\end{figure*}

Figs.\,\ref{fig:SAFARI_AGN}(a) and \ref{fig:SAFARI_AGN}(b) show the luminosity-redshift diagrams with the AGN that SAFARI and SMI { would} be able to detect at the various luminosities and redshifts using, respectively, the [\ion{O}{iv}]$\rm 25.9\micron$ and [\ion{Ne}{v}]$\rm 24.3\micron$ lines.  Figs.\,\ref{fig:SAFARI_SFG} and \ref{fig:SAFARI_SFG_PAH} show the luminosity-redshift diagrams with the SF galaxies that SMI and SAFARI { would} detect for the [\ion{Ne}{ii}]$\rm 12.8 \micron$ and [\ion{Ne}{iii}]$\rm 15.6 \micron$ lines (Fig. \ref{fig:SAFARI_SFG}(a) and Fig. \ref{fig:SAFARI_SFG}(b), respectively), and the PAH bands at $\rm 11.3 \micron$ and $\rm 17 \micron$ (Fig. \ref{fig:SAFARI_SFG_PAH}(a) and Fig. \ref{fig:SAFARI_SFG_PAH}(b), respectively). In Fig.\,\ref{fig:SAFARI_LMG}(a) and (b), we show the [\ion{S}{iv}]$\rm 10.5\micron$  and the [\ion{Ne}{iii}]$\rm 15.6\micron$ line predictions, respectively, for SF galaxies adopting the low-metallicity calibration. In all these figures the detection limits have been assigned at 5 $\sigma$. The figures show that the combination of SAFARI and SMI pointed spectroscopic observations { would} be able to detect fine-structure lines as well as PAH features in any of the three types of galaxy populations that have been considered. 

In other words, SPICA spectroscopy { would} be able to measure in a composite galaxy, which has both an active nucleus and star formation, the spectroscopic tracers of the two components. The deconvolution of the contribution of these components to the total luminosity of a galaxy { would} then be done through various techniques, including the use of the EW of the lines and features and of particular line ratios \citep[see, e.g.,][]{tommasin2010,diamondstanic2012} or through photoionization modeling \citep[see, e.g.,][]{spinoglio1992,spinoglio2005,pereirasantaella2010,zhuang2019}. Once the two components have been disentangled, we { would} be able to measure the SF and the BHA luminosity functions in the same population of galaxies. If the emitting gas metallicity is lower than solar, the mid-ionization lines of [\ion{S}{iv}] and [\ion{Ne}{iii}] would be relatively brighter. { SPICA-like} pointed spectroscopic observations { would} detect AGN well below the knee of the luminosity function with a total integration time of less than 5 hours and 10 hours, to detect the [\ion{O}{iv}]$\rm 25.9\micron$ and [\ion{Ne}{v}]$\rm 24.3\micron$ lines, respectively, at the M.-S. luminosities at any redshift below $z \lesssim 4$. For Star Forming galaxies, follow-up SPICA observations { would} detect the [\ion{Ne}{ii}]$\rm 12.8\micron$ line, always in less than 1 hour, in M.-S. galaxies, while the [\ion{Ne}{iii}]$\rm 15.6 \micron$ line can be detected with SMI only at z$<$1.5. The same Star Forming galaxies, however { would} always be detected through the PAH features in integration times much shorter that 1 hour. Finally the Low Metallicity galaxies { would} easily be detected at their M.-S. luminosities with the [\ion{Ne}{iii}]$\rm 15.6 \micron$ line and, up to z$\sim$2.5, with the [\ion{S}{iv}]$\rm 10.5\micron$ line. 
{Follow-up observations with the high resolution channels of the mid- to far-IR spectrometers onboard of a SPICA-type observatory, reaching resolving power of the order of a few thousands, { would} measure the profiles of the mid-IR fine structure lines to study the kinematics of the narrow-line regions in AGN, following the work done in the Local Universe with {\it Spitzer} \citep{dasyra2011} and also disentangle AGN from star formation spectroscopically.}
%\textcolor{red}{No mention is made of kinematics. This is puzzling, because both SAFARI and SMI-MR have adequate resolution for this to be done. referee 9.}

%\item 
With the pointed observations of the above defined sample of galaxies, {a SPICA-type observatory would} be able to determine, in addition to SFRs and instantaneous BHARs, the following physical properties of the galaxies as a function of cosmic age (redshift) and stellar mass: the metal abundances and outflow/infall occurrence. %
For this latter physical property in galaxies, we refer to \citet{gonzalezalfonso2017a}, where a complete assessment of the detection with SPICA of AGN feedback and feeding has been presented. 
%The evolution of these physical quantities as a function of cosmic time will tell us the  history of {\it the physical processes} taking place during galaxy formation and evolution. 
The study of the evolution of these quantities as a function of time will unveil the role of the most basic physical processes taking place in galaxy evolution throughout a large fraction (90\%) of cosmic history. 

\section{Discussion}\label{discuss}

The main goal of the mid- to far-IR %\textit{SPICA} 
cosmological surveys is to provide a complete characterization of the physical processes driving the growth of galaxies and BHs across the cosmic noon, covering all the relevant aspects such as the environment, the feedback from massive outflows, and the production of heavy elements and dust. The unique spectral mapping capabilities of a { SMI-type instrument would} deliver a complete 3-D view of the SF in the Universe by observing wide and deep cosmological volumes up to $z \sim 4$, from voids to cluster cores over 90\% of the cosmic time. Based on this survey a number of targets will be selected for dedicated follow-up IR spectroscopic observations, 
%with SAFARI and SMI, 
to provide answers to some of the most challenging questions in the field of galaxy evolution: \\

\noindent
\textit{What physical processes led to the peak of SF and BHA during the cosmic noon?}\\[0.1cm]
The global estimates of SFR and BHAR densities in galaxies across cosmic time show a parallel evolution with a sharp drop of the total energy released by these two phenomena, decreasing by a factor $\sim$30 from the peak at the cosmic noon till the present time \citep{madau2014}. %Disentangling the SF and BHA components in each galaxy of a statistically significant sample is essential to understand if and how the evolution of these two phenomena is connected. %, which are their characteristic timescales, and the delay between their peak activity. 
%\textit{SPICA} 
{ Mid- to far-IR spectroscopic} cosmological surveys { would} peer through the heavily opaque medium of M.-S. galaxies at high-z to provide a full census of the SFR (through low-ionization lines and PAH bands; \mbox{\citealt{tommasin2010}}, \mbox{\citealt{xie2019}}) and the instantaneous BHAR (using highly ionized species such as O$^{3+}$ and Ne$^{4+}$; \mbox{\citealt{tommasin2010}}, \mbox{\citealt{petric2011}}, \mbox{\citealt{diamondstanic2012}}, \mbox{\citealt{alonso-h2012}}) in the Universe up to $z \sim 4$, as shown in Fig.\,\ref{fig:SAFARI_sample}.
The measure of the SF and BHA components in the same large and representative sample of galaxies covering a wide range in cosmic time will allow us to understand the connections of SF with BHA as a function of stellar mass and environment and the relative role of these two components in galaxy growth and assembly. %A temporal delay between the SF and the BH activity peaks has already been shown both observationally \citep{wild2010} and theoretically \citep{hopkins2012} to be of the order of few hundreds million years.
\\

\noindent
\textit{Which physical processes are behind the quenching of SF in galaxies?}\\[0.1cm]
%Current simulations require a strong contribution of feedback from accreting BHs to reconcile the galaxy mass function with that of dark matter halos \citep{silk2012}, explain the migration of active SF galaxies in the blue cloud to the red and dead sequence \citep{heckman2014}, and the local scaling relations between the BH masses and host galaxy properties \citep{kormendy2013}. This feedback is mainly ascribed to quasar outflows and jets in AGN, however robust observational evidences probing such scenario at high-z are still missing.
{A SPICA-type observatory would} be able to detect massive molecular gas outflows through P-Cygni profiles in the OH\,119\,$\rm{\micron}$ and OH\,79\,$\rm{\micron}$ doublets, { as well as strong absorption profiles in the OH doublets at 65\,$\rm{\micron}$ and 84\,$\rm{\micron}$, up to $z \sim 2$,
while at lower redshift (z$<$1) also H$_2$O %lines %with $\lambda < 77\,\rm{\micron}$ 
and OH$^+$ lines would be detected %with SAFARI 
at high spectral resolution
\mbox{\citep{gonzalezalfonso2017, runco2020}}. 
}
Massive outflows are expected in bright galaxies above the M.-S., where strong feedback should regulate their growth by quenching the SF, driving these galaxies back to the M.-S. \citep[e.g.][]{tacchella2016}. 

An important issue in feedback is the relative role of AGN-driven winds versus starburst/SNe driven winds. Which is the most dominant process as a function of infrared luminosity L$_{IR}$ or stellar mass or environment? Is AGN driven feedback taking place mostly in galaxy groups/proto-cluster centers? Untangling the two (SNe driven vs AGN driven) is crucial and can be done by considering the energy balance that can be constrained by the measure of the outflow kinetic energy \citep{gonzalezalfonso2017a}, the AGN bolometric luminosity and the SFR \citep[see, e.g.][]{cicone2014,fiore2017}. %SNe driven wind can be detected through broad line components in highly-ionized lines such as those from [OIV] and [NeV].
%\textcolor{red}{why only mention to OH lines profile as made as probes of outflows? What about other species? referee 10.}

{ The ionized phase of AGN outflows would be traced, with the high-resolution channels spectrometers onboard a SPICA-like observatory, by highly ionized emission lines such as [OIV] 25.9$\mu$m \citep{pereirasantaella2010} %(Pereira-Santaella et al. 2010), 
[\ion{O}{iii}] 52 and 88$\mu$m, or [\ion{Ne}{v}] 14.3 and 24.3$\mu$m in nearby galaxies \citep{spoon2009, spoon2013}. %(Spoon \& Holt 2009, Spoon et al. 2013). 
While this phase represents a minor fraction of the total outflowing mass, its velocity is typically much higher ($>\times$5) when compared to the molecular phase, and can be driven farther away from the galaxy. Thus, the spatially resolved kinematic analysis of highly ionized gas in local, z $<$ 0.1 galaxies with SPICA allows for the determination of the amount of gas that can effectively escape the galactic potential well, contributing to the definitive quenching of the host. For z\,$>$\,0.3, nuclear outflows would be detected in the [\ion{O}{iv}] 25.9$\mu$m line with the sensitivity required to detect the OH doublets, which would be combined with the simultaneous observations of the [\ion{O}{iii}] lines to probe the ionization parameter of the outflowing ionized gas, the relationship with the ultra-fast outflows in X-rays and the molecular phase, including the stratification of the winds \citep[e.g., ][]{tombesi2015}, %(e.g. Tombesi et al. 2015), 
and their AGN/SB origin. }
\\

\noindent
\textit{What is the role and timescale of obscured BH-grow in shaping galaxy evolution?}\\[0.1cm]
Heavily obscured nuclei are also potential sources of AGN feedback. The buried phase is expected to precede the launch of massive winds of molecular gas during the quasar phase, following the evolutionary sequence suggested by OH doublet diagnostics in nearby analogs of these sources \mbox{\citep{gonzalezalfonso2017}}. Furthermore, deeply buried accreting BHs could be a substantial part of the AGN population without exceeding the observed cosmic X-ray background  (e.g. \citealt{gilli2007}, \citealt{treister2009}, \citealt{comastri2015}), their large obscuration makes them virtually invisible to current X-ray surveys. Since their discovery, these targets became a major science topic for future X-ray missions such as \textit{Athena} \mbox{\citep{nandra2013}}, which envisages surveys to reveal this population. {SPICA-type} cosmological surveys {would} provide a unique contribution in the field by revealing the dust reprocessed radiation from elusive sources in the \textit{Athena} surveys, i.e. Compton-Thick (CT) nuclei with the largest obscuring columns ($N_{\rm H} \gtrsim 10^{24.5}\rm{cm^{-2}}$) and moderate luminosities %($L_{\rm X}\sim 10^{42.5}\rm{erg\,s^{-1}} $) up to $z \sim 2$ 
(Barchiesi et al. 2021). 
%How many compton thick will we follow-up and how many will be find in a 2 deg2 or 10 deg2 survey? This is the most critical issue, since compton thick are relatively low surface density on the sky. They may require X-ray overlap.  and in that sense isn't really a SPICA goal by itself but good overlapping science for a different part of the yellow book.
In the high-luminosity regime ($L_{X} = 10^{44.0} \rm{erg\,s^{-1}}$), both \textit{Athena} and {a SPICA-like observatory would} detect CT AGN. This strong synergy is key to accurately reconstructing the accretion luminosity over cosmic time up to high redshift (z$\sim$4). Conversely, in the faint/moderate-luminosity regime ($L_{X} \sim 10^{42.3} \rm{erg\,s^{-1}}$) only {a SPICA-like observatory would} be able to recover a significant fraction of this elusive AGN population in the X-rays. The presence of a CT AGN would then be recognised, at least up to z$\sim$2, via the expected detection (based on Spitzer/IRS results at z$<$1) of a deep silicate absorption feature at 9.7$\mu$m in the SMI-LR spectra available for about 60\% of the photometrically detected sources (Barchiesi et al. 2021, submitted to PASA). { Mid- to far-IR spectroscopic follow-up} of these deeply buried nuclei { would} then measure the BHAR through the high-ionization fine-structure lines of [OIV] and [NeV] as well as the unique physical characterisation of the warm nuclear regions ($\gtrsim 300\, \rm{K}$) in the pre-quasar phase, e.g. through the detection of the OH absorption doublets at $65\, \rm{\micron}$ and $84\, \rm{\micron}$.\\
                                                          
\noindent
\textit{How did obscured SF build today's most massive clusters?}\\[0.1cm]
The build-up of the most massive galaxy clusters that we observe in the nearby Universe is directly linked to the formation of the large-scale structure in the Universe and the assembly of massive dark matter halos. Galaxies grow faster in dense environments, thus clusters are dominated by massive early-type galaxies already by the end of the cosmic noon ($z \lesssim 1$--$1.5$). To understand the formation of these structures requires finding proto-clusters at higher redshifts, which is challenging since the growing galaxies were still undergoing a dusty obscured SF phase at that epoch, and the systems were not yet virialized. This implies not only the need of deep IR observations, but also to obtain redshift information for individual galaxies over wide and deep cosmological volumes. In this aspect \textit{SPICA} is unique, since the SMI deep spectrophotometric surveys would provide a full tomographic view of clustering as a function of time up to $z \sim 3.5$ over wide (from one to fifteen square degrees) fields in the sky. %(see, e.g., Fig.\,\ref{fig:lumfun_pred}a). 
This 3-D view of the Universe, which can be carried out only with { a SPICA-like} deep spectrophotometric survey, will allow the identification of several hundreds to thousands of new proto-clusters that will serve as targets for follow-up spectroscopic studies.
%This has not been possible with the {\it Herschel} mission and is beyond the capability of {\it ALMA}, while {\it SPICA} will be the only instrument which will allow it.
\\

\noindent\textit{How were the dust-embedded earliest stages of elliptical galaxies and bulges of MW-like progenitors?}\\[0.1cm]
Massive elliptical galaxies at the center of galaxy clusters and the bulges of MW-like spirals were already in place by $z \sim 1$ \citep{perezgonzalez2008}. The rapid formation of these structures left a clear imprint in their content of heavy elements --\,metals\,-- which are byproducts of the SF and therefore scale with the stellar mass growth. Optical-based diagnostics show a drastic drop of about an order of magnitude in the metallicities of massive galaxies ($\gtrsim 10^{10}\, \rm{M_\odot}$) between $z \sim 2$ and $3$ {(\citealt{maiolino2008,mannucci2009,troncoso2014,wuyts2014,hunt2016,onodera2016,suzuki2020}; but see also \citealt{cullen2019}, \citealt{sanders2020a})}, which is in contrast with the large amounts of dust inferred for SF galaxies at that epoch \mbox{\citep{madau2014}}. Similar discrepancies are found in submillimeter galaxies at high-z and local analogs of dusty-obscured SF galaxies (\mbox{\citealt{santini2010}}, \mbox{\citealt{herreracamus2018}}). {Mid- to far-IR medium-resolution spectroscopy} would allow the use of metallicity diagnostics based on IR fine-structure lines, largely insensitive to extinction and gas temperature, to probe the true chemical age of galaxies during the cosmic noon \citep{fernandez2017}. {Additionally, IR metallicities would allow to probe variations in the chemical abundances, such as N/O ratios and alpha-enhancement, which provide unique information on the history of star formation of these galaxies \citep[e.g.][]{amorin2010,thomas1999}}.

\noindent
\textit{How obscured are the nuclei of galaxies and what are the dust properties in the AGN and starburst host galaxies?}\\[0.1cm]
An outcome of { mid-IR} spectroscopic observations will be the complete census of the shape and strength of the 9.7 $\mu$m silicate feature (and, to a lesser extent that at 18 $\mu$m) in the thousands of detected galaxies in the deep and ultra-deep surveys. The silicate features provide information on: {\it (i)} the obscuration: %the discrimination between type 1 and type 2 AGN can be assessed, in conjunction with other diagnostics, 
dust emission and absorption in the line of sight can be measured from the profile of the 9.7 $\mu$m feature, this can be combined with additional diagnostics to discriminate between type 1 and type 2 AGN and related to the X-ray obscuration \citep{lacaria2019}; {\it (ii)} the amount of silicate in the dust composition; {\it (iii)} as well as its properties (e.g. grain size distributions and crystallinity); {\it (iv)} the central wavelength, width, and relative strength of the 9.7 and 18 $\mu$m features carry information on the dust chemical composition and grain properties \citep{xie2017,fernandez2017}. Moreover, combining information from PAH and silicates provides insights into the physical conditions of the interstellar and circumnuclear medium in galaxies. Finally, the correlation among the silicate absorption in AGN and the hydrogen column density N$_{H}$ as measured from X-rays, even if often ascribable to host galaxy obscuration and not necessarily to a compact torus surrounding the AGN \citep{goulding2012}, might demonstrate the usefulness of silicates to detect and recognise heavily obscured AGN, i.e. the Compton thick objects.\\

\noindent
\textit{What are the properties of dust up to the epoch of reionization?}\\[0.1cm]
The origin and composition of dust in galaxies is one of the most challenging problems in chemical evolution studies \citep[][and references therein]{calura2008}.
Dust production originates in stars at the end of their lifetime (core collapse Supernovae (SNe), asymptotic giant branch (AGB) stars and red supergiants, and Wolf Rayet stars), 
when atomic metals in their enriched ejecta condense in dust seeds (\mbox{\citealt{sarangi2015}}, \mbox{\citealt{delooze2017}}, \mbox{\citealt{marassi2019}}) %(Sarangi et al. 2015, De Looze 2017, Marassi et al. 2019) 
and are released in the surrounding ISM, after surviving the local reverse shock  \citep{bocchio2014, bocchio2016}.%(Bocchio 2014, 2016). 
Depending on the ISM phase in which grains live, they can grow in cold/warm regions by accreting metals in gas phase \citep{hirashita2012}%(Hirashita 2012) 
or can suffer erosion by sputtering in hot gas \citep{tsai1995} %(Tsai 1995) 
or be completely destroyed by supernova shocks \citep{valiante2011}. %(Valiante 2011).  
It should be finally noted that episodes of star formation occurring in dust polluted gas remove part of the gas mass through astration. 
%when metals in their enriched ejecta condense in dust seeds, that later grow into dust grains in the ISM. On the other hand, dust destruction is primarily caused by SN shock waves and astration. 
The physical nature \citep{ceccarelli2018} %(Ceccarelli 2018) 
and  the efficiency and timescales of these processes are largely unconstrained and difficult to derive from micro-physical models, and dust properties and evolution remain largely unconstrained and difficult to measure. However, many recent hydrodynamical models phenomenologically incorporating dust formation from stellar sources and evolution in various ISM phases have shown that it is possible to reproduce the total estimated gas mass in a remarkable redshift range \citep{mckinnon2017, aoyama2018, gjergo2018, graziani2020}.%(McKinnon 2017, Aoyama 2018, Gjergo 2018, Graziani 2020) .
Dust production from AGB stars alone cannot certainly explain the large amounts of dust observed in galaxies at the reionization epoch \citep[$z = 8.4$,][]{laporte2017}, since the Universe was merely half billion years old and stars did not have enough time to evolve, but has to be accounted for with the production of dust seeds in supernova ejects.
%The efficiency and timescales of these processes are determined by micro-physical parameters that are largely unconstrained and difficult to measure (such as dust composition and structure, grain size distribution, sticking coefficients), and therefore we are still far from understanding the dust evolution in galaxies as a function of global parameters such as metallicity, SFR or gas mass. For instance, 
%dust production from AGB stars alone cannot explain the large amounts of dust observed in galaxies at the reionization epoch \citep[$z = 8.4$,][]{laporte2017}, since the Universe was merely a few billion years old and stars did not have enough time to evolve. 
The condensation of heavy elements in supernovae (SNe) ejecta is thought to be the main channel of dust production at high-z \citep{todini2001, bianchi2007, mancini2015, ferrara2016}. Still, the net production and composition of dust in SNe is unknown even in the nearest sources \citep{matsuura2019}, due to the lack of a sensitive far-IR spectroscopic observatory. { The estimated sensitivity %of \textit{SPICA} will 
of a space cryogenic IR telescope would} allow us to probe the dust properties and its composition from the local Universe to high redshift galaxies, through spectroscopic measurements of various solid-state features in the IR and the modelling of their spectral energy distribution.

\section{Synergies with \textit{JWST} and ALMA}\label{syn}

The \textit{SPace Infrared telescope for Cosmology and Astrophysics} \citep[\textit{SPICA},][]{swinyard2009, nakagawa2014, sibthorpe2015, roelfsema2018}, as described in the {\it preface} of this article, { would have} achieved a gain of over two orders of magnitude in spectroscopic sensitivity in the mid/far-IR compared to \textit{Herschel} and \textit{Spitzer}. It { would have} covered the large spectral gap between the \textit{James Webb Space Telescope} \citep[\textit{JWST},][]{gardner2006}, and what is currently observed at the Atacama Large Millimeter/submillimeter Array \citep[ALMA,][]{wootten2009}, allowing us to detect the very rich mid-IR rest frame spectra of galaxies at the key phase of their evolution, when the bulk of stars and SMBHs were assembled. 
%{\color{red} MOVE THIS SOMEWHERE ELSE
A { SPICA-like observatory would}  be complementary to \textit{JWST}, because it would observe the mid-IR spectra of galaxies at the comic noon, while \textit{JWST-MIRI} will obtain the same spectral range in the local Universe. Similarly, { it would} be complementary to ALMA, by covering an adjacent portion of the cosmic time, from z$\sim$1 to z$\sim$3, especially taking into account the science drivers of the ALMA 2030 Roadmap, which include ``tracing the cosmic evolution of key elements from the first galaxies (z$>$10) to the peak of star formation (z$\sim$2-4), by detecting their cooling lines'' \mbox{\citep{carpenter2020}}.

An example of how a { SPICA-like observatory} and ALMA will be complementary is suggested by the recent ALMA studies, which indicate that \textit{HST}-dark/NIR-dark, faint sub-mm galaxies are more ubiquitous than we thought before, i.e., although fainter than the classical SMGs, they exhibit a much higher space density \citep[e.g.][]{franco2018, yamaguchi2019, wang.t.2019, gruppioni2020}. These galaxies seem to be an important population as a progenitor of present-day massive elliptical galaxies, but their true redshift distributions are largely unexplored due to the difficulty of spectroscopy observations. A fraction of them is expected to be around \textit{z} $\sim 2 - 4$, and given the expected sensitivity of a { SPICA-like observatory}, they can be a good target for spectroscopic follow-up observations, while others can be uncovered by the proposed unbiased 3D survey (ultra-deep, see section \ref{sec:strategy}) via e.g., PAH 7.7/6.2$\mu$m features. Furthermore, co-growth of SMBH among such optically/NIR-dark faint submm galaxies is another important issue that is not yet fully addressed. We see that a { SPICA-like observatory would be} enough sensitive to detect [OIV]25.89$\mu$m up to $z = 4$ (see, e.g., Fig.\,\ref{fig:SAFARI_sample}b) from these optically/NIR-dark ALMA sources because they tend to be bright LIRGs/faint ULIRGs objects. 

\section{Summary}\label{sum}

In this work we have discussed the results { obtained for the preparation of the {\it SPICA} mission: in particular what can be achieved with} the deep spectroscopic 3-dimensional surveys { planned for the mid-IR camera-spectrometer} SMI in low spectral resolution over $1$--$15\, \rm{deg^2}$ fields, plus a hyper-deep survey of a fraction of square degree (0.033 deg$^2$). { We have also addressed the potential of pointed spectroscopic follow-up observations with mid- and far-IR grating spectrometers in the context of galaxy evolution studies.} Our main results can be summarized as follows:
\begin{enumerate}

\item The deepest survey that we have considered, the hyper-deep spectrophotometric survey of $120\, \rm{arcmin^2}$, would be able to detect star forming galaxies about two orders of magnitude below the knee of the luminosity function (LF) at various cosmic times up to z$\sim$4. It would also detect either low metallicity galaxies and AGN about one order of magnitude below the knee of the LF up to z$\sim$2.0 for LMG and up to z$\sim$3.0 for AGN.

\item The ultra-deep spectrophotometric surveys of $1\, \rm{deg^2}$ would be able to detect {\it spectroscopically} through PAH features all star forming galaxies detected {\it photometrically} in the deepest {\it Herschel} cosmological fields, down to one dex below the knee of their luminosity functions up to $z \sim 3.5$. These surveys can reach a depth well below the {\it Herschel} photometric limit, reaching spectroscopically star forming galaxies of $L \sim 10^{10}\, \rm{L_\odot}$ at $z \sim 2$, and $L < 10^{11}\, \rm{L_\odot}$ at $z \sim 3$.

\item The same ultra-deep spectrophotometric surveys would detect AGN, through high-ionization fine-structure lines, at the knee of their luminosity functions at each redshift up to $z \sim 3$.

%\item The same predictions show that the spectroscopic detections of AGN, through high-ionization fine-structure lines, have an higher limit of order of $0.5$--$0.8\, \rm{dex}$ more luminous than the population at the knee of the luminosity function. Still, these AGN will be detected through PAH bands produced by the star formation in their host galaxies.
\item In order to obtain a thorough spectroscopic characterization of the evolution of galaxies {\it in the 3-dimensional domain} at the cosmic noon ($1 < z < 3$) and beyond ($z \sim 4$) a suitable sample will need to cover the main physical parameters (Luminosity, SFR, BHAR, galaxy stellar mass) and -- even more importantly -- will need to select galaxies in both over-dense and sub-dense regions. Numerically, it will have to include about 800 -- 1200 galaxies. 

\item {Mid- to far-IR spectroscopic} observations of this sample of galaxies will trace galaxy evolution all the way down to $\sim 0.5\, \rm{dex}$ below the knee of the luminosity functions of star forming galaxies and AGN by measuring the SFR, the BHAR and the metallicity as a function of cosmic time. 

\item {Mid- to far-IR} spectroscopic follow-up observations will also be used to {characterize the physical properties of the ISM in low-metallicity environments, both in star forming galaxies and AGN at high-$z$}.
%A new calibration of the fine-structure line luminosities is derived for low metallicity galaxies, as a proxy for the ISM characteristics at high-$z$. \textit{SPICA} will measure the SFR in M.-S. star forming low metallicity galaxies through bright mid-ionisation lines such as [\ion{Ne}{iii}]$\rm 15\micron$ and [\ion{S}{iv}]$\rm 10.5\micron$.
\end{enumerate}

\vspace {0.1cm}

\begin{acknowledgements}
This paper is dedicated to the memory of Bruce Swinyard, who initiated the \textit{SPICA} project in Europe, but unfortunately died on 22 May 2015 at the age of 52. He was ISO-LWS calibration scientist, Herschel-SPIRE instrument scientist, first European PI of \textit{SPICA} and first design lead of SAFARI. We acknowledge the whole \textit{SPICA} Collaboration Team, as without its multi-year efforts and work this paper could not have been possible. We also thank the \textit{SPICA} Science Study Team appointed by ESA and the \textit{SPICA} Galaxy Evolution Working Group. LS and JAFO acknowledge financial support by the Agenzia Spaziale Italiana (ASI) under the research contract 2018-31-HH.0. AAH acknowledges support from grants PGC2018-094671-B-I00 (MCIU/AEI/FEDER,UE) and No. MDM-2017-0737 at Unidad de Excelencia Mar\'{\i}a de Maeztu- Centro de Astrobiolog\'{\i}a (INTA-CSIC). FJC acknowledges financial support from the Spanish Ministry MCIU under project RTI2018-096686-B-C21 (MCIU/AEI/FEDER/UE), cofunded by FEDER funds and from the Agencia Estatal de Investigaci\'on, Unidad de Excelencia Mar\'ia de Maeztu, ref. MDM-2017-0765. HD acknowledges financial support from the Spanish Ministry of Science, Innovation and Universities (MICIU) under the 2014 Ram\'on y Cajal program RYC-2014-15686 and AYA2017-84061-P, the later one co-financed by FEDER (European Regional Development Funds). AL acknowledges the support from Comunidad de Madrid through the Atracci\'on de Talento grant 2017-T1/TIC-5213. MPS acknowledges support from the Comunidad de Madrid, Spain, through Atracci\'on de Talento Investigador Grant 2018-T1/TIC-11035.
\end{acknowledgements}

\bibliographystyle{pasa-mnras}
\bibliography{spica_cosmological_surveys_rev}
%\onecolumn
%\appendix
%\begin{appendix}
%\section*{Survey expected detections}
%\input{counts_table_revised.tex}
%\end{appendix}

\begin{landscape}
\begin{table}[ht]
\caption{New calibration obtained for fine-structure lines and PAH features as a function of L$_{IR}$. }\label{tab:linecal}
\begin{center}
\resizebox{\linewidth}{!}{
\begin{tabular}{l|ccccl|ccccl|ccccl}
\bf Line/band ($\mu$m) &  & \bf AGN &  &  &  &  & \bf Star Formation &  &  &  &  & \bf Low Met. Galaxies &  &  &  \\[0.1cm]
  & $a$ $\pm$ $\delta a$ & $b$ $\pm$ $\delta b$ & N & $r$ & ref. & $a$ $\pm$ $\delta a$ & $b$ $\pm$ $\delta b$ & N & $r$ & ref. & $a$ $\pm$ $\delta a$ & $b$ $\pm$ $\delta b$ & N & $r$ & ref. \\ \hline
PAH 6.2 & 1.24$\pm$0.11 & -3.48$\pm$0.35 & 43 & 0.87 & (1) & 1.14$\pm$0.18 & -2.90$\pm$0.62 & 22 & 0.81 & (2) & -- & -- & -- & -- &  \\
{[NeVI]}7.65 & 1.06$\pm$0.22 & -3.59$\pm$0.75 & 8 & 0.87 & (3) & -- & -- & -- & -- &  & -- & -- & -- & -- &  \\
PAH 7.7 & 1.20$\pm$0.11 & -2.73$\pm$0.35 & 49 & 0.84 & (1) & 1.08$\pm$0.17 & -2.43$\pm$0.59 & 22 & 0.82 & (2) & -- & -- & -- & -- &  \\
PAH 8.6 & 1.06$\pm$0.10 & -2.73$\pm$0.33 & 43 & 0.85 & (1) & 1.17$\pm$0.28 & -3.48$\pm$0.96 & 21 & 0.69 & (2) & -- & -- & -- & -- &  \\
{[SIV]}10.51 & 1.02$\pm$0.09 & -3.96$\pm$0.32 & 74 & 0.79 & (4,5) & 1.25$\pm$0.66 & -5.81$\pm$2.19 & 9 & 0.55 & (6) & 0.87$\pm$0.07 & -2.83$\pm$0.12 & 32 & 0.92 & (7) \\
PAH 11.3 & 1.13$\pm$0.11 & -3.01$\pm$0.35 & 47 & 0.83 & (1) & 1.12$\pm$0.19 & -2.87$\pm$0.65 & 22 & 0.79 & (2) & -- & -- & -- & -- &  \\
{[NeII]}12.81 & 1.04$\pm$0.06 & -3.74$\pm$0.21 & 85 & 0.88 & (4,5) & 1.27$\pm$0.03 & -3.98$\pm$0.10 & 43 & 0.87 & (6,8) & 1.19$\pm$0.06 & -4.01$\pm$0.13 & 22 & 0.97 & (7) \\
{[NeV]}14.32 & 1.00$\pm$0.09 & -3.85$\pm$0.31 & 73 & 0.79 & (4,5) & -- & -- & -- & -- &  & -- & -- & -- & -- &  \\
{[NeIII]}15.56 & 1.09$\pm$0.09 & -3.96$\pm$0.30 & 87 & 0.81 & (4,5) & 1.03$\pm$0.04 & -3.95$\pm$0.12 & 43 & 0.74 & (6,8) & 1.01$\pm$0.04 & -2.93$\pm$0.08 & 30 & 0.97 & (7) \\
PAH 17 & 0.98$\pm$0.10 & -2.64$\pm$0.33 & 28 & 0.87 & (1) & 1.16$\pm$0.17 & -3.18$\pm$0.58 & 22 & 0.83 & (2) & -- & -- & -- & -- &  \\
{[SIII]}18.71 & 1.01$\pm$0.08 & -3.93$\pm$0.26 & 69 & 0.84 & (4,5) & 1.09$\pm$0.15 & -4.07$\pm$0.49 & 15 & 0.89 & (6) & 1.02$\pm$0.05 & -3.23$\pm$0.09 & 27 & 0.98 & (7) \\
{[NeV]}24.32 & 0.98$\pm$0.09 & -3.73$\pm$0.31 & 65 & 0.81 & (4,5) & -- & -- & -- & -- &  & -- & -- & -- & -- &  \\
{[OIV]}25.89 & 1.08$\pm$0.11 & -3.64$\pm$0.38 & 80 & 0.74 & (4,5) & 0.99$\pm$0.15 & -4.38$\pm$0.44 & 24 & 0.72 & (6,8) & 0.73$\pm$0.09 & -3.42$\pm$0.19 & 17 & 0.89 & (7) \\
{[SIII]}33.48 & 1.02$\pm$0.06 & -3.58$\pm$0.19 & 74 & 0.81 & (4,5) & 0.92$\pm$0.05 & -3.13$\pm$0.16 & 41 & 0.63 & (6,8) & 1.02$\pm$0.06 & -3.09$\pm$0.12 & 22 & 0.97 & (7) \\
{[SiII]}34.81 & 1.09$\pm$0.06 & -3.61$\pm$0.22 & 72 & 0.80 & (4,5) & 1.31$\pm$0.02 & -3.60$\pm$0.06 & 43 & 0.86 & (6,8) & 0.93$\pm$0.08 & -2.99$\pm$0.15 & 22 & 0.93 & (7)
\end{tabular}
}
\end{center}
%\begin{tablenotes}
      \small
Notes: For each class of objects, are reported the slope (a) and intercept (b) of the linear correlation with relative errors, the number of data from which the linear correlation was calculated (N) and the Pearson correlation coefficient (r). Galaxies from \citet{gallimore2010} are type 1 AGN (including those with hidden broad lines), in order to avoid the stronger star-formation activity in type 2 AGN contributing to the PAH emission. From \citet{tommasin2008,tommasin2010} we used galaxies dominated by the AGN component ($> 90\%$), as reported in these works. From \citet{bernardsalas2009} and \citet{goulding2009} we adopted all star-forming galaxies without high-excitation emission lines from [\ion{Ne}{v}]$_{\rm 14.3,24.3 \mu m}$ which is indicative of AGN activity.\\
References: (1) \citet{gallimore2010}, (2) \citet{brandl2006}, (3) \citet{sturm2002}, (4) \citet{tommasin2008}, (5) \citet{tommasin2010}, (6) \citet{bernardsalas2009}, (7) \citet{cormier2015}, (8) \citet{goulding2009}

%\end{tablenotes}

\end{table}
\end{landscape}

\begin{landscape}
\begin{table}[ht]
\caption{Total number of objects present in a 15 deg$^{2}$ area according to the luminosity functions by \citet{wang2019} in each redshift-luminosity bin (first line in bold-face). For each bin are also reported (in the second line) the fractions of detections of AGN through fine-structure lines (left), of star forming galaxies through PAH features (center) and of low metallicity galaxies through fine-structure lines.}\label{tab:deep_counts}
\centering
\begin{tabular}{lcccccccccccccccccccccccc}
	Log(L$_{IR}/$L$_{\odot}$) & \multicolumn{3}{c}{0-0.5} & \multicolumn{3}{c}{0.5-1.0} & \multicolumn{3}{c}{1.0-1.5} & \multicolumn{3}{c}{1.5-2.0} & \multicolumn{3}{c}{2.0-2.5} & \multicolumn{3}{c}{2.5-3.0} & \multicolumn{3}{c}{3.0-3.5} & \multicolumn{3}{c}{3.5-4.0} \\ \hline
	\multirow{2}{*}{13.0-13.5} & \multicolumn{3}{c}{-} & \multicolumn{3}{c}{-} & \multicolumn{3}{c}{-} & \multicolumn{3}{c}{-} & \multicolumn{3}{c}{-} & \multicolumn{3}{c}{-} & \multicolumn{3}{c}{-} & \multicolumn{3}{c}{-} \\
	&  &  &  &  &  &  &  &  &  &  &  &  &  &  &  &  &  &  &  &  &  &  &  &  \\
	\multirow{2}{*}{12.5-13.0} & \multicolumn{3}{c}{-} & \multicolumn{3}{c}{-} & \multicolumn{3}{c}{\textbf{3}} & \multicolumn{3}{c}{\textbf{12}} & \multicolumn{3}{c}{\textbf{23}} & \multicolumn{3}{c}{\textbf{45}} & \multicolumn{3}{c}{\textbf{27}} & \multicolumn{3}{c}{\textbf{14}} \\
	&  &  &  &  &  &  & 1 & 1 & 1 & 1 & 1 & 1 & 1 & 1 & 0.4 & 1 & 1 & - & 1 & 1 & - & 0.1 & 0.2 & - \\ \cline{23-25} 
	\multirow{2}{*}{12.0-12.5} & \multicolumn{3}{c}{-} & \multicolumn{3}{c}{\textbf{29}} & \multicolumn{3}{c}{\textbf{245}} & \multicolumn{3}{c}{\textbf{600}} & \multicolumn{3}{c}{\textbf{901}} & \multicolumn{3}{c}{\textbf{1353}} & \multicolumn{3}{c}{\textbf{873}} & \multicolumn{3}{c}{\textbf{573}} \\
	&  &  &  & 1 & 1 & 1 & 1 & 1 & 1 & 1 & 1 & 1 & 1 & 1 & - & 0.3 & 1 & - & - & 1 & - & - & 1 & - \\ \cline{17-22}
	\multirow{2}{*}{11.5-12.0} & \multicolumn{3}{c}{\textbf{18}} & \multicolumn{3}{c}{\textbf{1200}} & \multicolumn{3}{c}{\textbf{5752}} & \multicolumn{3}{c}{\textbf{9504}} & \multicolumn{3}{c}{\textbf{10279}} & \multicolumn{3}{c}{\textbf{11951}} & \multicolumn{3}{c}{\textbf{8325}} & \multicolumn{3}{c}{\textbf{6692}} \\
	& 1 & 1 & 1 & 1 & 1 & 1 & 1 & 1 & 1 & 0.6 & 1 & 0.3 & 0.1 & 1 & - & - & 1 & - & - & 1 & - & - & 0.6 & - \\ \cline{11-16}
	\multirow{2}{*}{11.0-11.5} & \multicolumn{3}{c}{\textbf{576}} & \multicolumn{3}{c}{\textbf{14817}} & \multicolumn{3}{c}{\textbf{45940}} & \multicolumn{3}{c}{\textbf{52468}} & \multicolumn{3}{c}{\textbf{40424}} & \multicolumn{3}{c}{\textbf{38830}} & \multicolumn{3}{c}{\textbf{28955}} &  &  &  \\
	& 1 & 1 & 1 & 1 & 1 & 1 & 0.1 & 1 & 0.4 & - & 1 & - & - & 1 & - & - & 1 & - & - & 0.3 & - &  &  &  \\ \cline{5-10}
	\multirow{2}{*}{10.5-11.0} & \multicolumn{3}{c}{\textbf{5599}} & \multicolumn{3}{c}{\textbf{65369}} & \multicolumn{3}{c}{\textbf{147861}} & \multicolumn{3}{c}{\textbf{129229}} & \multicolumn{3}{c}{\textbf{78949}} & \multicolumn{3}{c}{\textbf{}} &  &  &  &  &  &  \\
	& 1 & 1 & 1 & - & 1 & 1 & - & 1 & - & - & 1 & - & - & 0.6 & - &  &  &  &  &  &  &  &  &  \\ \cline{2-4}
	\multirow{2}{*}{10.0-10.5} & \multicolumn{3}{c}{\textbf{20937}} & \multicolumn{3}{c}{\textbf{140656}} & \multicolumn{3}{c}{\textbf{273947}} & \multicolumn{3}{c}{\textbf{206049}} & \multicolumn{3}{c}{\textbf{}} & \multicolumn{3}{c}{\textbf{}} &  &  &  &  &  &  \\
	& 1 & 1 & 1 & - & 1 & 0.2 & - & 0.8 & - & - & 0.2 & - &  &  &  &  &  &  &  &  &  &  &  &  \\
	\multirow{2}{*}{9.5-10} & \multicolumn{3}{c}{\textbf{42288}} & \multicolumn{3}{c}{\textbf{213956}} & \multicolumn{3}{c}{\textbf{}} & \multicolumn{3}{c}{\textbf{}} & \multicolumn{3}{c}{\textbf{}} &  &  &  &  &  &  &  &  &  \\
	& 0.2 & 1 & 1 & - & 0.2 & - &  &  &  &  &  &  &  &  &  &  &  &  &  &  &  &  &  &  \\
	\multirow{2}{*}{9-9.5} & \multicolumn{3}{c}{\textbf{63713}} & \multicolumn{3}{c}{\textbf{}} & \multicolumn{3}{c}{\textbf{}} &  &  &  &  &  &  &  &  &  &  &  &  &  &  &  \\
	& - & 0.8 & 0.8 &  &  &  &  &  &  &  &  &  &  &  &  &  &  &  &  &  &  &  &  & 
\end{tabular}
\end{table}
\end{landscape}
%\vspace{0.5cm}
\begin{landscape}
\begin{table}[ht]
\caption{Total number of objects present in a 1 deg$^{2}$ area according to the luminosity functions by \citet{wang2019} in each redshift-luminosity bin (first line in bold-face). For each bin are also reported (in the second line) the fractions of detections of AGN through fine-structure lines (left), of star forming galaxies through PAH features (center) and of   low metallicity galaxies through fine-structure lines.}\label{tab:ultradeep_counts}
\centering
\begin{tabular}{lcccccccccccccccccccccccc}
	%\hline
	Log(L$_{IR}/$L$_{\odot}$) & \multicolumn{3}{c}{0-0.5} & \multicolumn{3}{c}{0.5-1.0} & \multicolumn{3}{c}{1.0-1.5} & \multicolumn{3}{c}{1.5-2.0} & \multicolumn{3}{c}{2.0-2.5} & \multicolumn{3}{c}{2.5-3.0} & \multicolumn{3}{c}{3.0-3.5} & \multicolumn{3}{c}{3.5-4.0} \\ \hline
	\multirow{2}{*}{13.0-13.5} & \multicolumn{3}{c}{-} & \multicolumn{3}{c}{-} & \multicolumn{3}{c}{-} & \multicolumn{3}{c}{\textbf{1}} & \multicolumn{3}{c}{\textbf{2}} & \multicolumn{3}{c}{\textbf{3}} & \multicolumn{3}{c}{\textbf{2}} & \multicolumn{3}{c}{\textbf{1}} \\
	&  &  &  &  &  &  &  &  &  & 1 & 1 & 1 & 1 & 1 & 1 & 1 & 1 & - & 1 & 1 & - & 1 & 1 & - \\
	\multirow{2}{*}{12.5-13.0} & \multicolumn{3}{c}{-} & \multicolumn{3}{c}{\textbf{2}} & \multicolumn{3}{c}{\textbf{16}} & \multicolumn{3}{c}{\textbf{40}} & \multicolumn{3}{c}{\textbf{60}} & \multicolumn{3}{c}{\textbf{90}} & \multicolumn{3}{c}{\textbf{58}} & \multicolumn{3}{c}{\textbf{38}} \\
	&  &  &  & 1 & 1 & 1 & 1 & 1 & 1 & 1 & 1 & 1 & 1 & 1 & 1 & 1 & 1 & - & 1 & 1 & - & 0.2 & 1 & - \\ \cline{23-25} 
	\multirow{2}{*}{12.0-12.5} & \multicolumn{3}{c}{\textbf{1}} & \multicolumn{3}{c}{\textbf{80}} & \multicolumn{3}{c}{\textbf{383}} & \multicolumn{3}{c}{\textbf{634}} & \multicolumn{3}{c}{\textbf{685}} & \multicolumn{3}{c}{\textbf{797}} & \multicolumn{3}{c}{\textbf{555}} & \multicolumn{3}{c}{\textbf{446}} \\
	& 1 & 1 & 1 & 1 & 1 & 1 & 1 & 1 & 1 & 1 & 1 & 1 & 1 & 1 & - & 0.7 & 1 & - & 0.1 & 1 & - & - & 1 & - \\ \cline{17-22}
	\multirow{2}{*}{11.5-12.0} & \multicolumn{3}{c}{\textbf{38}} & \multicolumn{3}{c}{\textbf{988}} & \multicolumn{3}{c}{\textbf{3063}} & \multicolumn{3}{c}{\textbf{3498}} & \multicolumn{3}{c}{\textbf{2695}} & \multicolumn{3}{c}{\textbf{2589}} & \multicolumn{3}{c}{\textbf{1930}} & \multicolumn{3}{c}{\textbf{1787}} \\
	& 1 & 1 & 1 & 1 & 1 & 1 & 1 & 1 & 1 & 1 & 1 & 0.4 & 0.3 & 1 &  & - & 1 & - & - & 1 & - & - & 1 & - \\ \cline{11-16}
	\multirow{2}{*}{11.0-11.5} & \multicolumn{3}{c}{\textbf{373}} & \multicolumn{3}{c}{\textbf{4358}} & \multicolumn{3}{c}{\textbf{9991}} & \multicolumn{3}{c}{\textbf{8615}} & \multicolumn{3}{c}{\textbf{5263}} & \multicolumn{3}{c}{\textbf{4598}} & \multicolumn{3}{c}{\textbf{3560}} &  &  &  \\
	& 1 & 1 & 1 & 1 & 1 & 1 & 0.5 & 1 & 0.7 & - & 1 & - & - & 1 & - & - & 1 & - & - & 0.5 & - &  &  &  \\ \cline{5-10}
	\multirow{2}{*}{10.5-11.0} & \multicolumn{3}{c}{\textbf{1396}} & \multicolumn{3}{c}{\textbf{9377}} & \multicolumn{3}{c}{\textbf{18263}} & \multicolumn{3}{c}{\textbf{13737}} & \multicolumn{3}{c}{\textbf{7689}} & \multicolumn{3}{c}{\textbf{6544}} &  &  &  &  &  &  \\
	& 1 & 1 & 1 & 0.2 & 1 & 1 & - & 0.8 & - & - & 1 & - & - & 0.8 & - & - & 0.2 & - &  &  &  &  &  &  \\ \cline{2-4}
	\multirow{2}{*}{10.0-10.5} & \multicolumn{3}{c}{\textbf{2189}} & \multicolumn{3}{c}{\textbf{14264}} & \multicolumn{3}{c}{\textbf{26300}} & \multicolumn{3}{c}{\textbf{19079}} &  &  &  &  &  &  &  &  &  &  &  &  \\
	& 1 & 1 & 1 & - & 1 & 0.6 & - & 1 & - & - & 0.4 & - &  &  &  &  &  &  &  &  &  &  &  &  \\
	\multirow{2}{*}{9.5-10} & \multicolumn{3}{c}{\textbf{4248}} & \multicolumn{3}{c}{\textbf{19628}} &  &  &  &  &  &  &  &  &  &  &  &  &  &  &  &  &  &  \\
	& 0.4 & 1 & 1 & - & 0.4 & - &  &  &  &  &  &  &  &  &  &  &  &  &  &  &  &  &  &  \\
	\multirow{2}{*}{9-9.5} & \multicolumn{3}{c}{\textbf{5843}} &  &  &  &  &  &  &  &  &  &  &  &  &  &  &  &  &  &  &  &  &  \\
	& - & 0.8 & 1 &  &  &  &  &  &  &  &  &  &  &  &  &  &  &  &  &  &  &  &  & 
\end{tabular}
\end{table}
\end{landscape}

\begin{landscape}
\begin{table}[ht]
\caption{Total number of objects present in a 120 arcmin$^{2}$ area according to the luminosity functions by \citet{wang2019} in each redshift-luminosity bin (first line in bold-face). For each bin are also reported (in the second line) the fractions of detections of AGN through fine-structure lines (left), of star forming galaxies through PAH features (center) and of   low metallicity galaxies through fine-structure lines.}\label{tab:hyperdeep_counts}
\centering
\begin{tabular}{lcccccccccccccccccccccccc}
	Log(L$_{IR}/$L$_{\odot}$) & \multicolumn{3}{c}{0-0.5} & \multicolumn{3}{c}{0.5-1.0} & \multicolumn{3}{c}{1.0-1.5} & \multicolumn{3}{c}{1.5-2.0} & \multicolumn{3}{c}{2.0-2.5} & \multicolumn{3}{c}{2.5-3.0} & \multicolumn{3}{c}{3.0-3.5} & \multicolumn{3}{c}{3.5-4.0} \\ \hline
	\multirow{2}{*}{13.0-13.5} & \multicolumn{3}{c}{-} & \multicolumn{3}{c}{-} & \multicolumn{3}{c}{-} & \multicolumn{3}{c}{-} & \multicolumn{3}{c}{-} & \multicolumn{3}{c}{-} & \multicolumn{3}{c}{-} & \multicolumn{3}{c}{-} \\
	&  &  &  &  &  &  &  &  &  &  &  &  &  &  &  &  &  &  &  &  &  &  &  &  \\
	\multirow{2}{*}{12.5-13.0} & \multicolumn{3}{c}{-} & \multicolumn{3}{c}{-} & \multicolumn{3}{c}{\textbf{1}} & \multicolumn{3}{c}{\textbf{1}} & \multicolumn{3}{c}{\textbf{2}} & \multicolumn{3}{c}{\textbf{3}} & \multicolumn{3}{c}{\textbf{2}} & \multicolumn{3}{c}{\textbf{1}} \\
	&  &  &  &  &  &  & 1 & 1 & 1 & 1 & 1 & 1 & 1 & 1 & 1 & 1 & 1 & - & 1 & 1 & - & 1 & 1 & - \\ \cline{23-25} 
	\multirow{2}{*}{12.0-12.5} & \multicolumn{3}{c}{-} & \multicolumn{3}{c}{\textbf{3}} & \multicolumn{3}{c}{\textbf{13}} & \multicolumn{3}{c}{\textbf{21}} & \multicolumn{3}{c}{\textbf{23}} & \multicolumn{3}{c}{\textbf{27}} & \multicolumn{3}{c}{\textbf{19}} & \multicolumn{3}{c}{\textbf{15}} \\
	&  &  &  & 1 & 1 & 1 & 1 & 1 & 1 & 1 & 1 & 1 & 1 & 1 & 1 & 1 & 1 & - & 1 & 1 & - & 1 & 1 & - \\ \cline{17-22}
	\multirow{2}{*}{11.5-12.0} & \multicolumn{3}{c}{\textbf{1}} & \multicolumn{3}{c}{\textbf{33}} & \multicolumn{3}{c}{\textbf{102}} & \multicolumn{3}{c}{\textbf{117}} & \multicolumn{3}{c}{\textbf{90}} & \multicolumn{3}{c}{\textbf{86}} & \multicolumn{3}{c}{\textbf{64}} & \multicolumn{3}{c}{\textbf{60}} \\
	& 1 & 1 & 1 & 1 & 1 & 1 & 1 & 1 & 1 & 1 & 1 & 1 & 1 & 1 & 1 & 1 & 1 & - & 0.5 & 1 & - & - & 0.7 & - \\ \cline{11-16}
	\multirow{2}{*}{11.0-11.5} & \multicolumn{3}{c}{\textbf{12}} & \multicolumn{3}{c}{\textbf{145}} & \multicolumn{3}{c}{\textbf{333}} & \multicolumn{3}{c}{\textbf{287}} & \multicolumn{3}{c}{\textbf{175}} & \multicolumn{3}{c}{\textbf{153}} & \multicolumn{3}{c}{\textbf{119}} &  &  &  \\
	& 1 & 1 & 1 & 1 & 1 & 1 & 1 & 1 & 1 & 1 & 1 & 1 & 0.8 & 1 & 0.2 & 0.2 & 1 & - & - & 0.3 & - &  &  &  \\ \cline{5-10}
	\multirow{2}{*}{10.5-11.0} & \multicolumn{3}{c}{\textbf{47}} & \multicolumn{3}{c}{\textbf{313}} & \multicolumn{3}{c}{\textbf{609}} & \multicolumn{3}{c}{\textbf{458}} & \multicolumn{3}{c}{\textbf{256}} & \multicolumn{3}{c}{\textbf{218}} &  &  &  &  &  &  \\
	& 1 & 1 & 1 & 1 & 1 & 1 & 1 & 1 & 1 & 0.4 & 1 & 0.4 & - & 1 & - & - & 1 & - &  &  &  &  &  &  \\ \cline{2-4}
	\multirow{2}{*}{10.0-10.5} & \multicolumn{3}{c}{\textbf{94}} & \multicolumn{3}{c}{\textbf{475}} & \multicolumn{3}{c}{\textbf{877}} & \multicolumn{3}{c}{\textbf{636}} & \multicolumn{3}{c}{\textbf{350}} & \multicolumn{3}{c}{\textbf{296}} &  &  &  &  &  &  \\
	& 1 & 1 & 1 & 0.8 & 1 & 1 & - & 1 & 0.6 & - & 1 & - & - & 1 & - & - & 0.6 & - &  &  &  &  &  &  \\
	\multirow{2}{*}{9.5-10} & \multicolumn{3}{c}{\textbf{142}} & \multicolumn{3}{c}{\textbf{654}} & \multicolumn{3}{c}{\textbf{1194}} & \multicolumn{3}{c}{\textbf{861}} & \multicolumn{3}{c}{\textbf{472}} &  &  &  &  &  &  &  &  &  \\
	& 1 & 1 & 1 & - & 1 & 1 & - & 1 & - & - & 0.8 & - & - & 0.2 & - &  &  &  &  &  &  &  &  &  \\
	\multirow{2}{*}{9-9.5} & \multicolumn{3}{c}{\textbf{195}} & \multicolumn{3}{c}{\textbf{654}} & \multicolumn{3}{c}{\textbf{1612}} &  &  &  &  &  &  &  &  &  &  &  &  &  &  &  \\
	& 0.8 & 1 & 1 & - & 0.8 & - & - & 0.4 & - &  &  &  &  &  &  &  &  &  &  &  &  &  &  &  \\
	\multirow{2}{*}{8.5-9} & \multicolumn{3}{c}{\textbf{263}} &  &  &  &  &  &  &  &  &  &  &  &  &  &  &  &  &  &  &  &  &  \\
	& - & 1 & 1 &  &  &  &  &  &  &  &  &  &  &  &  &  &  &  &  &  &  &  &  & 
\end{tabular}
\end{table}
\end{landscape}

\begin{appendix}

\section*{Affiliations}
\vspace{0.1cm}
\affil{$^1$ Istituto di Astrofisica e Planetologia Spaziali (INAF--IAPS), Via Fosso del Cavaliere 100, I-00133 Roma, Italy}
\affil{$^2$ Dipartimento di Fisica, Universit\'a di Roma La Sapienza, P.le A. Moro 2, I-00185 Roma, Italy}
\affil{$^3$ Centro de Astrobiolog\'ia (CAB, CSIC-INTA), ESAC Campus, 28692 Villanueva de la Ca\~nada, Madrid, Spain}
\affil{$^4$ Spitzer Science Center, California Institute of Technology, Pasadena, CA 91125}
\affil{$^5$ Osservatorio di Astrofisica e Scienza dello Spazio (INAF--OAS), via Gobetti 93/3, I-40129 Bologna, Italy}
\affil{$^6$ Instituto de F\'isica de Cantabria (CSIC-Universidad de Cantabria), E-39005 Santander, Spain}
\affil{$^7$ Department of Physics and Astronomy, University of California, Irvine, CA 92697, USA}
\affil{$^8$ Instituto de Astrof\'isica de Canarias (IAC), E-38205 La Laguna, Tenerife, Spain }
\affil{$^9$ Universidad de La Laguna (ULL), Dept. de Astrof\'isica, Avd. Astrof\'isico Fco. S\'anchez s/n, E--38206 La Laguna, Spain}%
\affil{$^{10}$ Steward Observatory, University of Arizona, 933 North Cherry Avenue, Tucson, AZ 85721, USA}
\affil{$^{11}$ Laboratoire AIM-Paris-Saclay, CEA/DRF/Irfu, CNRS, Universit\'e Paris Diderot, CEA-Saclay, F-91191 Gif-sur-Yvette, France}
\affil{$^{12}$ Dipartimento di  Fisica e Astronomia, Universit\'a di Padova, vicolo dell'Osservatorio 3, I-35122 Padova, Italy}
\affil{$^{13}$ Universidad de Alcal\'a, Departamento de F\'sica y Matem\'aticas, Campus Universitario, 28871, Alcal\'a de Henares, Madrid, Spain}
\affil{$^{14}$ European Southern Observatory, Karl-Schwarzschild-Strasse 2, D-85748, Garching, Germany}
\affil{$^{15}$ Graduate School of Science, Nagoya University, Furo-cho, Chikusa-ku, Nagoya 464-8602, Japan}
\affil{$^{16}$ Institute of Astronomy, Graduate School of Science, The University of Tokyo, 2-21-1 Osawa, Mitaka, Tokyo 181-0015, Japan}
\affil{$^{17}$ Dark Cosmology Centre, Niels Bohr Institute, University of Copenhagen, Juliane Mariesvej 30, 2100 Copenhagen, Denmark}
\affil{$^{18}$ Department of Physics and Astronomy, UCLA, 430 Portola Plaza, Los Angeles, CA 90095, USA}
\affil{$^{19}$Institute of Space and Astronautical Science, Japan Aerospace Exploration Agency (JAXA), Sagamihara, Kanagawa 252-5210, Japan}
\affil{$^{20}$ Research Center for Space and Cosmic Evolution, Ehime University, Matsuyama 790-8577, Japan}
\affil{$^{21}$ Department of Physics \& Astronomy, Institute for Space Imaging Science, University of Lethbridge, Lethbridge, Alberta T1K 3M4, Canada}
\affil{$^{22}$ Centro de Astrobiolog\'ia (CSIC-INTA), Ctra. de Ajalvir, Km 4, 28850, Torrej\'on de Ardoz, Madrid, Spain}
\affil{$^{23}$ Dipartimento di Fisica e Astronomia, Universit\'a degli Studi di Bologna, Via P. Gobetti 93/2, I-40129 Bologna, Italy}
\affil{$^{24}$ SRON Netherlands Institute for Space Research, Verlengde Hereweg 140A, 9722 AK Groningen, Netherlands}
\affil{$^{25}$ University of Groningen, Kapteyn Astronomical Inst., PO Box 800, Landleven 12, 9700 AV Groningen, Netherlands}
\affil{$^{26}$ School of Physical Sciences, The Open University, Walton Hall, Milton Keynes, MK7 6AA, UK}
\affil{$^{27}$ Astronomical Institute, Tohoku University, 6-3 Aramaki, Aoba-ku, Sendai, Miyagi 980-8578, Japan}

\end{appendix}

\end{document}